\documentclass[twocolumn]{aastex631}
\usepackage{CLASSY_Preamble}

\begin{document}

\submitjournal{AASJournal ApJ}
\shorttitle{CLASSY~XIV}
\shortauthors{James et al.}

\title{CLASSY XIV: The Nitrogen Exception - Multi-Phase Enrichment and Feedback in High-$z$ Analogs \footnote{
Based on observations made with the NASA/ESA Hubble Space Telescope,
obtained from the Data Archive at the Space Telescope Science Institute, which
is operated by the Association of Universities for Research in Astronomy, Inc.,
under NASA contract NAS 5-26555.}}

\author[0000-0003-4372-2006]{Bethan L. James}
\affiliation{ESA for AURA \\
Space Telescope Science Institute \\
3700 San Martin Drive, Baltimore, MD 21218, USA}

\author[0000-0002-2764-6069]{Valentina Abril-Melgarejo}
\affiliation{Space Telescope Science Institute, 3700 San Martin Drive, Baltimore, MD 21218, USA}
\affiliation{LUX, Observatoire de Paris, Université PSL, CNRS, Sorbonne Université, Meudon, 92190, France}

\author[0000-0002-2644-3518]{Karla Z. Arellano-C\'{o}rdova}
\affiliation{Institute for Astronomy, University of Edinburgh, Royal Observatory, Edinburgh, EH9 3HJ, UK}

\author[0000-0001-9882-1576]{Adarsh Ranjan}
\affiliation{Space Telescope Science Institute, 3700 San Martin Drive, Baltimore, MD 21218, USA}

\author[0000-0002-8809-4608]{Kaelee S. Parker}
\affiliation{Department of Astronomy, The University of Texas at Austin, 2515 Speedway, Stop C1400, Austin, TX 78712, USA}
\affiliation{Cosmic Frontier Center, The University of Texas at Austin, Austin, TX 78712, USA}

\author[0000-0002-4153-053X]{Danielle A. Berg}
\affiliation{Department of Astronomy, The University of Texas at Austin, 2515 Speedway, Stop C1400, Austin, TX 78712, USA}
\affiliation{Cosmic Frontier Center, The University of Texas at Austin, Austin, TX 78712, USA}

\author[0000-0003-2589-762X]{Matilde Mingozzi}
\affiliation{ESA for AURA \\
Space Telescope Science Institute \\
3700 San Martin Drive, Baltimore, MD 21218, USA}

\author[0000-0003-4137-882X]{Alessandra Aloisi}
\affiliation{Space Telescope Science Institute, 3700 San Martin Drive, Baltimore, MD 21218, USA}
\affiliation{Astrophysics Division, Science Mission Directorate, NASA Headquarters, 300 E Street SW, Washington, DC 20546, USA}

\author[0000-0002-0302-2577]{John Chisholm}
\affiliation{Department of Astronomy, The University of Texas at Austin, 2515 Speedway, Stop C1400, Austin, TX 78712, USA}
\affiliation{Cosmic Frontier Center, The University of Texas at Austin, Austin, TX 78712, USA}

\author[0000-0003-1127-7497]{Timothy Heckman}
\affiliation{Center for Astrophysical Sciences, Department of Physics \& Astronomy, Johns Hopkins University, Baltimore, MD 21218, USA}

\author[0000-0002-6586-4446]{Alaina Henry}
\affiliation{Space Telescope Science Institute, 3700 San Martin Drive, Baltimore, MD 21218, USA}

\author[0000-0000-0000-0000]{Svea Hernandez}
\affiliation{ESA for AURA, Space Telescope Science Institute, 3700 San Martin Drive, Baltimore, MD 21218, USA}

\author[0000-0001-5538-2614]{Kristen B. W. McQuinn}
\affiliation{Rutgers University, Department of Physics and Astronomy, 136 Frelinghuysen Road, Piscataway, NJ 08854, USA}
\affiliation{Space Telescope Science Institute, 3700 San Martin Drive, Baltimore, MD 21218, USA}

\author[0000-0002-9217-7051]{Xinfeng Xu}
\affiliation{Center for Astrophysical Sciences, Department of Physics \& Astronomy, Johns Hopkins University, Baltimore, MD 21218, USA}

\author[0000-0000-0000-0000]{Chiaki Kobayashi}
\affiliation{Centre for Astrophysics Research, Department of Physics, Astronomy and Mathematics, University of Hertfordshire, Hatfield, AL10 9AB, UK}

\author[0000-0000-0000-0000]{\& THE CLASSY COLLABORATION}

\correspondingauthor{Bethan L. James} 
\email{bjames@stsci.edu}

\begin{abstract}
We present a first-of-its-kind analysis of the metal content across two interstellar medium (ISM) phases in a sample of 31 local star-forming galaxies from the COS Legacy Archive Spectroscopic SurveY (CLASSY), selected as analogues of high-$z$ systems. Using co-spatial UV absorption and optical emission-line spectroscopy, we compare abundances of N, O, S, and Fe in the low-ionization (neutral) and high-ionization (ionized) gas, providing a multi-phase view of enrichment shortly after the current starburst and over longer timescales when ejecta from previous episodes have cooled and mixed. We find that O and S, produced predominantly in short-lived massive stars, are well mixed between the two phases, with scatter reflecting local inhomogeneities. Fe, predominantly produced by Type Ia supernovae on $\sim$1 Gyr timescales, is higher in the neutral gas, reflecting either delayed mixing of older Fe-enriched material or preferential depletion of Fe from the ionized phase through dust formation in core-collapse supernova ejecta. N exhibits the largest phase offset, with N/H$_{ion}$ systematically $\sim$0.7 dex higher than N/H$_{neu}$, and the magnitude of this offset correlates with stellar mass, metallicity, star-formation rate, and most strongly with the ISM outflow velocity. N/O ratios in the ionized phase rise rapidly within 3–6 Myr relative to the neutral gas, consistent with N enrichment dominated by Wolf–Rayet stars rather than intermediate-mass AGB stars on longer timescales. These results demonstrate that localized stellar feedback, outflows, and phase-dependent mixing collectively regulate the chemical evolution of star-forming galaxies, providing key insight into the extreme N/O abundances recently observed in galaxies at cosmic dawn.
\end{abstract}

\keywords{Dwarf galaxies (416), Ultraviolet astronomy (1736), Galaxy chemical evolution (580), Galaxy spectroscopy (2171), High-redshift galaxies (734), Emission line galaxies (459)}

\section{Introduction} \label{sec:intro}
Chemical evolution, i.e., the production, distribution, and cycling of elements within galaxies, is a fundamentally four-dimensional problem. Observationally we know that metallicity varies spatially across galaxies \citep[e.g.,][]{Poetrodjojo:2018,James:2020,Mast:2014} and extends beyond them, as metals are ejected into the circumgalactic and intergalactic medium. A third dimension arises from the gas phase: metals reside in both cooler ($T\sim100$~K), neutral gas (\hi) and warmer ($T\sim10^4$~K), ionized gas (\hii). The latter is well traced by optical emission lines from  gas, while the former can only be probed through UV interstellar absorption lines. These UV diagnostics are crucial for quantifying the metals locked in the colder ISM - rather than the recently ejected, ionized material - and for connecting present-day enrichment to the star-forming gas reservoir that fuels future generations of stars \citep{Emerick:2020}. 

Elements are produced and injected into the ISM on different timescales depending on their nucleosynthetic origin, defining the fourth dimension of chemical evolution in a physical sense. $\alpha$-elements (O, S, Si) are synthesized predominantly in Type~II~SNe on $\sim$5–10~Myr timescales, whereas C and N arise mainly from AGB stars after $\gtrsim$100~Myr, and Fe-peak elements from Type~Ia~SNe on $\sim$1~Gyr timescales \citep{kobayashi06,Matteucci:2012,kobayashi20}. C and N can also be rapidly released by Wolf–Rayet (WR) stars within $\sim$5~Myr of a burst \citep{Crowther:2007}, whose strong winds expose and expel N- and C-enriched core material into the surrounding ISM.

Within these enrichment cycles, newly produced metals cool, mix, and transition between ionized and neutral phases. The relative abundances in each phase therefore encode the timescale since the most recent burst of star formation. Ionized gas reflects the immediate, localized enrichment, while the neutral phase integrates the chemical record of previous generations. Comparing these abundance patterns allows us to probe how metals are dispersed and homogenized across galaxies \citep{James:2018,Hernandez:2020,Abril-Melgarejo:2024}, providing direct constraints on the physical processes that drive chemical evolution.

Elemental abundance ratios—particularly those tracing elements produced on different timescales—act as powerful diagnostics of evolutionary stage. Ratios such as $\alpha$/Fe and N/O serve as “chemical clocks” distinguishing early, $\alpha$-enhanced systems from more evolved, N-enriched ones \citep{McWilliam:1997,Perez-Montero:2013,berg19a}. When measured in multiple phases, these ratios reveal both ongoing and past enrichment episodes: phase-dependent offsets indicate incomplete mixing, while their evolution with stellar age constrains the characteristic timescales over which newly synthesized material disperses through the ISM \citep{Abril-Melgarejo:2024}.

Simulations suggest that metal mixing in galaxies occurs over $\lesssim$40~Myr to $\sim$0.1–1~Gyr \citep[][and references therein]{Hirai:2017,Emerick:2020,Arabsalmani:2023}, depending on galaxy mass, ISM density, feedback energy, and star-formation rate. In low-mass systems, mixing efficiency increases during periods of high SFR, when stronger galactic outflows and enhanced turbulence promote redistribution \citep{Emerick:2020,Hunter:2022}. Observationally, metal-enriched outflows are a dominant factor shaping the distribution of metals, with outflow metallicities often 10–50 times higher than the ISM \citep{chisholm18b}. Low-mass galaxies, in particular, retain metals inefficiently \citep{McQuinn:2015b}, and a substantial fraction of their metals are found in the CGM or IGM \citep{Zheng:2024}. Conversely, localized conditions can delay mixing: strong density contrasts or turbulent mixing layers can trap newly ejected metals in confined regions, leading to highly localized N enrichment as seen in NGC~5253 and Mrk~996 \citep{Westmoquette:2013,james09,Kumari:2018}.

Despite their importance, direct multi-phase abundance comparisons remain rare because UV absorption studies require high S/N spectroscopy from space-based observatories. To date, only a few systems have been examined in this way. In some, the neutral and ionized gas metallicities agree within uncertainties \citep{Lebouteiller:2009,James:2018,Hernandez:2021}, while in others significant phase offsets are found. In the benchmark study of NGC~5253 by \citet{Abril-Melgarejo:2024}, N/H and O/H in the neutral gas were lower than in the ionized gas by 0.8 and 0.22~dex, respectively. By comparing clusters of different ages, they inferred a $\sim$10–15~Myr mixing timescale between phases—shorter than predicted by simulations—possibly due to active outflows enhancing chemical segregation.

With only a handful of such studies, the mechanisms and timescales governing metal exchange between gas phases remain poorly constrained. Addressing this requires a statistically significant sample of galaxies with spatially matched UV and optical spectroscopy. To this end, we employ data from the COS Legacy Archive Spectroscopic Survey (CLASSY; \citealt{berg:2022,james:2022}), which provides high-resolution UV spectra from HST/COS and co-spatial optical spectra from ground-based facilities. CLASSY targets were selected to mirror the properties of galaxies at $z>7$ \citep[e.g.,][]{steidel14,lefevre15,shapley15,mclure18,Simmonds:2024,Atek:2024}, spanning stellar masses $\log(M_\star/M_\odot)\!\sim\!6$–10, gas-phase metallicities $Z/Z_\odot\!\sim\!0.03$–1.2, SFRs $\log(\mathrm{SFR}/M_\odot\,\mathrm{yr}^{-1})\!\sim\!-3$–2, ionization levels $\mathrm{O}_{32}\!\sim\!-0.3$–1.4, and electron densities $10<n_e<1120$~cm$^{-3}$ . This spectroscopic atlas uniquely enables systematic studies of metal mixing over a broad parameter space that parallels high-$z$ systems, extending beyond individual clusters to integrated galaxy scales.

\begin{table}
    \centering
    {\begin{tabular}{ccc}
        Element & Neutral & Ionized\\
        \hline\hline
        H   & \hi\   & \hii\  (13.6)        \\
        N   & \Ni\  & [\nii]  (14.5)       \\
        O   & \oi  & [\oii] (13.6), [\oiii] (35.1)\\
        Si  & \sIii\ (8.15) &\sIiii\ (16.3), \sIiv\ (33.5)   \\
        P   & \pii\ (10.5) & \\
        S   & \sii\ (10.36)  & [\sii] (10.36), \fsiii\ (23.3) \\
        Fe  & \feii\ (7.90) & [\feii] (7.90), [\feiii] (16.2)
    \end{tabular}}
    \vspace{0.2em}
    \caption{Classification of neutral vs. ionized spectroscopic features measured within this study, where the delineation between the two phases is made according to the dominant ionization stage of hydrogen. The forbidden metal lines listed here are measured from optical emission lines whereas permitted metal lines are those measured from UV absorption lines. Numbers in parenthesis refer to the ionization potential energy required to create that feature in.}\label{tab:ions}
\end{table}

The paper is designed as follows: we provide an overview on the observations in Section~\ref{sec:obs}; details on absorption line fitting in Section~\ref{sec:method}; present the neutral gas abundances and ionization correction factors (ICFs) in \ref{sec:abund}; ionized gas abundances in Section~\ref{sec:ion_abund}; we discuss the comparison of multi-phase metals in Section~\ref{sec:discussion} and investigate properties affecting their distribution; and conclude our findings in Section~\ref{sec:conclusion}.

\section{Sample and Observations}\label{sec:obs}
\subsection{CLASSY}
CLASSY is a Hubble Space Telescope (HST) treasury program with the Cosmic Origins Spectrograph (COS), dedicated to creating the first high-resolution FUV spectral catalog of star-forming galaxies at $z\sim0$.  The targets have redshifts between $z$=0.002--~0.182, whose UV sources have Galaxy Evolution Explorer (GALEX) FUV magnitudes between $m_{FUV}=15.3-19.2$, ranging from 0\farcs11--1\farcs6 in diameter (as measured from the COS NUV acquisition images) and are primarily powered by star-formation \citep[as indicated by UV and optical emission line diagnostics][, hereafter \PVIII]{mingozzi:2024}. A complete description of CLASSY and its scientific goals can be found in \citep[][hereafter \PI]{berg:2022}, including full details of the targets within the sample.
While many properties are measured and presented here for the first time, we also draw on results from previous CLASSY publications, including: \Lya\ column densities from \citet[][hereafter \PVII]{Hu:2023}; outflow velocities from \citet[][hereafter \PIII]{Xu:22}; ionized gas abundances from \citet[][hereafter \PV]{arellano-cordova:2022} and \citet[][hereafter \PXII]{arellano-cordova:2025b}. In particular, we utilize several line properties from \citet[][hereafter\PXI]{Parker:2024}, who performed an in-depth analysis of the neutral gas properties (column density, covering fraction, and outflow velocity) of a selection of low-ionization state (LIS) interstellar UV absorption lines (\oi, \sIii, \sii, \cii, and \alii) of the CLASSY sample. 

\subsection{UV Spectroscopic Data}
Observation details of all the HST/COS datasets included within
CLASSY can be found in \citetalias{berg:2022}, including dataset IDs, gratings, central wavelength settings, position angles, exposure times and rest wavelength coverage. Technical details of the methodologies involved in extracting, reducing, aligning, and coadding of the CLASSY far-ultraviolet (FUV) and near-ultraviolet (NUV) spectra can be found in \citet[][hereafter \PII]{james:2022}. To summarize briefly here, CLASSY HLSP spectra consist of coadded G130M+G160M+G185M COS spectra covering 815--3200~\AA.  All CLASSY datasets were processed in a self-consistent way, in that they were all reduced via version 3.3.11 of the CalCOS pipeline. With regards to the homogeneity of the data reduction procedures employed, CLASSY contains data observed at different Lifetime Positions (LP) on the COS FUV detector (as detailed in table~2 of \citetalias{berg:2022}). All measurements reported in this paper were performed on the CLASSY HLSP coadded spectra provided on the CLASSY-MAST portal: \url{https://archive.stsci.edu/hlsp/classy}.

\subsection{Optical Spectroscopic Data} \label{sec:optical}
Optical spectra for each of the CLASSY targets comprises primarily from  APO/SDSS observations, with additional observations from LBT/MODS, VLT/MUSE/VIMOS, KECK/KCWI and MMT telescopes/instruments. Full details concerning the optical observations and their treatment are reported in \citetalias{berg:2022}, \citet[][hereafter \PIV]{mingozzi:2022}, and \citetalias{arellano-cordova:2022}. \citetalias{arellano-cordova:2022} compares the SDSS, LBT, and MUSE integrated spectra for the galaxies with multiple observations, demonstrating that flux calibration issues or aperture differences do not introduce significant discrepancies in the optical ISM
properties in terms of gas attenuation, density, temperature,
metallicity, and SFRs. 

A full suite of optical emission lines between 3700 and 9100~\AA\ were detected for each galaxy and fit using single or multi-component Gaussian profiles. Full details of the fitting methodology and optical line fluxes can be found in \citetalias{mingozzi:2022} and  \citetalias{arellano-cordova:2022}. Here, for consistency within the sample, we utilize optical properties derived from single, narrow component fits only. This component is also mostly associated with the bulk of ionized gas and provides the most relevant comparison with the ISM gas traced with the absorption line profiles discussed in Section~\ref{sec:method}.

\begin{deluxetable*}{lccc|lccc}
\tablecaption{List of ISM lines fitted within this study and their corresponding theoretical parameters}
\tablehead{
\colhead{Line ID} & \colhead{$\lambda_{lab}^a$ (\AA)} & \colhead{$f^b$} & \colhead{Reference} & 
\colhead{Line ID} & \colhead{$\lambda_{lab}^a$ (\AA)} & \colhead{$f^b$} & \colhead{Reference}} 
\startdata
\Lya\,     & 1215.6701 & 4.1640e-01 & 1 & \sii~1253   & 1253.8110 & 1.0880e-02 & 1  \\  
\Ni~1134.1 & 1134.1653 & 1.3420e-02 & 1 & \sii~1259   & 1259.5190 & 1.6240e-02 & 1  \\    
\Ni~1134.4 & 1134.4149 & 2.6830e-02 & 1 & \pii~1152   & 1152.8180 & 2.3610e-01 & 1  \\      
\Ni~1134.9 & 1134.9803 & 4.0230e-02 & 1 & \feii~1121  & 1121.9748 & 2.0200e-02 & 10 \\           
\Ni~1199   & 1199.5496 & 1.3280e-01 & 1 & \feii~1125  & 1125.4477 & 1.6000e-02 & 10 \\  
\Ni~1200.2 & 1200.2233 & 8.8490e-02 & 1 & \feii~1127  & 1127.0984 & 2.8000e-03 & 10 \\   
\Ni~1200.7 & 1200.7098 & 4.4230e-02 & 1 & \feii~1133  & 1133.6650 & 5.5000e-03 & 10 \\      
\oi~1302   & 1302.1685 & 4.8870e-02 & 1 & \feii~1142  & 1142.3656 & 4.2000e-03 & 10 \\      
\oi~1355   & 1355.5977 & 1.2480e-06 & 1 & \feii~1143  & 1143.2260 & 1.7700e-02 & 10 \\    
\sIii~1190 & 1190.4158 & 2.9300e-01 & 2 & \feii~1144  & 1144.9379 & 1.0600e-01 & 10 \\      
\sIii~1193 & 1193.2897 & 5.8400e-01 & 2 & \feii~1260  & 1260.5330 & 2.5000e-02 & 1  \\  
\sIii~1260 & 1260.4221 & 1.1800e-00 & 2 & \feiii~1122 & 1122.5260 & 7.8840e-02 & 1  \\  
\sIii~1304 & 1304.3702 & 8.6000e-02 & 6 & \feiii~1207 & 1207.0500 & 4.4230e-06 & 1  \\
\pii~1301  & 1301.8740 & 1.2710e-02 & 8 & \feiii~1214 & 1214.5660 & 4.2730e-04 & 1  \\      
\sii~1250  & 1250.5840 & 5.4530e-03 & 1 &             &           &            &    \\    
\enddata
\tablecomments{
\textsc{References.--}
(1)=\citet{Morton:1991}, (2)=\citet{Verner:1994}, (3)=\citet{Wiese:1996}, 
(4)=\citet{Morton:2003}, (5)=\citet{Majumder:2002}, (6)=\citet{Spitzer:1993},
(7)=\citet{Verner:1996}, (8)=\citet{Hibbert:1988}, (9)=\citet{Tayal:1995},
(10)=\citet{Howk:2000}, (11)=\citet{Zsargo:1998}. \\
$^a$ Vacuum wavelengths from \citet{Morton:1991}\\
$^b$ Oscillator strengths from references indicated in column (4).}\label{tab:fvals}
\end{deluxetable*}

\section{Absorption Line Fitting}\label{sec:method}
Absorption features in the CLASSY data were fitted following the methodology applied in \cite{James:2014}, \citet{Abril-Melgarejo:2024} and \citetalias{Parker:2024}. In this process, theoretical Voigt profiles were used to model the absorption lines using the Python package \texttt{VoigtFit} \citep{Krogager:2018}, which models absorption line profiles while taking into account the line spread function (LSF) of the observations by convolving the instrumental LSF with the intrinsic profile of the source. This package allows absorption line fits for each species to be optimized by simultaneously modeling different velocity components of both the target and any contaminating species from the target itself or other systems along the line of sight (e.g., the Milky Way).  For each galaxy we fitted the CLASSY coadded spectra using the appropriate LSF according to the LP of the G130M spectrum (listed in \citetalias{berg:2022}). While G160M and G185M spectral resolutions may be different, none of the ISM lines of interest to this study lie within those gratings. 

This study focuses on the comparison between neutral and ionized gas abundances. As such, only the elements who have species visible in both the UV and optical were included in our absorption profile fitting: nitrogen, sulfur, iron, and oxygen. In the UV, these species would be \Ni, \sii, \feii, \feiii\, and \oi. We also analyze \pii\ as it can be used to predict the \oi\ column density \citep{James:2018}, and \sIii\ and \nIii\ for dust depletion (Section~\ref{sec:depletion}). The optical emission-line counterparts for these species would be \fnii, \fsii\, \fsiii, \ffeiii, \foii, and \foiii. 
We also discard any galaxies for which the \Lya\ absorption profile was either not detected or unconstrained by \citetalias{Hu:2023}, resulting in a final sample of 31 galaxies. 

Before fitting each line, continuum normalization was performed using the continuum fits detailed in \citetalias{Hu:2023} and provided as a HLSP on the CLASSY-MAST portal. Owing to the low-redshift nature of our objects ($z<0.2$), MW ISM lines often contaminate the intrinsic ISM absorption of our individual objects. As such, a careful inspection of each target's species was made to identify all the MW absorption features before fitting.  A model was then created for each species within the target, detailing the velocity structure of the ISM (mostly single velocity components) and of any intervening MW ISM species. 
With this approach, we performed a simultaneous fit to all the lines of a species and, therefore, were able to solve for partial covering by using multiple lines with different values of $f\lambda$, where $f$ is the oscillator strength and $\lambda$ is the rest-frame wavelength of the absorption. 

In the case of \pii, there is only one absorption line detected.
Thus, in order to rule out partial covering, we compared the column density from the \pii\ fit to a strong line (i.e., high $f$) that is not saturated, such as \feii~$\lambda1144$. Since the product of $f$ and column density, $N(X)$, is directly proportional to the optical depth, and because the column density of \pii\ is $\sim$95 times weaker than that of \feii\ ($<N$(\pii)/$N($\feii)$>\sim$0.05) despite $f_{\rm PII}/f_{\rm FeII}$=1.5, we are safe to assume \pii\ is not saturated.  

\begin{figure*}
\begin{center}
    \includegraphics[width=0.9\textwidth, trim=0 0 0 0,  clip=yes]{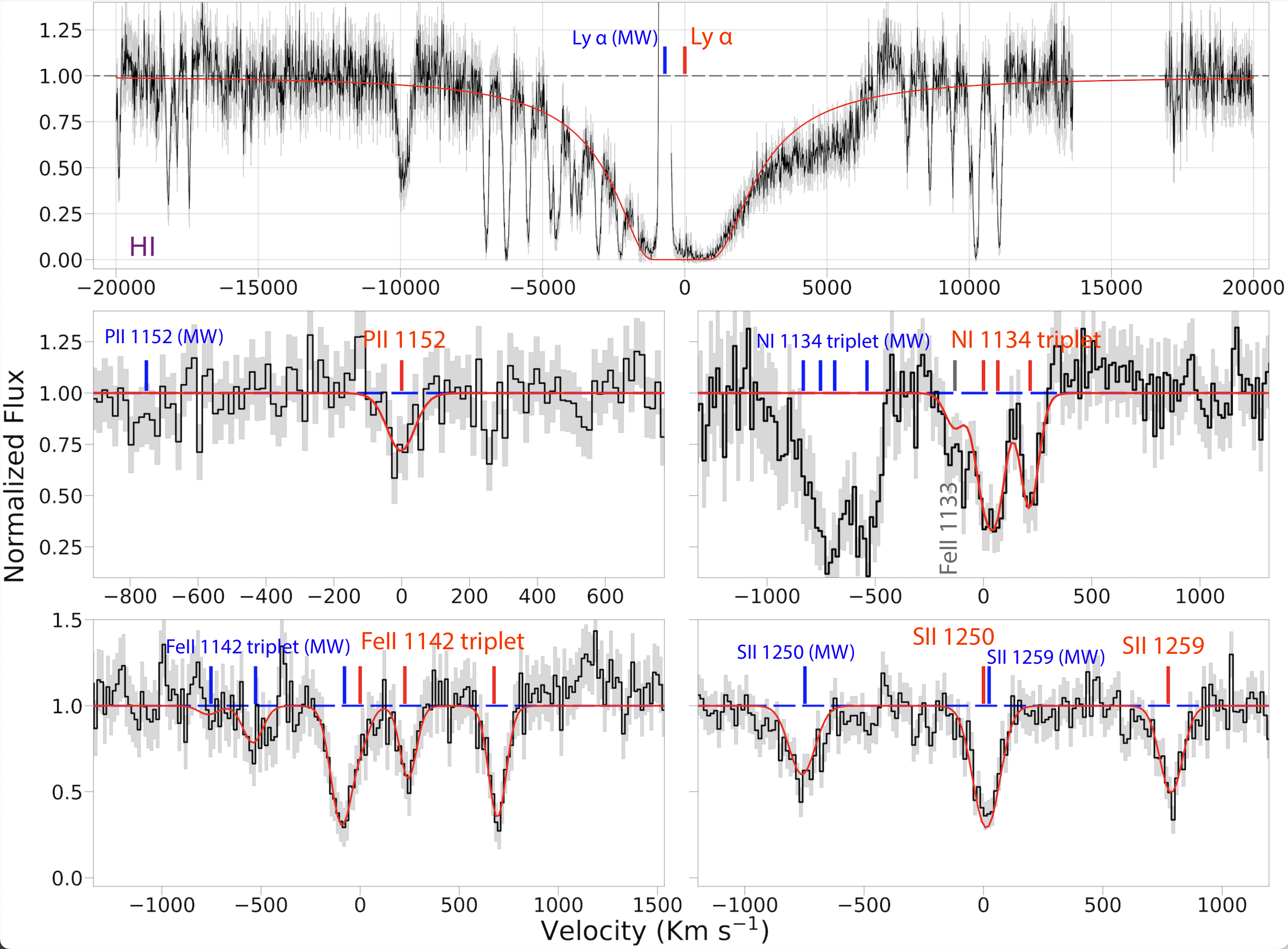}
        \caption{An example of the Voigt-fitting procedure employed within this study for J1225+6109. The top panel shows the CLASSY coadded spectrum up to $\sim1420$~\AA\ (in black, with errors in gray). Lower panels  show the fits for \pii~$\lambda$1152, the \Ni~$\lambda$1134 triplet (1134.1, 1134.4, 1134.9~\AA), \feii~$\lambda1133$, the \feii~$\lambda$1142 triplet (1142, 1143, and 1144~\AA) and \sii~$\lambda$1250, where the galaxy and MW components in each line complex are identified with red and blue vertical markers, respectively. Each panel is centered in velocity space on the lowest wavelength line being fitted for the target within that model. }
\label{fig:fit}
\end{center}
\end{figure*}

To enable an optimized fit, the velocity components ($V_{min}$) and $b$-parameters (i.e., the broadening of a line due to thermal and/or turbulent motion) of different species were not constrained to one another. The best-fit model profiles are estimated using a $\chi^2$ minimization approach, with output parameters including the line column density $N$(X), $b$-parameter, redshift and their uncertainties. An example result from this line-fitting procedure is shown in Figure~\ref{fig:fit}, which shows the best fit model for J1225+6109.

While there are numerous unresolved absorbing components along the line of sight within the COS aperture, the single velocity component fitting is deemed valid here, where the kinematical properties represent an average integration over a distribution of unresolved kinematics \citep[see][]{Clark:2024}. As originally shown by \citet{Jenkins:1986}, complex configurations with multiple unresolved absorbing components can be analyzed jointly by performing a single-velocity profile approximation even when the involved lines have variations in the $b$-parameter and saturation levels (with the exception of very strongly saturated lines). As such, the $b$-parameter in this sense represents the velocity dispersion of a species that is a combination of the thermal broadening, turbulence, and bulk motions (since it is a combination of several velocity components), while the column density is well-constrained.

One of the most challenging issues to deal with in absorption line fitting is saturation. Saturation can manifest itself in three ways: classical saturation, hidden saturation, or `partial covering'. While the former can be easy to identify because the equivalent width of the line increases while the line is optically thick (essentially 'bottoming-out' the line), the latter two are particularly difficult to decipher because while the line appears to be unsaturated, it actually is. This is because either multiple components from different absorbers along the line of sight are unresolved (hidden saturation), or the absorbing gas cloud only partially covers the background light source (partial covering), and the average column density results in an apparent unsaturated profile. 
Lines affected by classical saturation in this study are those with large oscillator strengths (e.g., \oi~$\lambda$1302 with $f=4.8870\times10^{-2} $, hence the need to predict $N$(\oi) using \pii\ and \sii, as discussed below). Hidden saturation is most easily identified by  fitting lines with different $f\lambda$ of the same species.  As such, in an effort to identify hidden saturation, careful tests were performed on each species where multiple lines for the same ion exist by separately fitting the weakest line and comparing the output with that obtained when fitting all lines together. In some cases, the line fit derived from the weakest line over-estimated the column density of the strongest lines, signaling that the strongest lines were suffering from hidden saturation / partial covering. In these cases, we adopted the column density derived from the weakest line only. If multiple lines for the same ion could be fitted by a single model (column density, $b-$parameter, and velocity), all lines were deemed unsaturated and not suffering from hidden saturation. A detailed description of classical and hidden saturation and their impact on UV absorption profiles can be found in \cite{James:2014} including an example of hidden saturation for the triplet \feii\ $\lambda$1142,$\lambda$1143,$\lambda$1144.

In cases where species were isolated (i.e., uncontaminated by either other species from the target itself or the MW), we utilized column density measurements from \citetalias{Parker:2024}, which only measured isolated or uncontaminated ISM profiles. In doing so, the methodology adopted by \citetalias{Parker:2024} differed slightly from that utilized here in that the $b$-parameter and velocity was held fixed between different species. When performing our line fits, we adopted the same $b$-parameter as \citetalias{Parker:2024} as an initial value and allowed it to vary. If this $b$-parameter provided a satisfactory fit to that particular line, as assessed via the $\chi^2$ value of the fit and a visual inspection of the model and residual fit, we adopted it as a fixed value. This was particularly helpful when fitting weak transitions, like
\pii~$\lambda$1152 \citep{Morton:1991}.

To check for consistency between the two methodologies, comparisons were made between the \sii~$\lambda$1253,1259 profile fits reported by \citetalias{Parker:2024} and those measured here for the same systems. In the majority of cases, differences of $0.5-1$~dex were found for $N$(\sii), which we attribute to the different $b$-parameters applied for many of the fits and, as noted in \PXI, their N(\sii) values were often inflated because they were not taking into account the contamination of \sii~$\lambda$1259 by \sIii~$\lambda$1260.

For the purpose of this study, we only adopt the \citetalias{Parker:2024} column densities derived from \oi~$\lambda$1302 line which, due to saturation issues, we use only to calculate a lower limit on O/H for a comparison with the O/H value derived from other species (discussed in Section~\ref{sec:abund}). We only use the \oi~$\lambda$1302 from \citetalias{Parker:2024} because this is the only line that overlaps between the two studies which is truly isolated.

The best fit parameters for each galaxy are listed in Table~\ref{tab:column_dens} within the Appendix, including column density, and $b$-parameter. 
Lines affected by saturation are listed with lower limit column densities. We also include the \hi\ column densities from \citetalias{Hu:2023} for completeness.

\section{Neutral Gas Abundances}\label{sec:abund}
Chemical abundances for the neutral gas were derived from the UV absorption lines of the low ionization states listed in Table~\ref{tab:ions} and were derived via the ratio of the ion column density with the \hi\ column density, i.e., $\log(X/{\rm H})=\log(N(X))-\log(N({\rm HI}))$. For calculating abundances relative to solar, we obtain solar abundance values from \citet{Asplund:2021}. These are listed in Table~\ref{tab:neutral}, alongside the abundances corrected for ionization levels along the line of sight, which we describe below in Section~\ref{sec:ICFs}.

Obtaining a true ``apples-to-apples" comparison of ionized gas and neutral gas abundances is of paramount importance for this study.  
Oxygen is traditionally used as a benchmark for the ionized gas metallicity of a galaxy as it is one of the strongest emitting species in the optical. Unfortunately, oxygen can be very challenging to constrain in the neutral ISM for two reasons: (i) the \oi~$\lambda$1302 line suffers from classical saturation due to its strength (and \oi~$\lambda$1355 is undetected) ; (ii) in low redshift objects, \oi~$\lambda$1302 is often contaminated by Geocoronal emission at $\sim 1304$~\AA. 

To circumvent the challenges associated with oxygen, we derive oxygen abundances following the methodology of \citet{James:2018} via the column densities of \sii\ and \pii. 
To describe briefly, \citet{James:2018} used relationships between the \oi, \sii\ and \pii\ column densities as robust proxies to estimate O/H abundances in the neutral gas of different environments (with regards to metallicity, star-formation rate, and $N(HI)$).

In Table~\ref{tab:oxygen} in the Appendix we list both the lower limit oxygen abundances derived from \oi~$\lambda$1302, which was inferred from \PXI, and oxygen abundances inferred from $N$(\sii) and $N$(\pii) proxies (log(O(P,S)/H)).
The later represents the average of log(O(P)) and log(O(S)). 
In the cases where $N$(\sii) or $N$(\pii) were not available, only one proxy relationship was used. On average log(O(P,S)/H) is greater than the lower limit on log(O/H) derived from \oi~$\lambda$1302 by $\sim$0.86~dex.

\subsection{Ionization Correction Factors}\label{sec:ICFs}

\begin{figure*}
\begin{center}
    \includegraphics[width=0.6\textwidth, trim=0 0 0 0,  clip=yes]{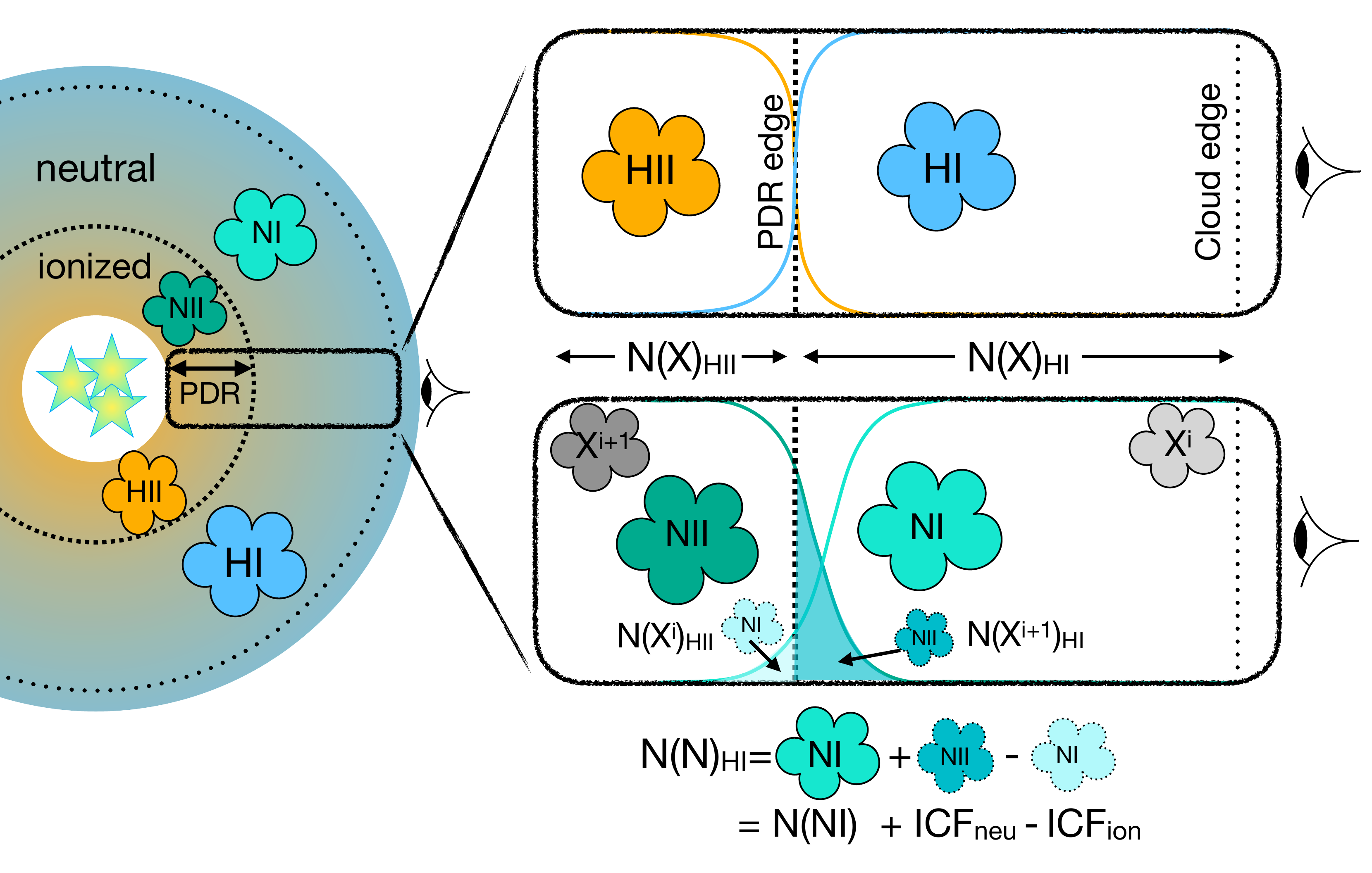}
        \caption{Schematic representation of the neutral-gas ionization correction factor modeling structure described in Section~\ref{sec:ICFs}, shown here using nitrogen as an example. The edges of the ionized and neutral gas zones are defined by  hydrogen (the photodisocciation region). A zoom in of a gas column, which contains both the PDR and cloud edge, is shown in the right panels, showing the ionization structure of \hii\ and \hi\ (top) and \nii\ and \Ni\ (bottom). 
        }
\label{fig:schematic}
\end{center}
\end{figure*}

In order to accurately determine the abundance of metals in the neutral ISM gas, the ionization state of the gas along the line of sight needs to be accounted for using ionization correction factors (ICFs), which depend on the gas structure of the system.  Note that neutral-gas ICFs are different from the ICFs used in ionized-gas chemical abundance calculations which instead account for missing ionic species in the ionized gas due to observational constraints (wavelength, resolution, depth etc). As such, we refer them as `neutral-gas ICFs' hereafter to avoid confusion between the two concepts. 

Neutral-gas ICFs were derived for each system following the methodology first described in \citet{James:2014} and incorporated by \citet{Hernandez:2020} and \citet{Abril-Melgarejo:2024}. In Appendix~\ref{sec:AICFs} we provide a full overview of the modeling concepts involved, which are designed to ensure that we correctly account for the total amount of the element within the neutral gas. A schematic representation of the ICF modeling structure is shown in Figure~\ref{fig:schematic}, where we illustrate how the two components of the neutral-gas ICFs (the neutral ICF, ICF$_{\rm neu.}$, and the ionized ICF, ICF$_{\rm ion.}$) are calculated according to a line-of-sight geometry. 
The dominant spectroscopic feature within the `neutral' ($N$(\hi)$>N$(\hii)) and `ionized' ($N$(\hii)$>N$(\hi)) zones of the ISM can be found in Table~\ref{tab:ions}. 

Overall, the size of ICF$_{\rm neu.}$ and ICF$_{\rm ion.}$ are comparable for most species and essentially cancel each other out to result in a very small ICF$_{\rm tot.}$ (see Equation A6 in the Appendix). The exceptions are nitrogen and oxygen, where ICF$_{\rm neu.}$ is larger than ICF$_{\rm ion.}$ by two orders of magnitude. This is because \Ni\ and \oi\ have comparable ionization potentials (IP = 14.53\,eV and 13.62\,eV, respectively) that are larger than hydrogen's (IP = 13.58\,eV) (see Table~\ref{tab:ions}).
Thus, N and O are in their neutral form within the neutral ISM (compared to \sii, \feii, and \pii, which are singly ionized). As such, we need to account for non-negligible amounts of \nii\ and \oii\ that lie outside the photodissociation region (PDR), and very small amounts of \Ni\ and \oi\ that lie within the PDR (essentially defined according to the IP of hydrogen), as illustrated in Fig.~\ref{fig:schematic}. This results in an overall positive ICF$_{\rm tot.}$ and net increase to account for the full amount of N and O within the `neutral' zone.

The total ionization correction factors (ICF$_{\rm total}$) computed for each galaxy are presented in Table~\ref{tab:ICFs} and used to compute the ICF-corrected abundances ($log(X/H)_{ICF}$) listed in Table~\ref{tab:neutral}.

\subsection{Dust Depletion}\label{sec:depletion}
The depletion $\Delta_X$ of an element $X$ is defined as the amount of element $X$ incorporated into the dust grains. Understanding dust depletion is essential to decipher both the origin of dust \citep[see, e.g.,][]{Mattsson:2019, Konstantopoulou:2022} and to estimate the total abundance of different elements within the gas \citep[see, e.g.,][]{DeCia:2016, DeCia:2018, Konstantopoulou:2024}. \citet{DeCia:2016} studied the dust depletion in the local universe (Milky Way) and high-redshift gas clouds (using Damped Lyman-$\alpha$ systems, or DLAs) and came up with a formalism that calculates the depletion of different metals onto dust independent of their gas-phase metallicity.  However, studies have shown that the overall depletion strength increases with metallicity, as more metals become available to condense onto dust grains \citep[e.g.,][]{jenkins09, DeCia:2018, Wiseman:2017}. At low metallicities, the reduced dust-to-gas ratio and slower grain growth lead to weaker depletions, particularly for volatile elements, while refractory elements (such as Fe, Ni, and Si) remain more strongly affected even in metal-poor environments.

We use the most updated version of the \citet{DeCia:2024} formalism (see their Equations 1 through 5) to estimate the dust depletion and calculate the depletion corrected abundance of different metals observed within our sample of CLASSY galaxies. Calculations are made using all available refractory elements within the CLASSY spectra using the column densities of \feii, \nIii, and \sIii, relative to \hi. The constants (A2$_x$ and B2$_x$) required for each metal to implement this depletion correction formalism were taken from \citet{Konstantopoulou:2022, Konstantopoulou:2024}. 

\begin{figure}
    \centering
    \includegraphics[width=1\linewidth]{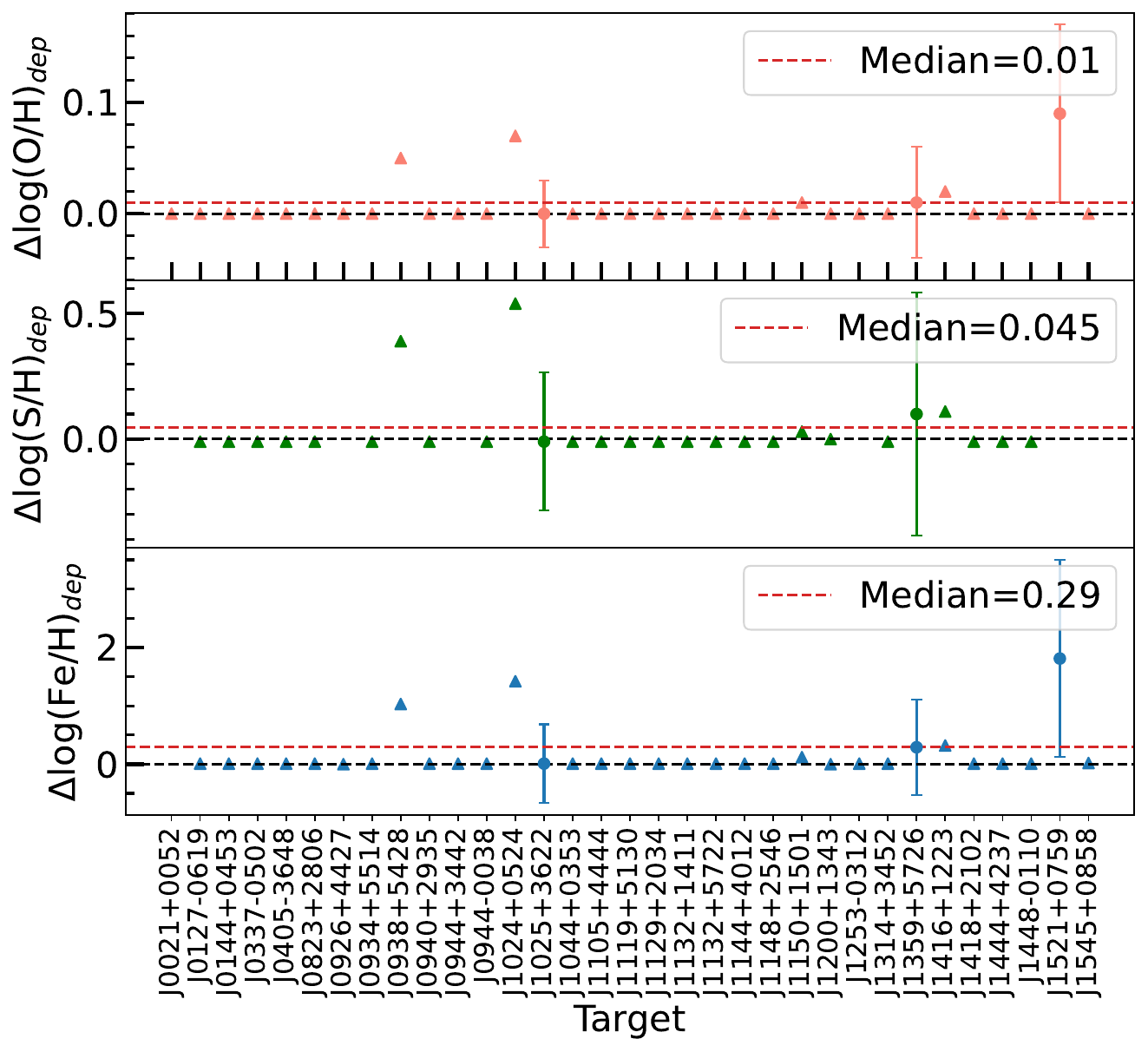}
    \caption{Depletion corrections for O, S and Fe for each of the galaxies within our sample. The dashed black and red horizontal lines in each panel show no depletion and average depletion value for each element, respectfully. Triangles represent lower limits.}
    \label{fig:depletion}
\end{figure}

Fig.~\ref{fig:depletion} shows the correction for depletion, $\Delta log(X/H)_{dep}$, required for different metals ($X$ = Fe, S, and O) for each galaxy in our sample. Nitrogen is not included here because it is not a refractory element and 
so does not deplete onto dust. For 28/31 of the galaxies within our sample, we only have lower limits for depletion (since the $v\sim0$\kms\ ISM component of silicon is saturated for these systems).

For the three galaxies with unsaturated silicon and available hydrogen column densities, the average depletion correction for elements Fe, S, and O are $\sim$0.7, 0.2 and 0.05, respectively. 
These average depletion levels are in agreement with previous studies, where Fe is known to deplete heavily onto dust grains, whereas S and O are far less depleted \citep[e.g.,][]{jenkins09,James:2018}, although typically, O should be slightly more depleted than S \citep{DeCia:2016}.
However, given the small sample size and large uncertainties associated with the average depletion corrections, we do not implement a depletion-correction in the remaining analysis.

	\begin{table*}
	\begin{center}
	\begin{scriptsize}
	\caption{Neutral gas abundances for N, O, S, and Fe, before and after applying neutral-gas ICFs (Appendix Table~\ref{tab:ICFs}). All abundances are expressed in terms of 12+log(X/H), where X is the element. }\label{tab:neutral}
	\begin{tabular}{lcccccccc}
	\hline\hline
	Target	&	log(N/H)$^{no ICF}_{neu}$			&	log(N/H)$_{neu}$		&	log(O$_{(P,S)}$/H)$^{no ICF}_{neu}$			&	log(O$_{(P,S)}$/H)$_{neu}$		&	log(S/H)$^{no ICF}_{neu}$			&	log(S/H)$_{neu}$			&	log(Fe/H)$^{no ICF}_{neu}$			&	log(Fe/H)$_{neu}$			\\
\hline															
	J0127-0619	&	6.07	$\pm$	0.29	&	6.07	$\pm$	0.29	&	7.80	$\pm$	0.40	&	7.82	$\pm$	0.40	&	6.34	$\pm$	0.27	&	6.35	$\pm$	0.27	&	5.76	$\pm$	0.29	&	5.76	$\pm$	0.29	\\
	J0144+0453	&	6.17	$\pm$	0.09	&	6.03	$\pm$	0.09	&	8.61	$\pm$	0.21	&	8.77	$\pm$	0.21	&	7.04	$\pm$	0.08	&	7.09	$\pm$	0.08	&	6.86	$\pm$	0.22	&	6.77	$\pm$	0.22	\\
	J0337-0502	&	5.02	$\pm$	0.02	&	5.02	$\pm$	0.02	&	6.60	$\pm$	0.15	&	6.61	$\pm$	0.15	&	5.15	$\pm$	0.03	&	5.16	$\pm$	0.03	&	5.51	$\pm$	0.04	&	5.51	$\pm$	0.04	\\
	J0405-3648	&	5.65	$\pm$	0.38	&	5.64	$\pm$	0.38	&	7.74	$\pm$	0.16	&	7.80	$\pm$	0.16	&	6.10	$\pm$	0.11	&	6.13	$\pm$	0.11	&	6.09	$\pm$	0.33	&	6.09	$\pm$	0.33	\\
	J0823+2806	&	6.67	$\pm$	0.11	&	6.67	$\pm$	0.11	&	7.99	$\pm$	0.04	&	8.00	$\pm$	0.04	&	6.42	$\pm$	0.04	&	6.43	$\pm$	0.04	&	6.19	$\pm$	0.34	&	6.19	$\pm$	0.34	\\
	J0934+5514	&	5.30	$\pm$	0.06	&	5.29	$\pm$	0.06	&	7.29	$\pm$	0.20	&	7.33	$\pm$	0.20	&	5.72	$\pm$	0.05	&	5.74	$\pm$	0.05	&	5.57	$\pm$	0.06	&	5.57	$\pm$	0.06	\\
	J0938+5428	&	6.48	$\pm$	0.09	&	6.26	$\pm$	0.09	&	8.67	$\pm$	0.16	&	8.78	$\pm$	0.16	&	6.92	$\pm$	0.11	&	6.88	$\pm$	0.11	&	6.39	$\pm$	0.14	&	6.24	$\pm$	0.14	\\
	J0940+2935	&	5.46	$\pm$	0.05	&	5.46	$\pm$	0.05	&	7.66	$\pm$	0.40	&	7.67	$\pm$	0.40	&	6.12	$\pm$	0.38	&	6.13	$\pm$	0.38	&	6.16	$\pm$	0.41	&	6.17	$\pm$	0.41	\\
	J0944+3442	&	5.60	$\pm$	0.06	&	5.60	$\pm$	0.06	&	7.69	$\pm$	0.13	&	7.71	$\pm$	0.13	&	\ldots			&	\ldots			&	5.22	$\pm$	0.21	&	5.22	$\pm$	0.21	\\
	J0944-0038	&	5.62	$\pm$	0.07	&	5.62	$\pm$	0.07	&	7.90	$\pm$	0.23	&	7.91	$\pm$	0.23	&	5.78	$\pm$	0.21	&	5.79	$\pm$	0.21	&	6.14	$\pm$	0.38	&	6.14	$\pm$	0.38	\\
	J1024+0524	&	5.88	$\pm$	0.12	&	5.79	$\pm$	0.12	&	8.21	$\pm$	0.21	&	8.32	$\pm$	0.21	&	6.58	$\pm$	0.09	&	6.59	$\pm$	0.09	&	5.64	$\pm$	0.16	&	5.57	$\pm$	0.16	\\
	J1025+3622	&	5.65	$\pm$	0.23	&	5.61	$\pm$	0.23	&	8.14	$\pm$	0.21	&	8.18	$\pm$	0.21	&	6.57	$\pm$	0.21	&	6.61	$\pm$	0.21	&	6.86	$\pm$	0.23	&	6.85	$\pm$	0.23	\\
	J1044+0353	&	5.62	$\pm$	0.54	&	5.62	$\pm$	0.54	&	7.24	$\pm$	0.63	&	7.24	$\pm$	0.63	&	5.62	$\pm$	0.44	&	5.62	$\pm$	0.44	&	5.73	$\pm$	0.45	&	5.73	$\pm$	0.45	\\
	J1105+4444	&	5.48	$\pm$	0.11	&	5.48	$\pm$	0.11	&	7.53	$\pm$	0.17	&	7.54	$\pm$	0.17	&	6.02	$\pm$	0.12	&	6.02	$\pm$	0.12	&	5.82	$\pm$	0.16	&	5.82	$\pm$	0.16	\\
	J1119+5130	&	5.45	$\pm$	0.10	&	5.43	$\pm$	0.10	&	7.94	$\pm$	0.19	&	8.01	$\pm$	0.19	&	6.42	$\pm$	0.12	&	6.45	$\pm$	0.12	&	6.87	$\pm$	0.50	&	6.87	$\pm$	0.50	\\
	J1129+2034	&	6.37	$\pm$	0.09	&	6.37	$\pm$	0.09	&	8.11	$\pm$	0.12	&	8.14	$\pm$	0.12	&	6.63	$\pm$	0.09	&	6.65	$\pm$	0.09	&	6.16	$\pm$	0.10	&	6.16	$\pm$	0.10	\\
	J1132+1411	&	6.64	$\pm$	0.03	&	6.59	$\pm$	0.03	&	8.83	$\pm$	0.15	&	8.91	$\pm$	0.15	&	7.45	$\pm$	0.09	&	7.50	$\pm$	0.09	&	6.75	$\pm$	0.07	&	6.73	$\pm$	0.07	\\
	J1132+5722	&	5.54	$\pm$	0.05	&	5.54	$\pm$	0.05	&	7.48	$\pm$	0.23	&	7.50	$\pm$	0.23	&	6.04	$\pm$	0.09	&	6.05	$\pm$	0.09	&	5.68	$\pm$	0.09	&	5.68	$\pm$	0.09	\\
	J1144+4012	&	6.56	$\pm$	0.11	&	6.49	$\pm$	0.11	&	8.60	$\pm$	0.09	&	8.62	$\pm$	0.09	&	7.03	$\pm$	0.09	&	7.05	$\pm$	0.09	&	7.12	$\pm$	0.35	&	7.09	$\pm$	0.35	\\
	J1148+2546	&	5.68	$\pm$	0.05	&	5.68	$\pm$	0.05	&	7.92	$\pm$	0.11	&	7.97	$\pm$	0.11	&	6.26	$\pm$	0.05	&	6.28	$\pm$	0.05	&	6.43	$\pm$	0.08	&	6.43	$\pm$	0.08	\\
	J1150+1501	&	6.41	$\pm$	0.26	&	6.41	$\pm$	0.26	&	7.98	$\pm$	0.15	&	8.02	$\pm$	0.15	&	6.48	$\pm$	0.04	&	6.50	$\pm$	0.04	&	5.86	$\pm$	0.12	&	5.87	$\pm$	0.12	\\
	J1225+6109	&	6.01	$\pm$	0.22	&	6.01	$\pm$	0.22	&	7.70	$\pm$	0.15	&	7.73	$\pm$	0.15	&	6.17	$\pm$	0.05	&	6.19	$\pm$	0.05	&	5.92	$\pm$	0.12	&	5.92	$\pm$	0.12	\\
	J1253-0312	&	6.37	$\pm$	1.18	&	6.37	$\pm$	1.18	&	7.55	$\pm$	0.07	&	7.59	$\pm$	0.07	&	\ldots			&	\ldots			&	6.19	$\pm$	0.19	&	6.19	$\pm$	0.19	\\
	J1314+3452	&	6.27	$\pm$	0.04	&	6.26	$\pm$	0.04	&	8.28	$\pm$	0.07	&	8.35	$\pm$	0.07	&	6.70	$\pm$	0.04	&	6.74	$\pm$	0.04	&	6.07	$\pm$	0.06	&	6.08	$\pm$	0.06	\\
	J1359+5726	&	\ldots			&	\ldots			&	7.35	$\pm$	0.48	&	7.37	$\pm$	0.48	&	5.82	$\pm$	0.42	&	5.83	$\pm$	0.42	&	5.43	$\pm$	0.30	&	5.43	$\pm$	0.30	\\
	J1416+1223	&	7.28	$\pm$	3.45	&	6.92	$\pm$	3.45	&	8.52	$\pm$	0.80	&	8.49	$\pm$	0.80	&	7.10	$\pm$	0.10	&	6.97	$\pm$	0.10	&	7.11	$\pm$	0.20	&	6.88	$\pm$	0.20	\\
	J1418+2102	&	5.55	$\pm$	0.05	&	5.55	$\pm$	0.05	&	7.45	$\pm$	0.12	&	7.47	$\pm$	0.12	&	5.97	$\pm$	0.06	&	5.99	$\pm$	0.06	&	5.85	$\pm$	0.18	&	5.85	$\pm$	0.18	\\
	J1444+4237	&	5.32	$\pm$	0.06	&	5.32	$\pm$	0.06	&	7.44	$\pm$	0.27	&	7.44	$\pm$	0.27	&	6.06	$\pm$	0.18	&	6.06	$\pm$	0.18	&	5.94	$\pm$	0.72	&	5.94	$\pm$	0.72	\\
	J1448-0110	&	5.74	$\pm$	0.04	&	5.74	$\pm$	0.04	&	7.53	$\pm$	0.17	&	7.54	$\pm$	0.17	&	6.01	$\pm$	0.04	&	6.02	$\pm$	0.04	&	5.88	$\pm$	0.08	&	5.88	$\pm$	0.08	\\
	J1521+0759	&	6.06	$\pm$	0.81	&	5.82	$\pm$	0.81	&	8.43	$\pm$	0.17	&	8.56	$\pm$	0.17	&	\ldots			&	\ldots			&	6.05	$\pm$	0.12	&	5.88	$\pm$	0.12	\\
	J1545+0858	&	5.15	$\pm$	0.06	&	5.15	$\pm$	0.06	&	7.28	$\pm$	0.40	&	7.31	$\pm$	0.40	&	\ldots			&	\ldots			&	5.43	$\pm$	0.11	&	5.44	$\pm$	0.11	\\
	\tableline
	\end{tabular}
	\end{scriptsize}
	\end{center}
	\end{table*}

%=======================================
\section{Ionized Gas Abundances}\label{sec:ion_abund}

	\begin{table}
	\begin{center}
	\begin{scriptsize}
	\caption{Ionized gas abundances after applying ICF corrections. N/H, O/H, and S/H are compiled from \PXII, \PIV, and \PIV, respectfully, while Fe/H are from this work.}\label{tab:ionized}
	\begin{tabular}{lccccc}
	Target 	&	log(N/H)$_{ion}$			&	log(O/H)$_{ion}$			&	log(S/H)$_{ion}$			&	log(Fe/H)$_{ion}$			\\
	\hline\hline		
	J0127-0619	&	6.66	$\pm$	0.14	&	7.52	$\pm$	0.08	&	6.08	$\pm$	0.03	&	\ldots			\\
	J0144+0453	&	5.83	$\pm$	0.22	&	7.45	$\pm$	0.08	&	\ldots			&	\ldots			\\
	J0337-0502	&	5.91	$\pm$	0.06	&	7.23	$\pm$	0.01	&	5.43	$\pm$	0.01	&	5.36	$\pm$	0.07	\\
	J0405-3648	&	5.98	$\pm$	0.15	&	7.28	$\pm$	0.07	&	5.54	$\pm$	0.03	&	5.36	$\pm$	0.15	\\
	J0823+2806	&	7.16	$\pm$	0.16	&	8.25	$\pm$	0.10	&	6.55	$\pm$	0.02	&	6.18	$\pm$	0.04	\\
	J0934+5514	&	\ldots			&	7.09	$\pm$	0.02	&	\ldots			&	5.29	$\pm$	0.19	\\
	J0938+5428	&	7.02	$\pm$	0.15	&	8.26	$\pm$	0.08	&	6.59	$\pm$	0.03	&	5.99	$\pm$	0.07	\\
	J0940+2935	&	6.23	$\pm$	0.20	&	7.98	$\pm$	0.23	&	6.15	$\pm$	0.05	&	\ldots			\\
	J0944+3442	&	6.67	$\pm$	0.19	&	7.66	$\pm$	0.15	&	6.27	$\pm$	0.11	&	\ldots			\\
	J0944-0038	&	6.69	$\pm$	0.11	&	7.83	$\pm$	0.02	&	6.16	$\pm$	0.02	&	5.65	$\pm$	0.06	\\
	J1024+0524	&	6.41	$\pm$	0.13	&	7.80	$\pm$	0.05	&	6.13	$\pm$	0.01	&	5.97	$\pm$	0.05	\\
	J1025+3622	&	6.77	$\pm$	0.14	&	8.13	$\pm$	0.08	&	6.41	$\pm$	0.03	&	5.74	$\pm$	0.12	\\
	J1044+0353	&	6.07	$\pm$	0.14	&	7.55	$\pm$	0.02	&	5.43	$\pm$	0.01	&	5.47	$\pm$	0.25	\\
	J1105+4444	&	6.77	$\pm$	0.15	&	8.23	$\pm$	0.07	&	6.47	$\pm$	0.02	&	6.10	$\pm$	0.04	\\
	J1119+5130	&	5.99	$\pm$	0.16	&	7.59	$\pm$	0.08	&	5.86	$\pm$	0.05	&	5.73	$\pm$	0.33	\\
	J1129+2034	&	6.82	$\pm$	0.19	&	8.30	$\pm$	0.12	&	6.54	$\pm$	0.01	&	5.79	$\pm$	0.58	\\
	J1132+1411	&	6.79	$\pm$	0.15	&	8.24	$\pm$	0.07	&	6.50	$\pm$	0.02	&	5.81	$\pm$	0.06	\\
	J1132+5722	&	5.91	$\pm$	0.15	&	7.34	$\pm$	0.09	&	5.86	$\pm$	0.05	&	5.35	$\pm$	0.28	\\
	J1144+4012	&	7.45	$\pm$	0.12	&	8.65	$\pm$	0.08	&	\ldots			&	0.00	$\pm$	0.00	\\
	J1148+2546	&	6.62	$\pm$	0.14	&	8.01	$\pm$	0.04	&	6.27	$\pm$	0.01	&	5.97	$\pm$	0.05	\\
	J1150+1501	&	6.63	$\pm$	0.17	&	8.14	$\pm$	0.10	&	6.43	$\pm$	0.01	&	5.97	$\pm$	0.36	\\
	J1225+6109	&	6.37	$\pm$	0.15	&	8.02	$\pm$	0.05	&	6.21	$\pm$	0.02	&	5.91	$\pm$	0.40	\\
	J1253-0312	&	7.09	$\pm$	0.12	&	8.02	$\pm$	0.03	&	6.54	$\pm$	0.01	&	6.17	$\pm$	0.03	\\
	J1314+3452	&	6.76	$\pm$	0.18	&	8.27	$\pm$	0.11	&	6.49	$\pm$	0.01	&	5.89	$\pm$	0.39	\\
	J1359+5726	&	6.62	$\pm$	0.14	&	7.99	$\pm$	0.07	&	6.27	$\pm$	0.02	&	5.94	$\pm$	0.06	\\
	J1416+1223	&	7.31	$\pm$	0.06	&	8.15	$\pm$	0.04	&	6.47	$\pm$	0.11	&	6.09	$\pm$	0.10	\\
	J1418+2102	&	6.05	$\pm$	0.15	&	7.51	$\pm$	0.03	&	5.60	$\pm$	0.00	&	5.40	$\pm$	0.21	\\
	J1444+4237	&	6.19	$\pm$	0.20	&	7.38	$\pm$	0.09	&	\ldots			&	\ldots			\\
	J1448-0110	&	6.71	$\pm$	0.14	&	8.07	$\pm$	0.03	&	6.43	$\pm$	0.01	&	6.11	$\pm$	0.04	\\
	J1521+0759	&	7.24	$\pm$	0.11	&	8.67	$\pm$	0.07	&	7.04	$\pm$	0.25	&	6.62	$\pm$	0.16	\\
	J1545+0858	&	6.35	$\pm$	0.13	&	7.75	$\pm$	0.02	&	6.05	$\pm$	0.00	&	5.84	$\pm$	0.20	\\														
	\tableline
	\end{tabular}
	\end{scriptsize}
	\end{center}
	\end{table}																
Ionized gas abundances were taken from \citetalias{mingozzi:2022}, \citetalias{arellano-cordova:2022}, and \citetalias{arellano-cordova:2025b}, which contain full details of the calculations and ionized-gas ICF corrections, and from this work for Fe/H.
We include Fe in this multi-phase abundance study for two reasons - firstly, Fe/H is well measured in the neutral gas (owing to the \feii$\lambda\lambda$1142,43,44 triplet). Secondly, Fe provides important insight into the long-term chemical evolutionary status of the CLASSY galaxies, as Fe is produced predominantly by SNe Ia on timescales longer than $\sim$1\,Gyr 
\citep[e.g.,][]{kobayashi20}. 

To determine the ionic abundances of Fe$^{2+}$ in the ionized gas phase, we used the dust-corrected fluxes of [\feiii]~\W4658 for 25 CLASSY galaxies reported in \citetalias{mingozzi:2022} that have both neutral and ionized Fe/H abundances and followed the procedure presented in \citetalias{arellano-cordova:2025b}. 
Multiple Fe emission lines are seen in \ion{H}{2} regions, such as [\feii], [\feiii], [\ion{Fe}{4}], and [\ion{Fe}{5}]. 
However, we used [\feiii]~\W4658 alone to characterize the abundance of Fe, as [\feii] can be affected by fluorescence and and [\ion{Fe}{4}] is inherently weak \citep[e.g.,][]{rodriguez05, rodriguez99, berg21a, Mendez-delgado:2024}.  
We calculated the Fe$^{+2}$/H$^+$ ionic abundance using \texttt{PyNeb} with the transition probabilities from \citet{quinet96} and collision strengths from \citet{Zhang96}. 
We adopted physical conditions derived from ions with similar ionization potentials as [\feiii] (16.2 eV), namely $T_{\rm e}$[\oii] (or $T_{\rm e}$[\nii]) and $n_{\rm e}$[\ion{S}{2}]). 

One of the most critical aspects to derive the total abundances of Fe (and other elements) is the selection of appropriate ionized-gas ICFs \citep[e.g.,][]{rodriguez05, izotov06, berg21a, Mendez-delgado:2024}. In \citet[][]{Mendez-delgado:2024}, the authors analyzed the performance of the Fe ICF using a sample of star-forming regions with measurements of different ionization states of Fe (i.e., Fe$^{+}$/H$^{+}$, Fe$^{2+}$/H$^{+}$, Fe$^{3+}$/H$^{+}$, and Fe$^{4+}$/H$^{+}$). The total Fe/H abundance derived without using an ICF (i.e., a `robust' Fe/H obtained by summing all of the observed states) was then compared with the Fe/H values obtained by applying the Fe ICFs from \citet{rodriguez05, izotov06}, that utilize the observed Fe$^{2+}$/H$^{+}$. \citet{Mendez-delgado:2024} found that the Fe/H abundances derived using the ICF from \citet{rodriguez05} show much better agreement with the `robust' Fe/H than the Fe/H abundances derived from the ICF of \citet{izotov06}, with differences of up to 0.2 dex relative to the `robust' Fe/H values.
 In this study, we used the Fe ionized-gas ICF proposed by \citet{rodriguez05} to take into account the contribution of Fe$^{+}$ and Fe$^{3+}$. 
 Additionally, we stress that Fe can be significantly depleted onto dust grains, making it difficult to mitigate any bias derived only from the ICF. In Table~\ref{tab:ionized} we list the total CLASSY abundances of Fe/H and other elements used in this work. 

\section{Results} \label{sec:results}
To summarize this work so far, all absorption-line abundances presented here include corrections for unobserved ionization stages. In practice, this means each species is accounted for in both the neutral and ionized zones as defined by hydrogen. Since ions with ionization potentials below 13.6 eV (e.g., \sii, \feii, \pii, see Table~\ref{tab:ions}) can arise in \ion{H}{2} regions and PDRs, their non-neutral contribution has been removed through ICF$_{\rm ion.}$. Conversely, ions with IP $> 13.6$ eV that reside outside the \ion{H}{2} region have been added to the neutral budget via ICF$_{\rm neu.}$. Although many species are not truly neutral in the `neutral zone', we refer to the resulting quantities as neutral-gas abundances because they trace the metals associated with neutral hydrogen.

In this section we directly compare the ICF-corrected neutral gas abundances (X/H$_{neu}$, Table~\ref{tab:neutral}) with the (also ICF-corrected) ionized gas abundances (X/H$_{ion}$, Table~\ref{tab:ionized}), firstly as a one-to-one comparison and then in a nucleosynthetic context by assessing the multi-phase abundance ratios (N/O, S/O, $\alpha$/Fe) as a function of  metallicity (i.e., 12+log(O/H) for the ionized gas). 

\subsection{Neutral vs.\ Ionized Gas Abundances}

\begin{figure*}[ht!]
\centering

% ---------- Row 1 ----------
\begin{minipage}[b]{0.48\textwidth}
    \centering
    \includegraphics[width=\textwidth]{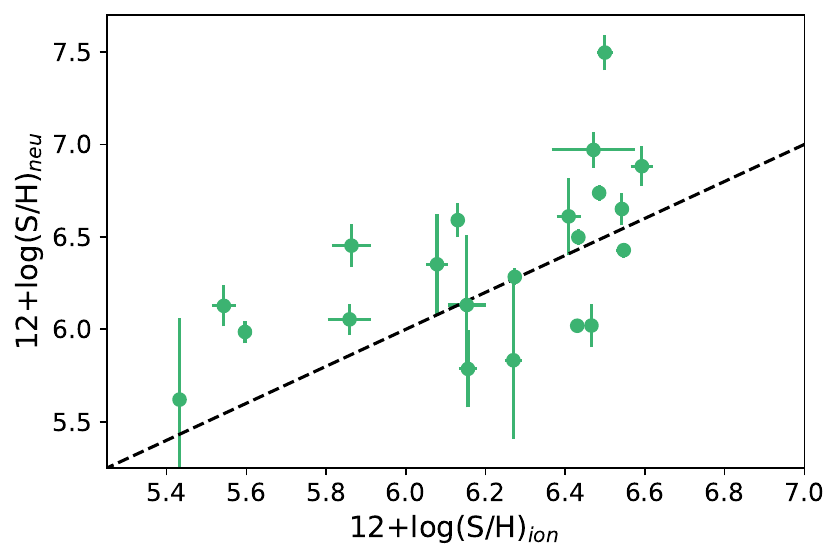}
    \textbf{(a) Sulfur}
\end{minipage}%
\hfill
\begin{minipage}[b]{0.48\textwidth}
    \centering
    \includegraphics[width=\textwidth]{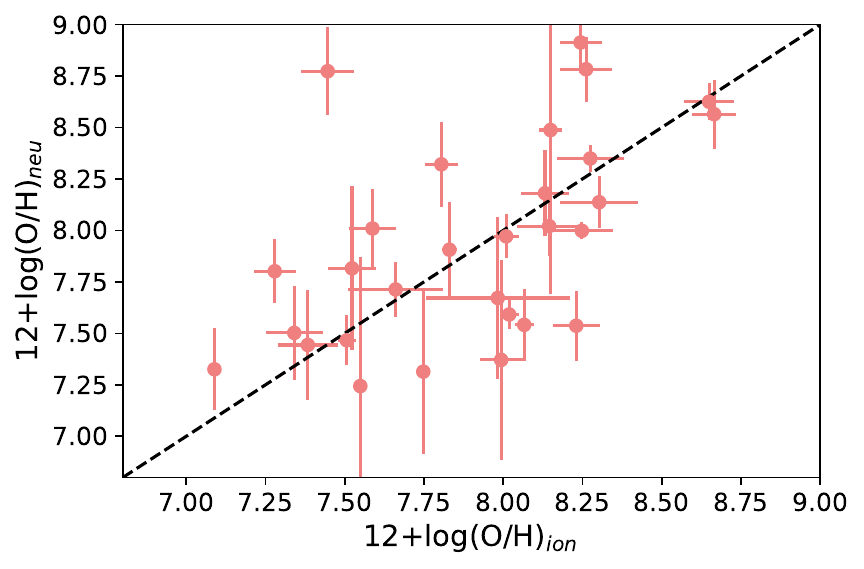}
    \textbf{(b) Oxygen}
\end{minipage}

\vspace{1em}

% ---------- Row 2 ----------
\begin{minipage}[b]{0.48\textwidth}
    \centering
    \includegraphics[width=\textwidth]{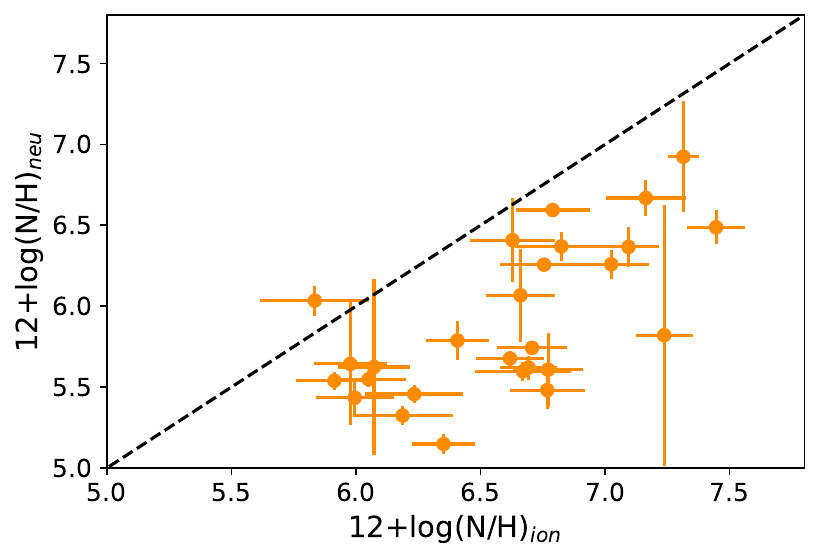}
    \textbf{(c) Nitrogen}
\end{minipage}%
\hfill
\begin{minipage}[b]{0.48\textwidth}
    \centering
    \includegraphics[width=\textwidth]{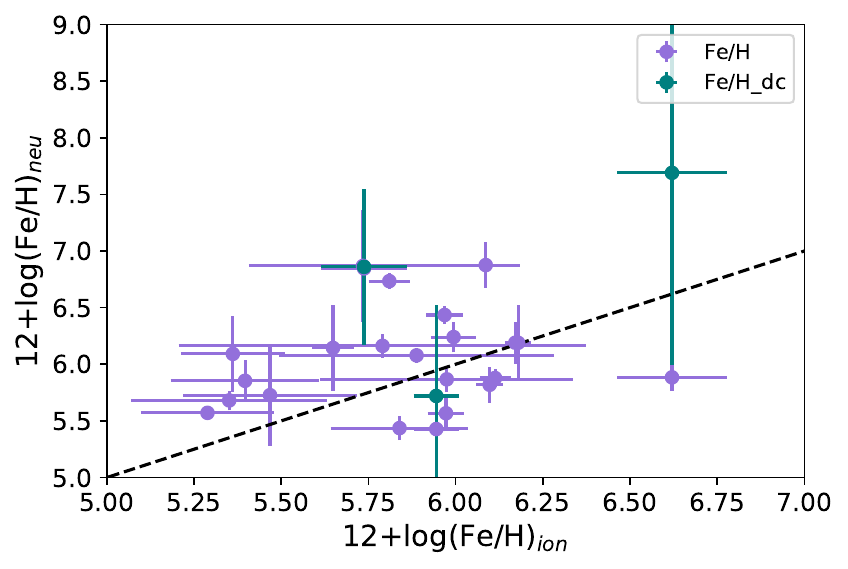}
    \textbf{(d) Iron}
\end{minipage}

\caption{
Ionized vs.\ neutral gas abundances, with a 1:1 ratio overlaid (black dashed line).
Both neutral- and ionized-gas abundances are ICF-corrected, as detailed in
Sections~\ref{sec:ICFs} and \ref{sec:ion_abund}.
For neutral-gas Fe, we include both Fe/H$_{neu}$ and the depletion-corrected
values (Fe/H$_{dc}$) for comparison.}
\label{fig:XY}
\end{figure*}

In Figures ~\ref{fig:XY}a-d we plot the ionized and neutral gas abundances for nitrogen, oxygen, sulfur and iron, along with the 1:1 ratio (black dashed line). We define the offset in abundance between the two phases as $\Delta X/{\rm H} = X/{\rm H}_{\rm neu.}-X/{\rm H}_{\rm ion.}$, such that a positive $\Delta(X/{\rm H})$ implies a higher abundance of $X$ in the neutral gas. All $\Delta X/$H values are listed in Table~\ref{tab:offsets}.

\subsubsection{Oxygen \& Sulfur}
For $\alpha$-elements O and S, which are mostly produced on short timescales (5--100\,Myr) via core collapse \citep[Type II SNe; e.g.,][]{kobayashi20}, there is a good agreement between the two phases. The average offset for sulfur is small, albeit with a relatively large amount of scatter ($\left<\Delta S/H\right>= 0.13 \pm 0.36 $). 
Similarly, oxygen shows a good agreement between the phases and large amount of scatter ($\left<\Delta O/H\right>= 0.01 \pm 0.43 $), which may be due to the fact that O/H$_{neu.}$ is derived from a combined average of sulfur and phosphorus using the relation described above. Another contributing factor to the amount of scatter seen here is the lack of dust depletion corrections, which would typically affect O more than S in low-metallicity environments  \citep{jenkins09,DeCia:2016}. 

\subsubsection{Iron}
For iron, which is produced primarily in Type~Ia SNe on $\sim$1 Gyr timescales ($\sim$60\%\ contribution) and to a lesser extent in core-collapse SNe on 5–100~Myr timescales ($\sim$40\%), the CLASSY galaxies show Fe/H$_{\rm neu.} > $Fe/H$_{\rm ion.}$, with $\langle\Delta{\rm Fe/H}\rangle = 0.21 \pm 0.48$. The large dispersion indicates that, on average, Fe is broadly well mixed between the two phases. In Figure~\ref{fig:XY}d we show both the ICF and ICF+depletion-corrected Fe/H values for the neutral gas, applying the depletion factor to the three galaxies for which it can be constrained (Fe/H$_{\rm dc}$). This correction increases $\langle\Delta{\rm Fe/H}\rangle$ to $0.66 \pm 0.62$. If we were to adopt the average Fe depletion correction of $\sim$0.7 dex for all galaxies (see Section~\ref{sec:depletion}), the offset between the phases would increase further to $\langle\Delta{\rm Fe/H}\rangle \sim 0.9$. Considering that 28/31 galaxies have only lower-limit depletion corrections, it is likely that Fe/H$_{\rm neu.}$ exceeds Fe/H$_{\rm ion.}$ in most systems.

The deficit of Fe in the ionized gas likely reflects ongoing dust depletion and grain formation in core-collapse SNe ejecta, which efficiently lock Fe into dust while leaving $\alpha$-elements such as O and S largely in the gas phase \citep[e.g.,][]{rodriguez05,Matsuura2011,DeLooze:2017}. These results suggest that feedback and dust processing regulate the apparent Fe abundance in starburst environments, delaying the return of Fe to the gas phase until grains are destroyed and metals are fully mixed.

\subsubsection{Nitrogen}
A rather different distribution is seen for N in Figure ~\ref{fig:XY}a, with N/H$_{\rm ion.}$ being consistently higher than N/H$_{\rm neu.}$. 

The mean ICF$_{\rm tot.}$ value for N/H$_{\rm neu.}$ is $\sim$0.05, which is far smaller than the average offset between the two phases of $\left<\Delta N/H\right>= -0.69 \pm 0.38$. N is mostly produced in intermediate-mass AGB stars \citep[4–7 \Msol, over $>100$ Myr timescales;][]{kobayashi11}, with a small portion produced in Type II SNe \citep[][]{kobayashi20}. WR stars can also be a primary source of nitrogen production in galaxies over $\sim$3--5\,Myr timescales.

WR stars are massive stars that have shed their outer hydrogen layers, exposing a helium core enriched with nitrogen created as a biproduct of the CNO cycle.
This nitrogen is then expelled into the surrounding ISM via strong stellar winds, significantly contributing to the overall nitrogen abundance in a galaxy. While WR features have been definitively detected in only four CLASSY galaxies (J0127$-$0619, J1129+2034, J1157+3220, and J1225+6109), and may remain undetected in the other low-metallicity systems due to the weakness of the spectral signature, WR stars can still be a source of significant N enrichment due to the timescales involved with chemical mixing.   

We further explore factors that may contribute to the offset in nitrogen abundance in Section~\ref{sec:nitrogen}.

	\begin{table}
	\begin{center}
	\begin{scriptsize}
	\caption{Offset between neutral and ionized gas abundances, where $\Delta X/H = log(X/H_{neu})-log(X/H_{ion})$.}\label{tab:offsets}
	\begin{tabular}{lcccc}
	\hline\hline
	Target	&	$\Delta$N/H			&	$\Delta$O/H			&	$\Delta$S/H			&	$\Delta$Fe/H			\\
	\hline														
	J0127-0619	&	-0.59	$\pm$	0.32	&	0.29	$\pm$	0.41	&	0.27	$\pm$	0.28	&	\ldots			\\
	J0144+0453	&	0.20	$\pm$	0.24	&	1.33	$\pm$	0.23	&	\ldots			&	\ldots			\\
	J0337-0502	&	-0.89	$\pm$	0.06	&	-0.62	$\pm$	0.15	&	-0.28	$\pm$	0.03	&	0.15	$\pm$	0.08	\\
	J0405-3648	&	-0.33	$\pm$	0.41	&	0.52	$\pm$	0.17	&	0.58	$\pm$	0.12	&	0.73	$\pm$	0.37	\\
	J0823+2806	&	-0.50	$\pm$	0.20	&	-0.25	$\pm$	0.11	&	-0.12	$\pm$	0.05	&	0.01	$\pm$	0.34	\\
	J0934+5514	&	\ldots			&	0.24	$\pm$	0.20	&	\ldots	&	0.28	$\pm$	0.20	\\
	J0938+5428	&	-0.77	$\pm$	0.18	&	0.52	$\pm$	0.18	&	0.29	$\pm$	0.11	&	0.25	$\pm$	0.15	\\
	J0940+2935	&	-0.78	$\pm$	0.20	&	-0.31	$\pm$	0.46	&	-0.02	$\pm$	0.38	&	\ldots			\\
	J0944+3442	&	-1.07	$\pm$	0.20	&	0.05	$\pm$	0.20	&	\ldots			&	\ldots			\\
	J0944-0038	&	-1.07	$\pm$	0.14	&	0.08	$\pm$	0.23	&	-0.37	$\pm$	0.21	&	0.50	$\pm$	0.38	\\
	J1024+0524	&	-0.62	$\pm$	0.18	&	0.52	$\pm$	0.21	&	0.46	$\pm$	0.09	&	-0.41	$\pm$	0.17	\\
	J1025+3622	&	-1.16	$\pm$	0.27	&	0.05	$\pm$	0.22	&	0.20	$\pm$	0.21	&	1.11	$\pm$	0.26	\\
	J1044+0353	&	-0.45	$\pm$	0.56	&	-0.31	$\pm$	0.63	&	0.19	$\pm$	0.44	&	0.26	$\pm$	0.51	\\
	J1105+4444	&	-1.29	$\pm$	0.19	&	-0.69	$\pm$	0.19	&	-0.45	$\pm$	0.12	&	-0.28	$\pm$	0.16	\\
	J1119+5130	&	-0.56	$\pm$	0.19	&	0.42	$\pm$	0.21	&	0.59	$\pm$	0.13	&	1.13	$\pm$	0.59	\\
	J1129+2034	&	-0.45	$\pm$	0.21	&	-0.17	$\pm$	0.17	&	0.11	$\pm$	0.09	&	0.37	$\pm$	0.59	\\
	J1132+1411	&	-0.20	$\pm$	0.15	&	0.67	$\pm$	0.16	&	1.00	$\pm$	0.10	&	0.92	$\pm$	0.09	\\
	J1132+5722	&	-0.37	$\pm$	0.16	&	0.16	$\pm$	0.24	&	0.19	$\pm$	0.10	&	0.33	$\pm$	0.29	\\
	J1144+4012	&	-0.96	$\pm$	0.16	&	-0.02	$\pm$	0.12	&	\ldots			&	\ldots			\\
	J1148+2546	&	-0.94	$\pm$	0.14	&	-0.04	$\pm$	0.11	&	0.01	$\pm$	0.05	&	0.47	$\pm$	0.09	\\
	J1150+1501	&	-0.22	$\pm$	0.31	&	-0.12	$\pm$	0.18	&	0.07	$\pm$	0.04	&	-0.11	$\pm$	0.38	\\
	J1225+6109	&	-0.36	$\pm$	0.27	&	-0.30	$\pm$	0.16	&	-0.02	$\pm$	0.05	&	0.02	$\pm$	0.41	\\
	J1253-0312	&	-0.73	$\pm$	1.19	&	-0.43	$\pm$	0.08	&	\ldots			&	0.02	$\pm$	0.19	\\
	J1314+3452	&	-0.50	$\pm$	0.18	&	0.07	$\pm$	0.13	&	0.25	$\pm$	0.04	&	0.19	$\pm$	0.40	\\
	J1359+5726	&	\ldots			&	-0.62	$\pm$	0.49	&	-0.44	$\pm$	0.42	&	-0.52	$\pm$	0.31	\\
	J1416+1223	&	-0.39	$\pm$	3.45	&	0.34	$\pm$	0.80	&	0.50	$\pm$	0.14	&	0.79	$\pm$	0.22	\\
	J1418+2102	&	-0.50	$\pm$	0.16	&	-0.04	$\pm$	0.12	&	0.39	$\pm$	0.06	&	0.46	$\pm$	0.28	\\
	J1444+4237	&	-0.86	$\pm$	0.21	&	0.06	$\pm$	0.28	&	\ldots			&	\ldots			\\
	J1448-0110	&	-0.96	$\pm$	0.15	&	-0.52	$\pm$	0.18	&	-0.41	$\pm$	0.04	&	-0.23	$\pm$	0.09	\\
	J1521+0759	&	-1.42	$\pm$	0.82	&	-0.10	$\pm$	0.18	&	\ldots			&	-0.74	$\pm$	0.20	\\
	J1545+0858	&	-1.20	$\pm$	0.14	&	-0.43	$\pm$	0.40	&	\ldots			&	-0.40	$\pm$	0.22	\\	
	\tableline
	\end{tabular}
	\end{scriptsize}
	\end{center}
	\end{table}																	
    
    %%%ABUNDANCE RATIOS----===============
\subsection{Multi-phase abundance ratios: N/O, S/O, $\alpha/Fe$}\label{sec:abund_ratios}
\begin{figure*}[ht!]
\centering

% ---------- Row 1 ----------
\begin{minipage}[b]{0.48\textwidth}
    \centering
    \includegraphics[width=\textwidth]{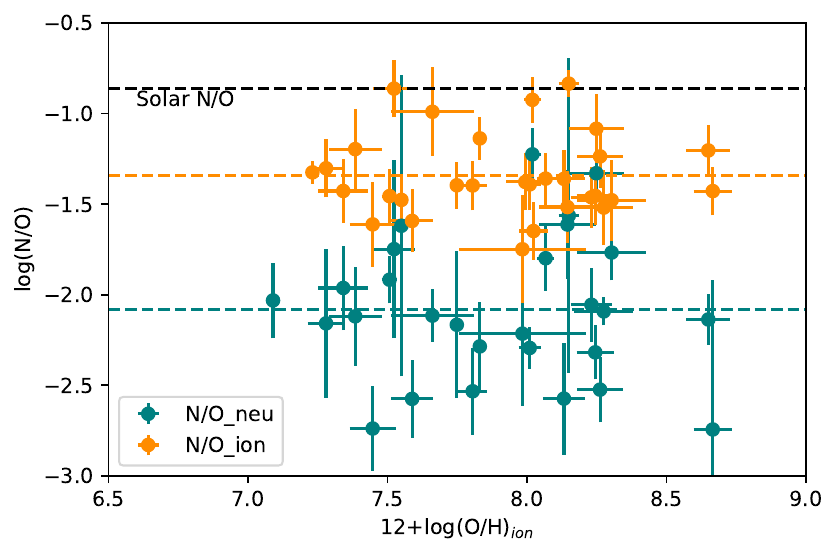}
    \textbf{(a) N/O}\label{fig:N/O}
\end{minipage}%
\hfill
\begin{minipage}[b]{0.48\textwidth}
    \centering
    \includegraphics[width=\textwidth]{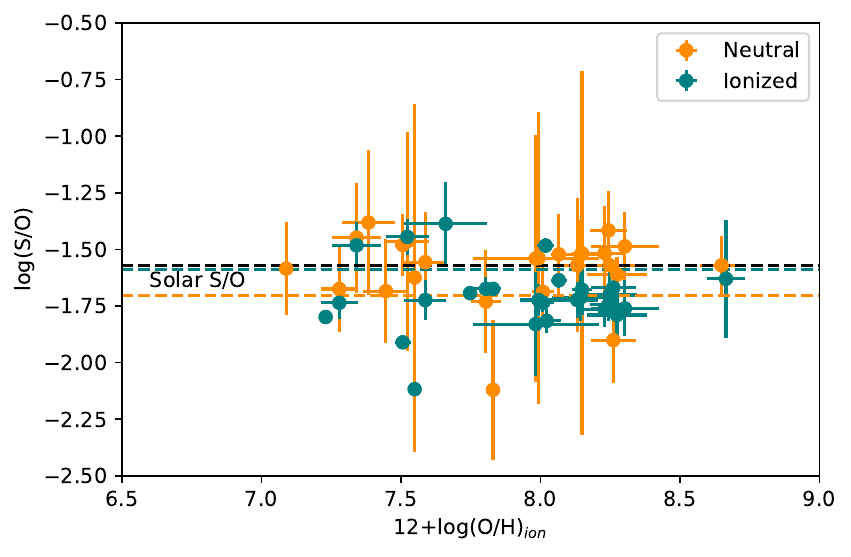}
    \textbf{(b) S/O}\label{fig:S/O}
\end{minipage}

\vspace{1em}

% ---------- Row 2 ----------
\begin{minipage}[b]{0.48\textwidth}
    \centering
    \includegraphics[width=\textwidth]{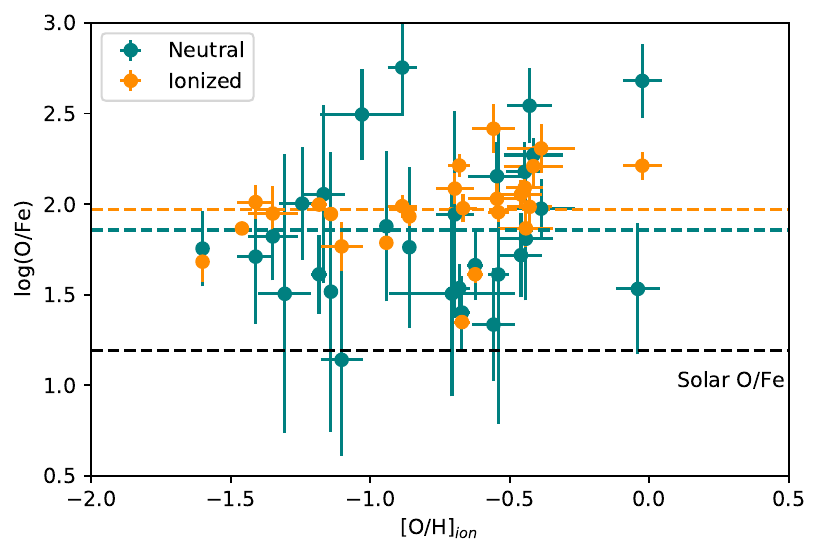}
    \textbf{(c) O/Fe}\label{fig:O/Fe}
\end{minipage}%
\hfill
\begin{minipage}[b]{0.48\textwidth}
    \centering
    \includegraphics[width=\textwidth]{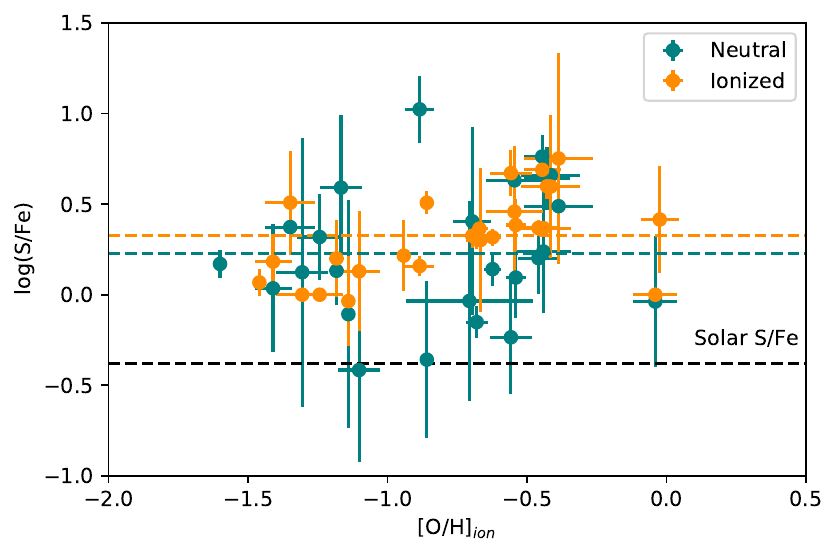}
    \textbf{(d) S/Fe}\label{fig:S/Fe}
\end{minipage}

\caption{\small Abundance ratios for both the ionized and neutral gas as a function of the metallicity (i.e., oxygen abundance of the ionized gas). In the bottom two panels the $x$-axis shows metallicity relative to solar, [O/H]=log(O/H)-log(O/H)$_{\odot}$. In each panel we show the respective solar ratio value as a black dashed line, with the average values as orange and teal dashed lines for the ionized and neutral gas, respectively.}
\label{fig:X/O}
\end{figure*}

Different elements are created as part of different nucleosynthetic processes and in stars with different stellar masses, on different timescales. As such, assessing trends in elemental abundances relative to O will be a direct result of how different elements are synthesized in stars and therefore provides clues about the age and metallicity of the stellar population that has enriched the interstellar gas. In particular, comparing the abundance ratio within the ionized and neutral phases of the gas can be a powerful tool to understand if/how the nucleosynthesis yields, which begin mixing in the ionized gas, are also echoed in the neutral gas (provided that dust affects the neutral and ionized gas phases in a similar manner). In Figure~\ref{fig:X/O}a-d we show the neutral and ionized gas abundances of N/O, S/O, O/Fe and S/Fe within that specific phase as a function of the ionized phase metallicity. 

\subsubsection{N/O vs. O/H}
Nitrogen is nucleosynthetically produced as a biproduct of the CNO cycle, which consumes carbon and oxygen as catalysts. 
At low metallicities, there is very little C and O in the gas from which stars form, and so only stars massive enough to produce their own seed C and O can trigger N production via the CNO cycle. 
Because this production only depends on the initial H and He within the star, the production of N/O is metallicity independent, or {\it primary}.
At higher metallicities, the initial C and O can sufficiently seed the CNO cycle such that N/O increases with increasing O/H.
Further, N and O are produced and released by stars of different masses, on different time scales, essentially behaving as a chemical evolutionary `clock' \citep[e.g.,][and references therein]{vanzee06a,berg19a,Perez-Montero:2013} of recent star formation. 
The complexity of the N/O--O/H relationship lies in the fact that N can be produced in both massive stars (M$>$8\Msol) \textit{and} intermediate mass stars \citep[e.g., AGB stars][]{kobayashi11}. In the latter, N may not be released until up to $>100$~Myr \citep{kobayashi20}, which is a significant delay compared to O being released by Type II SNe after $\sim10-40$\,Myr.  

\citetalias{arellano-cordova:2025b} showed that CLASSY galaxies follow the canonical N/O--OH relationship for the ionized gas.
At low metallicities, N/O remains fairly constant at $-1.5<\log(\rm N/O)<-1.0$, consistent with the plateau seen in other ionized-gas samples \citep[log(N/O)$\sim-1.4$ e.g.,][]{berg19a}. 
However, at 12+log(O/H)$\sim$8, CLASSY shows a large scatter in log(N/O) of up to 1.5 dex which \citetalias{arellano-cordova:2025b} attributes to a combination of density, star-formation, and outflows tracing different gas conditions.

Interestingly, as we now compare with the neutral gas, in Figure~\ref{fig:N/O} we can see a distinct offset between the two phases, with $\left<N/O_{\rm ion.}\right>= -1.34 \pm 0.25 $ and $\left<N/O_{\rm neu.}\right>= -2.12 \pm 0.38 $. Since there is a good agreement between O/H$_{\rm ion.}$ and O/H$_{\rm neu.}$ (Figure~\ref{fig:X/O}), the process that produces more N in the ionized gas needs to be on shorter timescales than the release of O by SNeII. Instead, it confirms that the higher levels of N/H in the ionized gas relative to the neutral must be related to other processes which we explore in Section~\ref{sec:discussion}. In particular, here we can see that while the typical `N/O plateau' is seen in the ionized gas (\citetalias{arellano-cordova:2025b}), the plateau in the neutral gas appears to have a much larger dispersion ($\sigma(\rm N/O)_{\rm neu.}=0.38$), which could be a reflection on the different levels of enrichment experienced by previous star-formation episodes in these systems.

\subsubsection{S/O vs. O/H}
While both S and O are created as part of the $\alpha-$capture process, S also has contributions from Type 1a SNe \citep{Matteucci:2005}. In Figure~\ref{fig:S/O}, we show the S/O values for the neutral and ionized gas as a function of metallicity. For both phases, the S/O values are aligned with the solar value, with averages of $\left<S/O_{\rm ion.}\right>= -1.71 \pm 0.16 $ and $\left<S/O_{\rm neu.}\right>= -1.59 \pm 0.15 $ for the ionized and neutral phases, respectively. The fact that O and S abundances for the neutral and ionized gas abundances scale in-sync with one another suggests that the Type 1a contribution to S enrichment may be negligible in CLASSY galaxies, or yet to be seen since the age of the current starbursts are all $<$1\,Gyr (see Appendix Section~\ref{sec:SPS_Age}). 

\subsubsection{$\alpha/Fe$ vs. O/H}
Iron is predominantly produced in Type~Ia SNe ($>$70\%), with smaller contributions from Type~II SNe and AGB stars \citep{kobayashi20,Matteucci:2012}.  
Consequently, Fe enrichment occurs over long timescales ($\sim$1~Gyr), whereas $\alpha$-elements such as O and S are injected into the ISM rapidly, on $\sim$5–10~Myr timescales.  
The $\alpha$/Fe abundance ratio therefore provides insight into a galaxy’s star formation history and chemical evolution, with higher $\alpha$/Fe indicating shorter/more recent star formation timescales, typically observed in young galaxies or outer regions of galaxies \citep{Tolstoy:2009,kobayashi20,Velichko:2024}.  
This arises because $\alpha$-elements are primarily produced in the early stages of a star-formation episode and are more prevalent in low-metallicity environments, whereas Fe accumulates later, particularly from Type~Ia SNe in more evolved populations.

In Figures~\ref{fig:O/Fe} and \ref{fig:S/Fe}, we show O/Fe and S/Fe for the neutral and ionized gas as a function of the ionized gas metallicity ([O/H]).  
Both phases exhibit super-solar ratios:  
$\left<O/Fe_{\rm neu.}\right>= 1.86 \pm 0.40$, $\left<O/Fe_{\rm ion.}\right>= 1.97 \pm 0.22$;  
$\left<S/Fe_{\rm neu.}\right>= 0.23 \pm 0.36$, $\left<S/Fe_{\rm ion.}\right>= 0.33 \pm 0.22$.  
These elevated $\alpha$/Fe ratios are consistent with the starburst-dominated, UV-bright, nature of CLASSY galaxies (\citetalias{berg:2022}).  
The similarity between the neutral and ionized phases suggests that both retain the imprint of previous star formation episodes older than $\sim$1~Gyr, allowing sufficient time for Type~Ia SNe to enrich Fe.  This also signifies that any $\alpha$-enriched material from the most recent starburst has not had sufficient time to fully mix into either phase.The small dispersion between the neutral and ionized phases, together with the age-dependent abundance offsets discussed in Section~\ref{sec:age}, indicates that any rapid, galaxy-wide mixing of recently produced $\alpha$-elements is limited.

Comparisons to other studies reinforce this interpretation.  
For example, \citet{Velichko:2024} analyzed $\alpha$/Fe in 110 DLAs, identifying a high-$\alpha$ plateau ([\,$\alpha$/Fe$] \sim 0.38$) followed by an $\alpha$-knee at [M/H]$\sim-1$, corresponding to the onset of significant Type~Ia Fe enrichment.  
CLASSY systems show a somewhat higher plateau for S/Fe ($\sim$0.66 relative to solar, with $\left<{\rm S/Fe}\right>\sim0.28$ for the two phases and $\log({\rm S/Fe})_\odot=-0.38$), likely reflecting top-heavy IMFs and high star formation efficiencies that enhance early $\alpha$-element production.  
As with DLAs, lower-mass, low-metallicity galaxies tend to reach the $\alpha$-knee at lower metallicities; given the low-mass, low-metallicity nature of CLASSY, we would similarly expect an $\alpha$-knee around [O/H] $\sim -1$.  
However, the diverse star formation histories and ages in our sample introduce substantial scatter, preventing a precise determination of the knee location.

%=====================MIXING=========
\section{Discussion} \label{sec:discussion}
The goal of this study is to understand whether an offset exists between amount of metals locked into the neutral and ionized gas within star-forming galaxies and investigate factors that would contribute to such an offset. To this end, we explore the abundance offsets presented in Section~\ref{sec:results} as a function of various galaxy properties, and finally focus on the behavior of nitrogen between gas phases. 

\begin{figure*}[ht!]
\centering

% ---------- Row 1 ----------
\begin{minipage}[b]{0.48\textwidth}
    \centering
    \includegraphics[width=\textwidth]{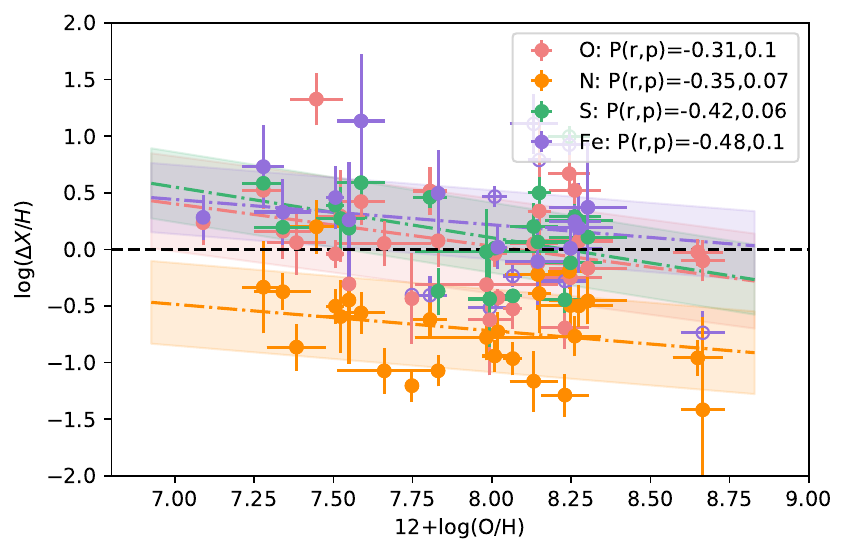}
    \textbf{(a) $\Delta X/H$ vs. metallicity}
\end{minipage}%
\hfill
\begin{minipage}[b]{0.48\textwidth}
    \centering
    \includegraphics[width=\textwidth]{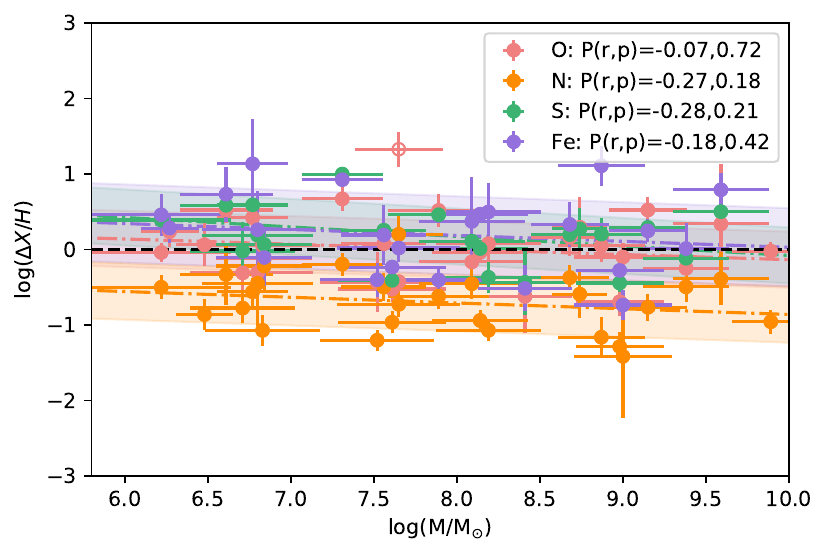}
    \textbf{(b) $\Delta X/H$ vs. stellar mass}
\end{minipage}

\vspace{1em}

% ---------- Row 2 ----------
\begin{minipage}[b]{0.48\textwidth}
    \centering
    \includegraphics[width=\textwidth]{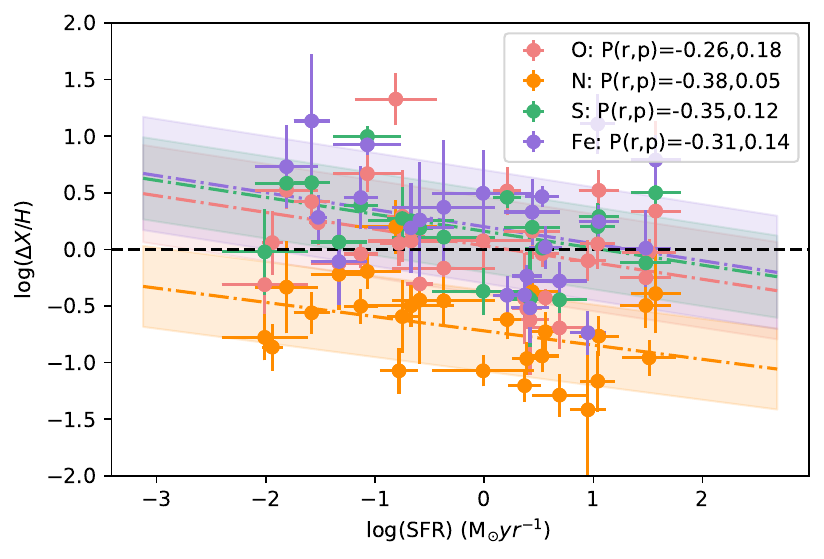}
    \textbf{(c) $\Delta X/H$ vs. star-formation rate}
\end{minipage}%
\hfill
\begin{minipage}[b]{0.48\textwidth}
    \centering
    \includegraphics[width=\textwidth]{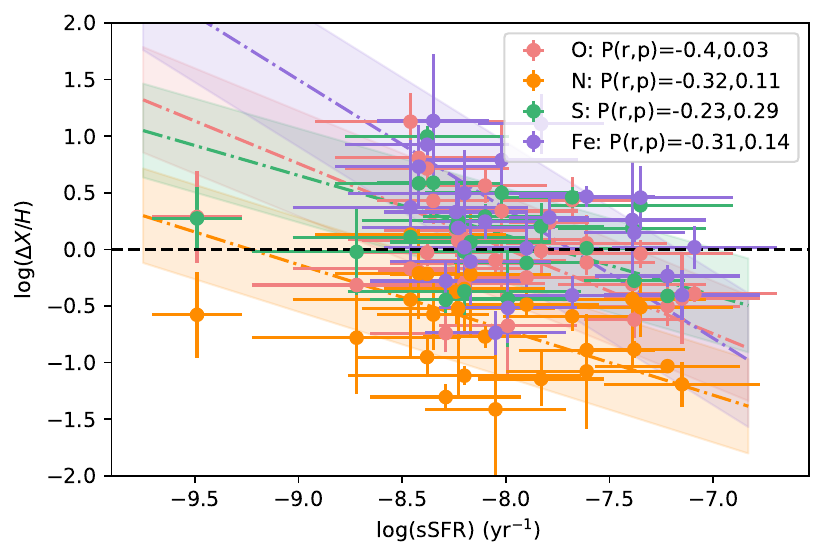}
    \textbf{(d) $\Delta X/H$ vs. specific star-formation rate (sSFR)}
\end{minipage}

\vspace{1em}

% ---------- Row 3 ----------
\begin{minipage}[b]{0.48\textwidth}
    \centering
    \includegraphics[width=\textwidth]{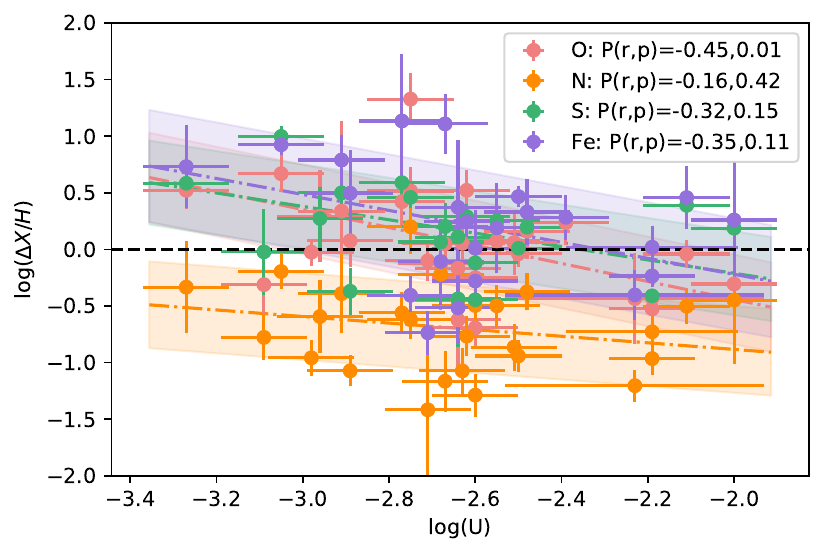}
    \textbf{(e) $\Delta X/H$ vs. ionization parameter}
\end{minipage}%
\hfill
\begin{minipage}[b]{0.48\textwidth}
    \centering
    \includegraphics[width=\textwidth]{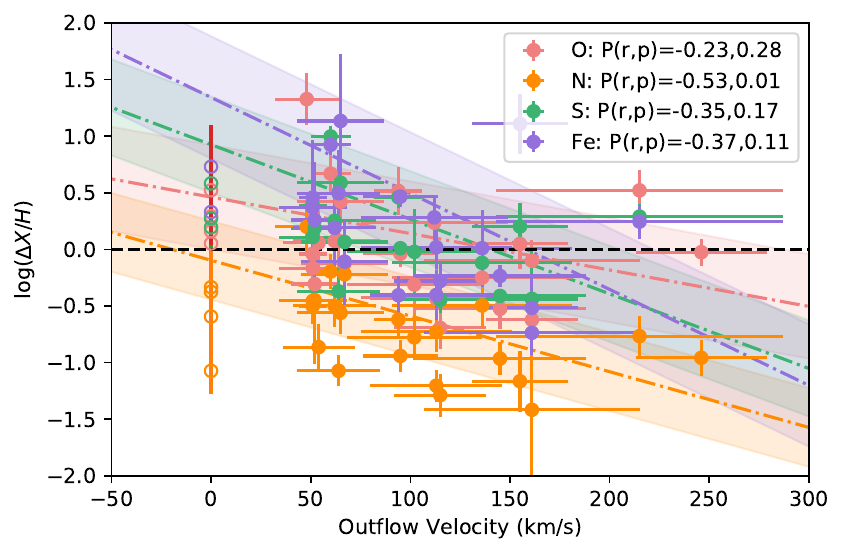}
    \textbf{(f) $\Delta X/H$ vs. ISM outflow velocity ($V_{out}$)}
\end{minipage}

\caption{\small Neutral - ionized gas abundance offset ($\Delta X/H$) as a function of different galaxy properties, where $\Delta X/H = X/H_{neutral}-X/H_{ionized}$, such that a positive $\Delta(X/H)$ implies a higher abundance of $X$ in the neutral gas. The Pearson correlation coefficients for each linear fit are provided in the plot legend. Open circles in panel (f) denote galaxies with no detected outflow and are not included in the linear fit.}
\label{fig:delta}
\end{figure*}

\begin{figure}
\begin{center}
    \includegraphics[width=0.5\textwidth, trim=0 0 0 0,  clip=yes]{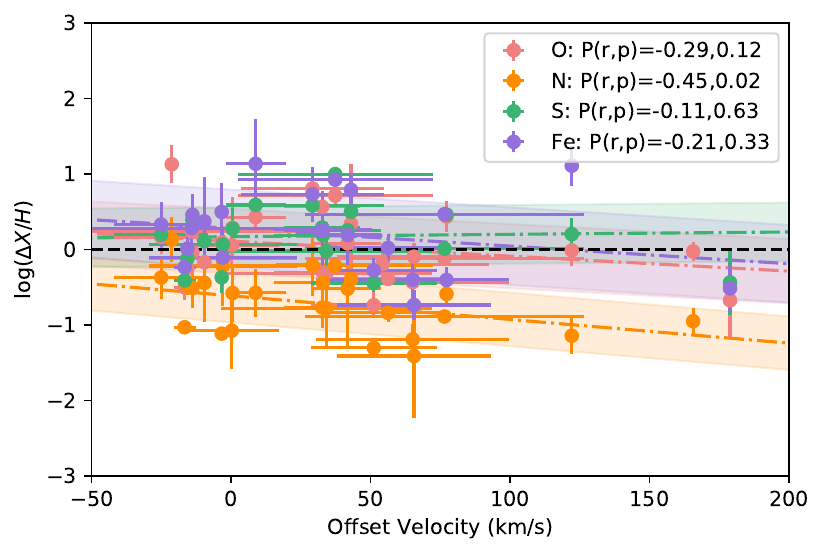}
        \caption{ $\Delta X/H$ as a function of the ISM velocity ($V_{min}$) derived by \citetalias{Parker:2024}.}
\label{fig:delta_Vmin}
\end{center}
\end{figure}

\begin{figure*}
    \centering
    \includegraphics[width=1\linewidth]{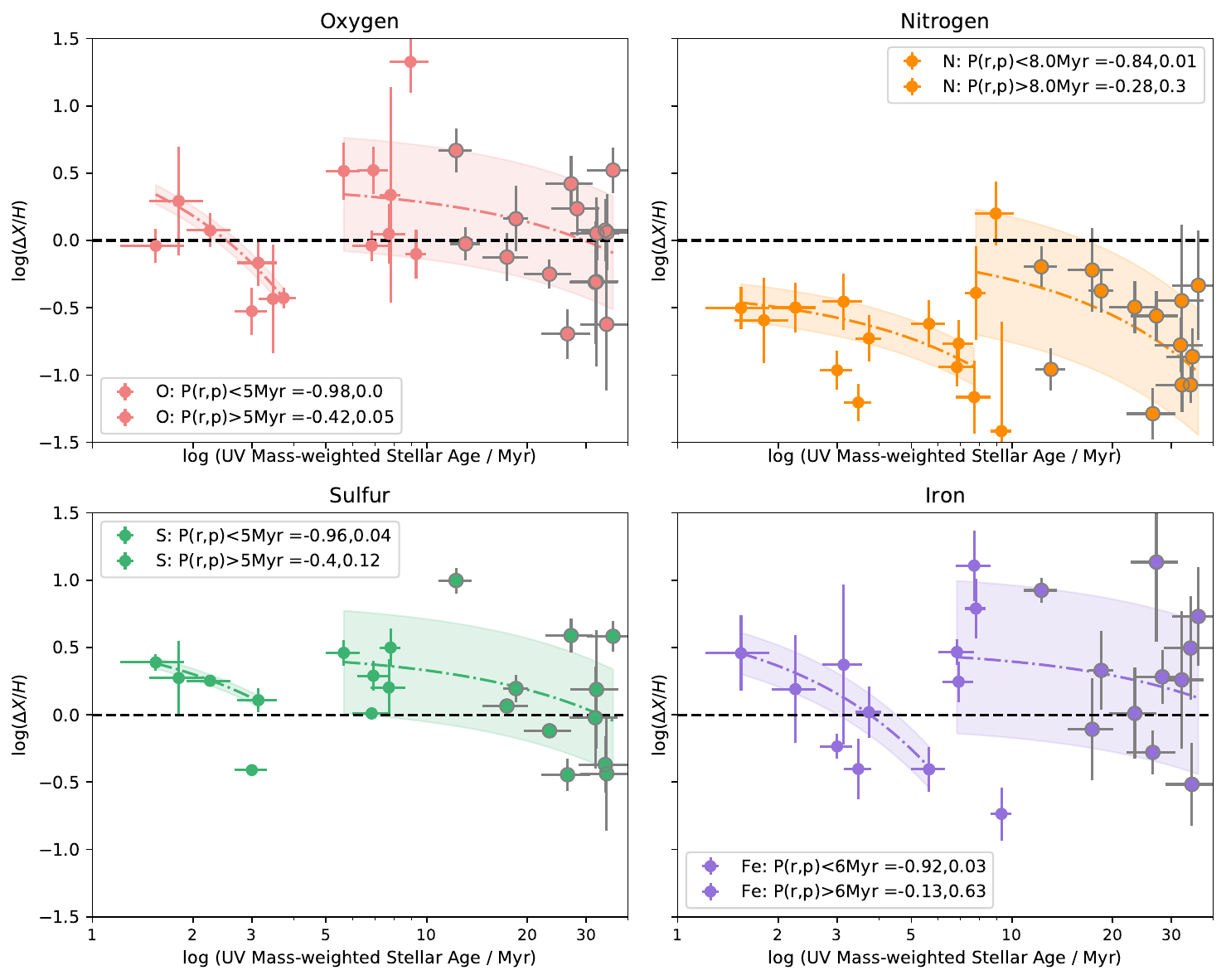}
    \caption{Neutral $-$ ionized gas abundance offset ($\Delta X/H$) as a function of mass-weighted stellar population age (Section~\ref{sec:age}), where $\Delta X/{\rm H} = X/{\rm H}_{\rm neu.}-X/{\rm H}_{\rm ion.}$, such that a positive $\Delta(X/{\rm H})$ implies a higher abundance of $X$ in the neutral gas. Linear regression fits are shown in each panel (dot-dashed lines and shaded regions), along with the corresponding Pearson correlation coefficients, where piece-wise linear fits were performed according to age break-points indicated in the legend. Galaxies classified as multi-burst systems by \citetalias{Parker:2025} are outlined in gray.}
    \label{fig:stellar_age}
\end{figure*}

\subsection{What Controls Gas Mixing Between the Phases?} \label{sec:mixing}
The relative abundances between the neutral and ionized gas phases trace the efficiency of metal transport and mixing in galaxies.  
In low-mass systems, feedback from massive stars, supernovae, and stellar winds injects newly synthesized elements into the ISM, but the degree to which these metals cool and mix into the neutral phase remains uncertain \citep{peeples11,tumlinson17}.  
If metals remain confined to the ionized medium, subsequent generations of stars will form from metal-poor gas, delaying chemical enrichment and sustaining large-scale inhomogeneities \citep{Emerick:2020, James:2020,decia21}.  
Conversely, efficient mixing on short timescales would homogenize the ISM and erase phase-dependent differences.  
To investigate these processes, we examine how the abundance offsets ($\Delta$N/H, $\Delta$O/H, $\Delta$S/H, $\Delta$Fe/H) vary with global galaxy properties and internal conditions (Figures~\ref{fig:delta}a--f).  
Linear relations were fitted using the \texttt{ltsfit} algorithm \citep{Cappellari:2013a}, and correlation strengths were quantified with the Pearson coefficient ($r$) and significance ($p$).

\subsubsection{Metallicity}
Figure~\ref{fig:delta}a shows the abundance offsets as a function of ionized-gas metallicity (12+log(O/H)).  
All four elements (O, N, S, Fe) show weak decreasing correlations with metallicity ($-0.48\le r \le -0.31$).  
O, S, and Fe exhibit modestly higher neutral-phase abundances at low metallicities (12+log(O/H)$\lesssim8.25$), whereas N is offset by roughly 1~dex lower than the other elements, indicating consistently higher ionized-phase abundances. This trend is surprising as no significant connection is expected between the amount of metals in a galaxy and relative distribution of metals between the neutral and ionized phases. 
However, the production mechanisms of the elements considered here may play a role.

High N/O in the ionized phase has been proposed to result from low effective yields of 
oxygen, where $\alpha-$elements are preferentially ejected from ionized regions via core-collapse SNe
(CCSNe, or SNe II) \citep[e.g.,][]{berg19a}.
Such events inject large amounts of O, S, and Fe into hot, metal-rich outflows that may escape low-mass galaxies or remain in the circumgalactic medium \citep[e.g.,][]{MacLow:1999,Recchi:2008,Emerick:2018}, temporarily depleting these elements from the warm ISM. In contrast, nitrogen released by slower, less energetic sources such as AGB stars and WR winds is more readily retained, as these stars are longer-lived, more spatially dispersed, and deposit their material into denser, cooler regions where mixing is efficient. This differential retention naturally leads to an enhancement of N/O in the ionized phase.

\subsubsection{Stellar Mass}
Galaxy mass can influence metal retention and mixing.  
Dwarf galaxies retain only $\sim$5--10\% of their metals in the ISM and stars \citep{McQuinn:2015b}, but whether the neutral and ionized phases are affected differently is unclear.  
Outflow properties derived from emission and absorption lines (\sIii, \sIiii, \sIiv) suggest that broad components trace the same outflowing gas \citep{Xu:25}, implying similar retention in both phases.  
Stellar mass estimates from \citetalias{berg:2022} (Figure~\ref{fig:delta}b) show little correlation with abundance offsets ($P(p)>0.2$), suggesting that mass has little effect on metal mixing between the phases. 

\subsubsection{Star-Formation Rate}
Feedback from massive stars can eject metals into the ionized medium and affect the relative abundances between phases \citep[e.g.,][]{Jecmen:2023, Hopkins:2019}.  
Figure~\ref{fig:delta}c shows abundance offsets as a function of SFR.  
While O, S, and Fe show moderate, non-significant trends ($P(p)=0.12$--0.18), nitrogen exhibits a strong negative correlation ($P(r,p)=-0.38,0.05$), with higher SFR systems showing enhanced ionized-phase nitrogen.  
This likely reflects the short timescales ($\sim$10~Myr) of recent starburst-driven enrichment, consistent with WR stars as the primary source of N.  

For specific SFR (Figure~\ref{fig:delta}d), oxygen shows a statistically significant negative correlation ($P(r,p)=-0.4,0.03$), consistent with enhanced Type~II SNe enrichment in high-sSFR systems.  
S and Fe follow similar, weaker trends, indicating that starburst intensity modulates the relative enrichment of ionized gas.

\subsubsection{Ionization Parameter}
The ionization parameter ($U$, derived in, \citetalias{mingozzi:2022}) traces the strength of the radiation field and the relative ionization state of the gas.  
Figure~\ref{fig:delta}e shows that higher log($U$) correlates with smaller abundance offsets, especially for oxygen ($P(r,p)=-0.45,0.01$), consistent with harder radiation fields driving a larger fraction of metals into the ionized phase.  
Correlations are weaker for N ($P(r,p)=-0.16,0.42$) and moderate for Fe ($P(r,p)=-0.35,0.11$) and S ($P(r,p)=-0.32,0.15$), reflecting differences in ionization potentials and depletion.

\subsubsection{Gas Kinematics}
Outflows and ISM velocities can affect metal mixing. Figure~\ref{fig:delta}f shows abundance offsets versus outflow velocity ($v_{out}$) measured from the broad component of the \sIii\ absorption lines in \citetalias{Xu:22}. All elements show a negative trend with outflow velocity, although only for nitrogen is this correlation highly statistically significant: faster ISM velocities correspond to higher ionized-phase N/H. For elements such as N, O, and S, which are predominantly produced by massive stars, these trends are consistent with enrichment driven by stellar feedback and outflows, and the observed abundance patterns in the ionized gas reflect the expected signature of recent massive-star activity. Figure~\ref{fig:delta_Vmin} shows similar trends using the average ISM velocity ($v_{min}$) derived from the centroid of the main, narrow ISM component of the absorption line fits. These results indicate that nitrogen enrichment is particularly sensitive to gas kinematics, likely because outflows delay mixing of WR-ejected material with the neutral phase, which we discuss further in Section~\ref{sec:nitrogen}.

\subsubsection{Stellar Population Age} \label{sec:age}
Stellar population age strongly influences element release and mixing timescales. In this context,  
UV-based mass-weighted ages are particularly useful as they sample both the young burst populations responsible for rapid WR and Type~II SNe enrichment and older populations where metals have equilibrated between phases.  
By comparison, UV light-weighted ages trace only the most recent burst, while optical ages sample only older Gyr populations (as detailed in Appendix Section~\ref{sec:SPS_Age} and shown in Table~\ref{tab:ages}). Utilizing UV mass-weighted ages,  
Figure~\ref{fig:stellar_age} shows complex trends: offsets decrease from positive to negative until $\sim$5--8~Myr, rise sharply by $\sim$1~dex, then decline gradually, stabilizing by $\sim$20~Myr.  
Each element required a bi-modal linear fit separating `young' ($<5$--8~Myr) and `old' ($>5$--10~Myr) populations.  
The correlation is strongest in young populations ($P(p)=0.01$--0.1) and weaker in older ones, particularly for N and Fe ($P(p)=0.3,0.63$).  

All elements except nitrogen show early ionized-phase enhancements due to Type~II SNe, followed by cooling and mixing that equilibrates O, S, and Fe by $\sim$20--30~Myr.  
Nitrogen remains overabundant in the ionized gas, peaking within $\sim$10~Myr, consistent with WR star yields \citep{henry00}.  
We also indicate which systems are classified as `multi-burst' \citep[][hereafter \PXIII]{Parker:2025}, whose stellar continua were best fit with multiple bursts separated by $>20$~Myr.  
These correspond to older populations where $\Delta$X/H$\sim0$, confirming that sufficient time allows metals to homogenize between phases for S, O, and Fe, but not N.  

\vspace{3mm}
\noindent
Overall, the analysis of abundance offsets reveals a complex, element-dependent picture of chemical mixing in low-mass star-forming galaxies.  
Oxygen, sulfur, and iron are generally well mixed across all systems, with modest ionized-phase enhancements in galaxies with higher sSFRs, harder radiation fields, or younger stellar populations, reflecting prompt enrichment from Type~II SNe and rapid SN-driven turbulent mixing.  
Nitrogen, in contrast, consistently exhibits higher abundances in the ionized gas, with strong dependencies on metallicity, SFR, sSFR, ionization parameter, stellar age, and outflow velocity.  
The UV-based mass-weighted stellar ages provide a particularly powerful diagnostic, sampling both the recent starburst populations responsible for WR-driven nitrogen enrichment and the older stellar populations where metals have equilibrated between phases.  
Multi-burst systems \citetalias{Parker:2025} show $\Delta$X/H$\sim0$, confirming that sufficient time and dynamical mixing allow metals to homogenize.  
These results underscore the importance of both stellar population age and gas dynamics in shaping the phase-dependent chemical structure of low-mass galaxies, and provide a framework for interpreting observed abundance offsets in terms of the timescales of feedback, enrichment, and mixing.

\begin{figure*}
\begin{center}
    \includegraphics[width=0.7\textwidth, trim=0 0 0 0,  clip=yes]{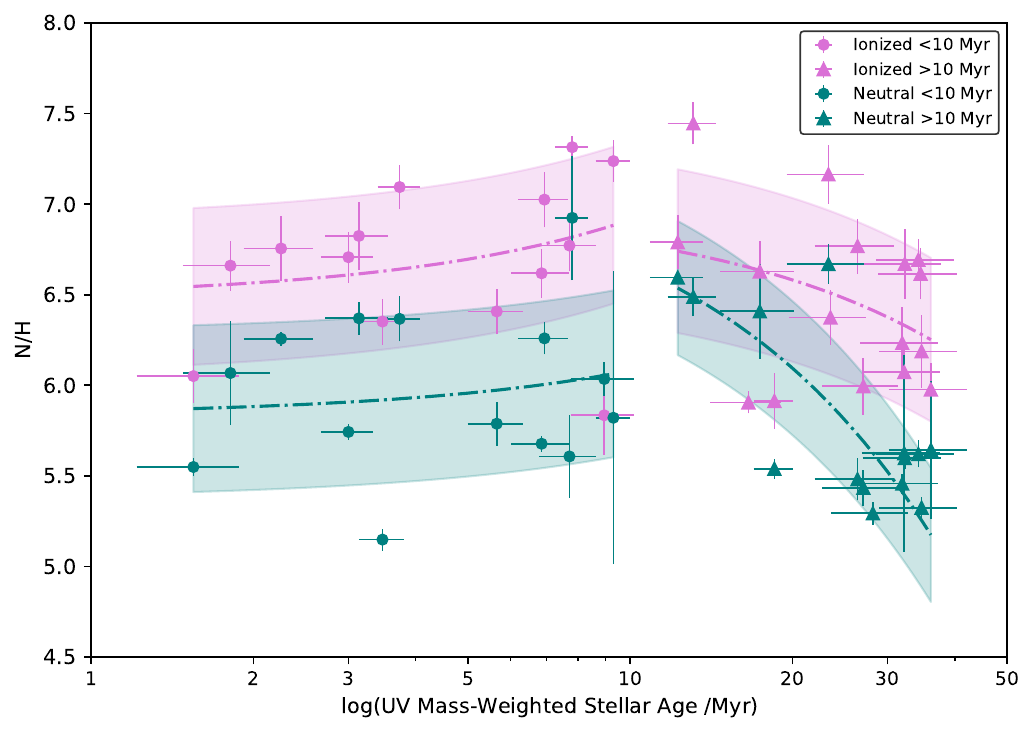}
        \caption{Nitrogen abundances for the neutral and ionized gas as a function of stellar age. Here we separate the sample at a stellar age of $\sim10$\,Myr to highlight the two-piece correlation with age shown in Figure~\ref{fig:stellar_age}.}
\label{fig:N_stellar_age}
\end{center}
\end{figure*}

\begin{figure*}[ht!]
\centering

% ---------- Row 1 ----------
\begin{minipage}[b]{0.49\textwidth}
    \centering
    \includegraphics[width=\textwidth]{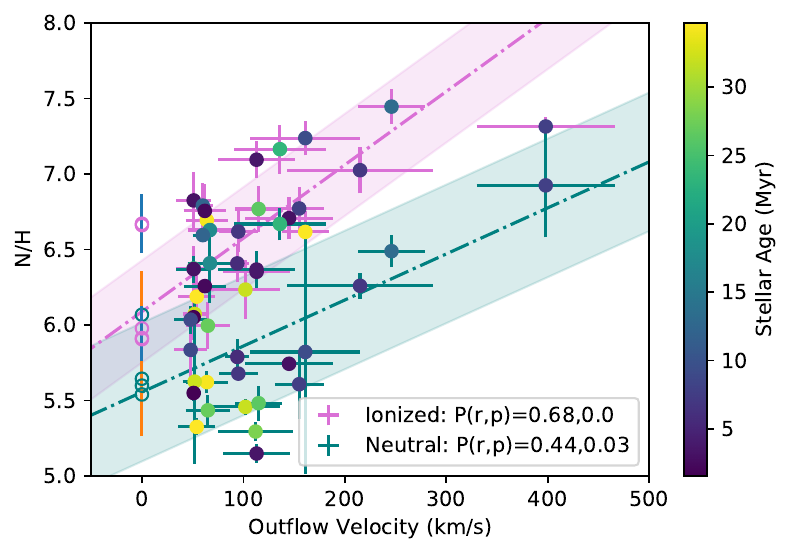}
    \textbf{(a) }
\end{minipage}%
\hfill
\begin{minipage}[b]{0.49\textwidth}
    \centering
    \includegraphics[width=\textwidth]{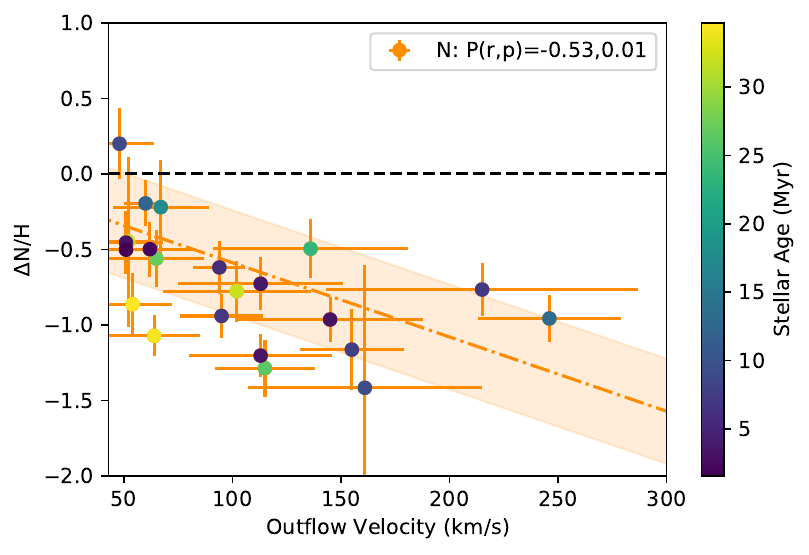}
    \textbf{(b) }
\end{minipage}

\caption{\small Nitrogen abundances for the neutral and ionized gas as a function of outflow velocity. In panel (a) we show the absolute abundance (N/H) separately for each phase, while in panel (b) we show the difference between the two phases ($\Delta$N/H). Both panels are color-coded by mass-weighted stellar age.}
\label{fig:N_outflows}
\end{figure*}

\subsection{Why is Nitrogen Behaving Differently?} \label{sec:nitrogen}

As shown in Section~\ref{sec:mixing}, several global and internal parameters correlate with an enhancement in the amount of nitrogen in the ionized gas relative to the neutral phase.  Increasing stellar mass, star-formation rate, outflow velocity, and stellar population age (particularly up to $\sim$10~Myr) all correspond to a larger ionized-gas offset for N/H.  Regardless of the property considered, nitrogen is consistently more abundant in the ionized phase than in the neutral phase, unlike O, S, and Fe.  This systematic behavior implies that the production, release, and transport of nitrogen differ fundamentally from those of $\alpha$- and iron-peak elements in the ISM of star-forming galaxies.

To disentangle the processes controlling this behavior, Figure~\ref{fig:N_stellar_age} shows N/H in both the ionized and neutral phases as a function of stellar age.  The absolute abundance of nitrogen increases slightly in both phases for $<10$~Myr, subsequently decreasing slightly for the ionized-gas but strongly for the neutral-gas, indicating that the total amount of nitrogen is not solely governed by the timescale on which it was produced.  However, Figure~\ref{fig:N_stellar_age} demonstrates that the \emph{relative} abundance, $\Delta$(N/H), depends strongly on stellar age within the first $\sim$10~Myr of a burst, implying that the enrichment source itself---rather than the global evolutionary timescale---drives the observed offsets. In other words, the phase differences reflect the immediate chemical imprint of massive-star feedback rather than gradual, global mixing processes. Once galaxies evolve beyond $\sim$10~Myr, $\Delta$(N/H) decreases and the phases approach chemical equilibrium, consistent with the end of the Wolf--Rayet (WR) phase and subsequent mixing of the ejecta.  We did investigate whether there was a correlation between stellar age and outflow velocity, but found none, suggesting that efficient mixing in older systems is not simply the result of weaker feedback.

If WR stars dominate nitrogen production in these systems, their fast, ionized stellar winds should both enrich and dynamically disturb the surrounding gas. Figure~\ref{fig:N_outflows}a shows a strong correlation between both N/H$_{ion}$ and N/H$_{neu}$ with $v_\mathrm{out}$ ($P(r,p)=0.68,0.01$ and $0.44, 0.03$, respectively), while the offset $\Delta$N/H exhibits an even stronger dependence (Figure~\ref{fig:N_outflows}b; see also Figure~\ref{fig:delta}f). These trends indicate that the same winds responsible for nitrogen injection also drive the observed ISM kinematics, with higher-velocity systems manifesting a larger fraction of nitrogen in the ionized phase. A similar relationship between $v_\mathrm{out}$ and N/O was reported by \citetalias{arellano-cordova:2025b}, who found more elevated N/O ratios in systems with stronger outflows. In this scenario, nitrogen-rich, high-velocity winds inhibit efficient cooling and mixing with the neutral medium, thereby maintaining the observed phase-dependent abundance offsets.

\begin{figure*}[ht!]
\centering

% ---------- Row 1 ----------
\begin{minipage}[b]{0.49\textwidth}
    \centering
    \includegraphics[width=\textwidth]{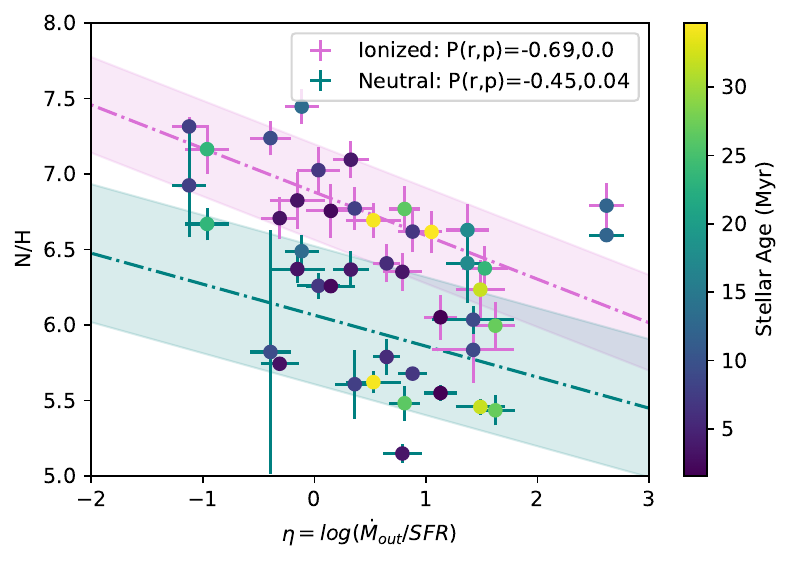}
    \textbf{(a)}
\end{minipage}%
\hfill
\begin{minipage}[b]{0.49\textwidth}
    \centering
    \includegraphics[width=\textwidth]{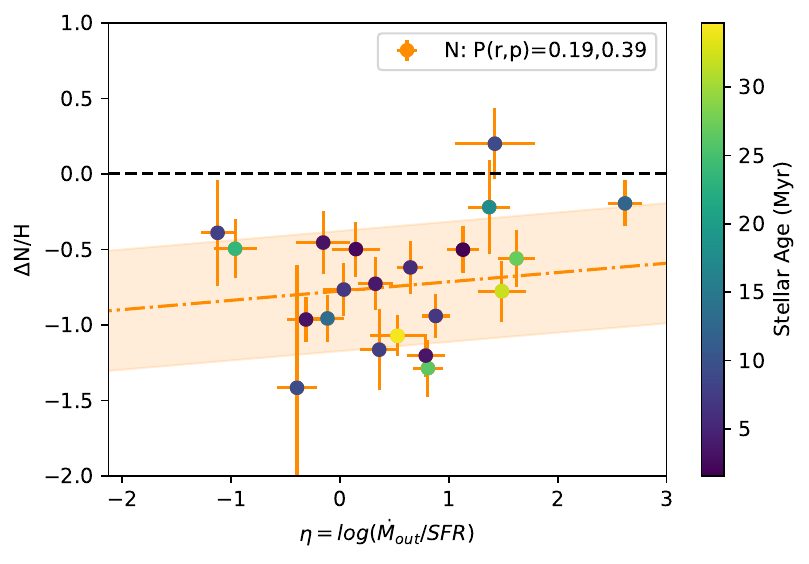}
    \textbf{(b)}
\end{minipage}

\caption{\small Nitrogen abundances for the neutral and ionized gas as a function of mass outflow factor, $\eta$. In panel (a) we show the absolute abundance (N/H) separately for each phase, while in panel (b) we show the difference between the two phases ($\Delta$N/H).}
\label{fig:N_eta}
\end{figure*}

We further examine the role of feedback strength using the mass-loading factor, $\eta$, defined as the ratio of mass outflow rate ($\dot{M}_{\rm out}$) to SFR, as measured by \citetalias{Xu:22}.  As shown in Figure~\ref{fig:N_eta}, both ionized and neutral N/H decrease strongly with increasing $\eta$ ($P(r,p)=-0.69,0.01$ and $-0.45,0.04$, respectively). If we consider N/H to be a proxy for the overall metal content (and hence mass) of the galaxy, the correlation is consistent with low-mass galaxies exhibiting more efficient outflows.  This behavior accords with the findings of \citetalias{Parker:2025}, who showed that systems with low $\eta$ display higher N/O in the ionized gas, implying that inefficient winds allow nitrogen to remain and cool within the ISM.  Conversely, large $\eta$ systems lose a greater fraction of their freshly produced metals, leading to lower overall abundances in both phases.  Although one might expect $\Delta$N/H to correlate with $\eta$, such a relationship is not observed, indicating that feedback-driven metal loss affects both phases similarly, whereas the persistence of nitrogen in the ionized medium is instead tied to its mode of ejection and ionization structure.

Overall, the data reveal that nitrogen remains overabundant in the ionized gas under all conditions examined.  While this could suggest that the cooling and mixing timescales for nitrogen exceed the stellar ages sampled here, Figure~\ref{fig:N_stellar_age} demonstrates that the gas becomes chemically well mixed on $\sim$10--35~Myr timescales for some systems.  The magnitude of $\Delta$N/H increasing with ISM velocity therefore implies that in systems with strong outflows, a portion of the ionized nitrogen may be expelled before cooling and recombining into the neutral phase.  The opposite trend with stellar mass---larger offsets in more massive, metal-rich systems with larger outflow velocities \citetalias{Xu:22}---may instead reflect ionization-structure effects: since the ionization potential of N is higher than that of H, a smaller fraction of nitrogen will reside in the neutral state.  We note that removing the neutral-phase ICF$_{\mathrm{tot}}$ corrections would increase $\Delta$N/H further, reinforcing that this offset is real and not an artifact of ionization modeling.

\begin{figure*}
    \centering
    \includegraphics[width=0.8\linewidth]{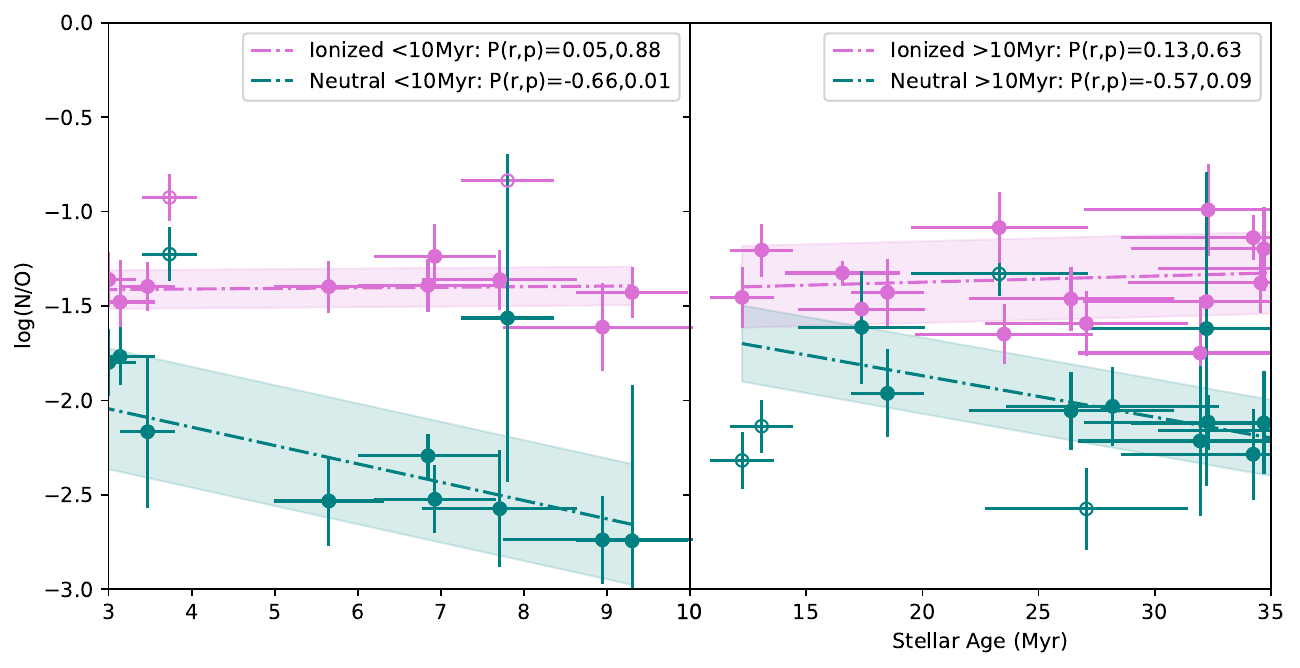}
    \caption{N/O as a function of stellar age. Here we separate the sample at a stellar age of $\sim10$\,Myr to highlight the two-piece correlation with age shown in Figure~\ref{fig:stellar_age}. Linear regression fits are
shown in each panel (dot-dashed lines and shaded regions), along with the corresponding Pearson correlation coefficients, where
separate fits were performed on the $<$ 10 Myr and $>$ 10 Myr populations. Empty symbols were removed from the fit via the sigma-clipping algorithm.}
    \label{fig:NO_age}
\end{figure*}

\begin{figure*}
    \centering
    \includegraphics[width=0.7\linewidth]{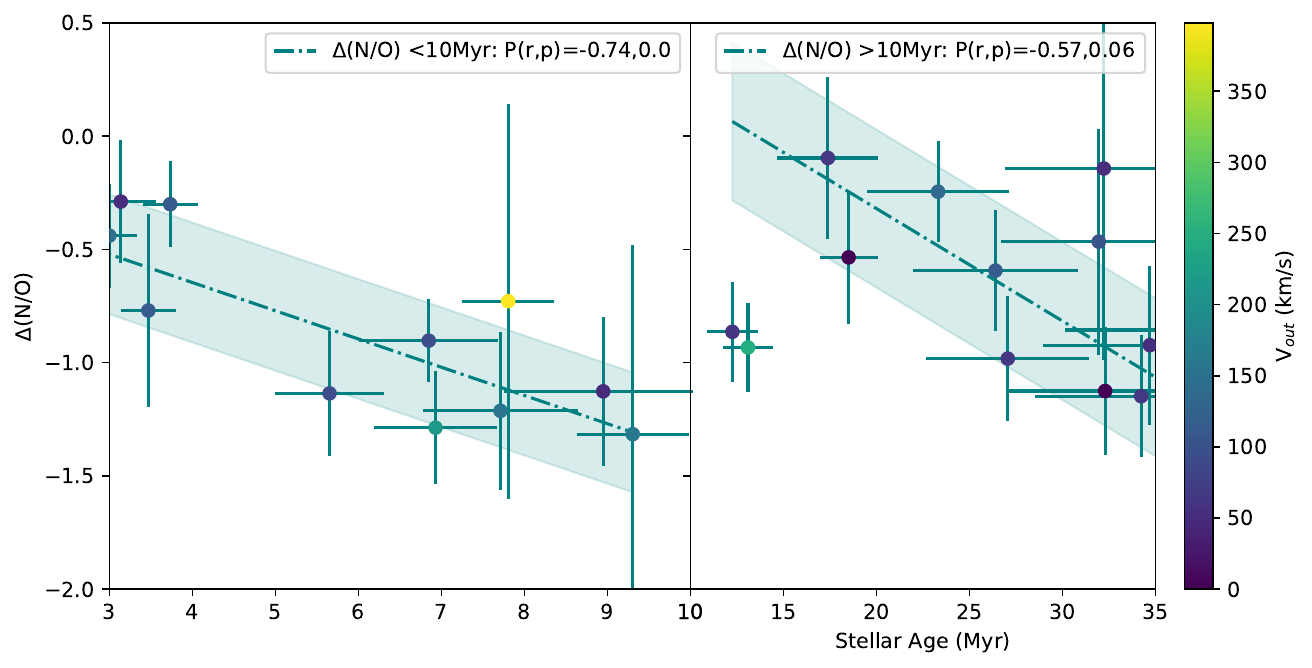}
    \caption{$\Delta(\rm N/O)$ ($={\rm N/O}_{\rm neu.} - {\rm N/O}_{\rm ion.}$) as a function of stellar population age, color-coded by outflow velocity (V${out}$). Here we separate the sample at a stellar age of $\sim10$\,Myr to highlight the two-piece correlation with age shown in Figure~\ref{fig:stellar_age}.}
    \label{fig:dNO_age}
\end{figure*}

To determine whether the observed nitrogen excess arises from recent enrichment or delayed mixing, we examine the N/O ratio in both phases as a function of stellar age (Figure~\ref{fig:NO_age}). For the youngest systems ($<$10~Myr), N/O${_{ion}}$ remains approximately constant with age, while N/O${_{neu}}$ decreases significantly ($P(p)=-0.66,0.01$) and displays substantial scatter—consistent with freshly ejected, nitrogen-rich material from WR stars that has not yet mixed into the cooler ISM. At older ages ($>$10~Myr), N/O${_{ion}}$ stays roughly constant, but N/O${_{neu}}$ rises by nearly 1 dex before gradually declining again with increasing age ($P(p)=-0.57, 0.09$). The nitrogen seen in the neutral gas represents material expelled during earlier bursts that has cooled and mixed into the ISM, while the ionized phase traces additional N enrichment from WR stars during the most recent burst over a timescale of $\sim$3 Myr. Since oxygen from core-collapse SNe is released on longer $\sim$10 Myr timescales, this introduces a natural time lag in the production and mixing of nitrogen relative to oxygen, which manifests as the evolving N/O ratio across both phases. The evolution of the phase offset $\Delta$(N/O) (Figure~\ref{fig:dNO_age}) supports this scenario: a strong anti-correlation at young ages ($P(p)=0.01$) reflects the WR enrichment phase, after which $\Delta$(N/O) increases by $\sim$1.5 dex as the ISM becomes chemically homogenized, then gradually declines as mixing continues on longer timescales.

A remaining puzzle is that only four CLASSY galaxies exhibit detectable WR spectral features, despite the nitrogen signatures implying WR-driven enrichment.  The discrepancy likely reflects observational and temporal limitations: WR features are short-lived ($\lesssim3$~Myr) and can be diluted in globally integrated spectra \citep{Lagos:2012,James:2013b,Kumari:2018}. Even after the WR phase ends, their nitrogen-rich ejecta can persist in the ionized ISM, explaining why the chemical imprint outlasts the direct spectral signature.  Alternatively, other prompt enrichment channels, such as rapidly rotating massive stars or binary interactions, could contribute to early nitrogen release, similar to those proposed for high-redshift galaxies.

To summarize, the combined evidence from stellar age, kinematics, and mass loading demonstrates that nitrogen traces feedback-regulated enrichment in the ionized phase. Massive stars rapidly inject nitrogen through stellar winds and subsequent supernovae, driving outflows that both enrich and dynamically shape the surrounding ISM. The close link between N/H, outflow velocity, and mass-loading factor suggests that stellar winds—particularly from WR stars—may play a more significant role in powering galactic outflows than previously thought. Indeed, as noted by \citet{murray05}, winds from massive stars can supply a momentum flux comparable to that of supernovae \citep[see also][]{leitherer99}, supporting the view that stellar winds alone can launch substantial outflows prior to the onset of SNe-driven feedback. While canonical models attribute the bulk of feedback energy to Type~II supernovae \citep[e.g.,][]{heckman15}, our results imply that massive stars dominate the earliest and most chemically influential stages of outflow development. On $\sim$10~Myr timescales, subsequent mixing redistributes these metals into the neutral phase. Taken together, these findings point to a more prominent role for stellar winds in regulating both the dynamics and chemical evolution of galaxies—an idea that warrants further exploration through simulations coupling enrichment, feedback, and multiphase gas transport.

\subsubsection{N-enrichment at high-$z$}
As discussed in Section~\ref{sec:mixing}, the relationship between stellar population age, ionized outflows, and gas-phase abundances provides a direct window into the cycle of metal enrichment and redistribution.  
In particular, Figure~\ref{fig:N_outflows} illustrates how systems with strong outflows and young, burst-dominated stellar populations show elevated N/H in the ionized phase, implying recent nitrogen enrichment by massive stars whose winds have not yet mixed with the cooler gas.  
These local signatures of feedback-driven N enrichment provide a crucial reference point for interpreting the ``nitrogen-loud'' galaxies that JWST is now revealing at high redshift.

Nitrogen is an intriguing element that has become a focus of several recent studies, as JWST is revealing high-$z$ systems with extremely high N/O ratios.  
The first of these was GNz11, a galaxy at $z\sim10.6$ with unusually bright rest-frame UV emission lines \citep{Bunker:2023}, showing a level of nitrogen significantly higher than that found in local star-forming galaxies of comparable metallicity (log(N/O)$=-0.38$).  
Several more ``nitrogen-loud'' galaxies have since been discovered, including CEERS-1019 \citep[log(N/O)$=-0.38$;][]{Finkelstein:2022,Tang:2023}, GHZ2/GLASS-z12 \citep[log(N/O)$=-0.29$;][]{Castellano:2024}, GN-z9p4 \citep[log(N/O)$=-0.59$;][]{Schaerer:2024}, and RXCJ2248-4431 \citep[log(N/O)$=-0.39$;][]{Topping:2024}.  
How these systems achieved such high N/O ratios within such short evolutionary timescales—only $\sim$440\,Myr after the Big Bang in the case of GNz11—is puzzling, challenging theoretical stellar yields and galactic chemical evolution, particularly in the early Universe \citep[see][for a review]{Nandal:2024}.  

Several explanations have been proposed for the large amount of nitrogen in their ionized gas, including signatures of globular cluster precursors, massive star winds, runaway stellar collisions, tidal disruption events, or the presence of an AGN or accreting black hole \citep{Cameron:2023,Charbonnel:2023,Maiolino:2024,Senchyna:2024}.  
Chemical-evolution models have also managed to reproduce these high N/O levels via two main pathways: (1) rapid chemical enrichment in a dual-starburst model, where WR stars dominate during the second burst \citep{Kobayashi:2024}, and (2) the contribution of very massive ($\sim$2000\,\Msol) Pop~III stars with ejecta rich in $^4$He, $^1$H, and $^{14}$N \citep{Nandal:2024}.  
However, spectral signatures of WR stars have not yet been observed in GNz11 despite strong \niv~$\lambda\lambda$1483,86, \civ~$\lambda\lambda$1548,50, and \heii~$\lambda$1640 emission \citep[with a possible broad WR spectral signature;][]{Maiolino:2024}.  
If these features originated from WR winds, comparable strength should also appear in \niv~$\lambda$1718 and \ion{N}{5}~$\lambda\lambda$1238,40 \citep{Senchyna:2021}.  
In contrast, in the Sunburst Arc ($z=2.37$), \citet{Rivera-thorsen:2024} detected clear WR signatures but concluded that the extreme and rapid nitrogen enrichment ($\log(\rm N/O)=-0.74$) could not be explained by binary WR models alone, likely requiring an additional population of very massive stars (VMSs).  
The link between VMSs ($>100$\,\Msol) and N-enriched galaxies has also been proposed by \citet{Vink:2024}, as such stars contribute to both N-enrichment and strong \heii\ emission.  

The nitrogen trends observed in CLASSY provide an essential low-redshift analogue for understanding these chemical conditions now being uncovered at cosmic dawn.
In our nearby starbursts, high N/H ratios in the ionized phase are often associated with the presence of WR stars - despite the lack of shortly-lived WR spectral features - and their enriched winds, which inject nitrogen into the ISM and drive metal-loaded outflows out of the galactic plane (i.e., beyond the neutral ISM), that delay subsequent mixing with the neutral gas reservoir.
This combination of localized stellar feedback and incomplete mixing naturally explains the offsets between ionized and neutral gas metallicities (see Figures~\ref{fig:N_outflows},\ref{fig:N_eta}), as well as the broad, nitrogen-enriched UV emission components (\ion{N}{4}], \ion{N}{5}, \ion{He}{2}) observed in both local and high-redshift starbursts \citep[e.g.,][]{Senchyna:2021, Rivera-thorsen:2024, Maiolino:2024}.
Such conditions offer a clear physical pathway for the extreme N/O ratios seen in early galaxies: nitrogen enrichment dominated by WR winds, combined with inefficient metal retention and mixing in the turbulent ISM of rapidly assembling systems.

If the nitrogen emission in high-$z$ ``nitrogen-loud'' systems is predominantly nebular, the CLASSY galaxies offer direct low-redshift analogues for understanding their chemical conditions.
Our analysis shows that WR-driven winds in young, burst-dominated systems can enrich the ionized gas in nitrogen while simultaneously driving outflows that expel this material before it can cool and mix with the neutral ISM.
In this framework, the extreme N/O ratios observed at high redshift naturally arise from localized, rapid enrichment coupled with inefficient mixing, rather than from delayed chemical evolution.
Ionized outflows are known to be widespread at high redshift, with 25–40\% of $3<z<9$ galaxies in the JADES survey showing signatures of outflowing gas \citep{Carniani:2024}; this fraction is likely a lower limit, as the relatively low spectral resolution of current data ($R\sim1000$) may obscure weaker or narrower wind components.
The absence of conclusively detected broad outflows in the nitrogen-loud galaxies may therefore reflect observational limitations rather than intrinsic differences in feedback.
Together, these findings suggest that the high N/O levels in early galaxies are best explained by nitrogen-rich material ejected into the ISM by WR stars within the most recent $\sim$3–5\,Myr starburst, consistent with the dual-burst scenario of \citet{Kobayashi:2024}.

The local–high-$z$ connection seen in nitrogen thus provides a unified picture: both regimes trace young, burst-dominated galaxies where stellar feedback and outflow dynamics regulate the pace of chemical mixing.
The CLASSY galaxies, with their intense radiation fields, elevated ionization parameters, and fast, metal-rich winds, represent nearby analogues of the rapidly assembling systems now accessible with JWST.
Understanding these processes in the local Universe is therefore essential for interpreting the emerging population of nitrogen-loud galaxies at $z>7$, and for disentangling the roles of stellar yields, feedback-driven gas dynamics, and timescale-dependent mixing in shaping galactic chemical evolution across cosmic time.

\section{Conclusions}\label{sec:conclusion}
We have presented a comprehensive multi-phase chemical analysis of 31 nearby star-forming galaxies from the CLASSY survey, which serve as analogues of high-redshift systems owing to their low metallicities, compact morphologies, and intense radiation fields.  
By directly comparing the metal content of the ionized and neutral gas phases using matched UV and optical diagnostics, we quantified how key elements (N, O, S, and Fe) are distributed, mixed, and transported within the ISM, and identified the stellar and galactic conditions that regulate this process.

Neutral-gas abundances were derived from Voigt-profile fits to the UV absorption lines of \Ni, \feii, \sii, \oi, and \pii, corrected for ionization and line-of-sight contamination using tailored \textsc{cloudy} models.  
Ionized-gas abundances were taken from \citetalias{mingozzi:2022}, \citetalias{arellano-cordova:2024}, and \citetalias{arellano-cordova:2025b}, with Fe/H$_{\rm ion.}$ newly derived here from \feiii.  
Our combined dataset enables a robust, element-by-element comparison of the ionized and neutral metal reservoirs across a wide dynamic range in mass, metallicity, and star formation rate.  
The main results are summarized below.

\begin{enumerate}
    \item Oxygen and sulfur show consistent abundances between the neutral and ionized phases, indicating efficient large-scale mixing of $\alpha$-elements produced in short-lived massive stars (core-collapse SNe). The remaining intrinsic scatter in both phases reflects local variations in recent enrichment and regions where metals have not yet fully homogenized, revealing small-scale chemical inhomogeneities in the ISM that have not yet mixed on $\sim$10~Myr timescales.
    
   \item Iron is modestly higher in the neutral phase, even without a full correction for dust depletion, which remains constrained only as a lower limit for all but three galaxies. This likely reflects a combination of enrichment from previous star formation episodes—since Fe from Type~Ia SNe requires $\sim$1,Gyr to accumulate and mix into the cooler ISM—and preferential depletion of Fe onto dust in the ionized phase, where core-collapse SNe shocks can efficiently lock Fe into grains.
    
    \item Nitrogen exhibits the largest phase offset, with N/H$_{\rm ion.}$ systematically higher than N/H$_{\rm neu.}$ by $\sim$0.7\,dex in 35/36 systems. The ionized enhancement is consistent with fresh nitrogen from Wolf–Rayet (WR) stars and massive star winds on $\sim$3–5\,Myr timescales, prior to cooling and mixing with the neutral gas.
    
    \item While N/O$_{\rm ion.}$ lies on the canonical low-metallicity plateau, N/O$_{\rm neu.}$ is systematically lower, driving an average offset of $\sim$0.8\,dex. In contrast, S/O ratios agree between phases, consistent with their shared nucleosynthetic origin. Both O/Fe and S/Fe ratios are super-solar ($\sim$0.7\,dex) and consistent between phases, reflecting $\alpha$-enhanced enrichment from recent bursts of star formation.
    
    \item The abundance offsets $\Delta(X/H) = X/H_{\rm neu.} - X/H_{\rm ion.}$ correlate most strongly with stellar population age and gas kinematics. Within $\sim$10\,Myr of a burst, the ionized gas becomes progressively enriched in O, S, and Fe produced by core-collapse SNe relative to the neutral medium, after which these elements begin to cool and mix. In contrast, Fe from Type Ia SNe, released on $\sim$1\,Gyr timescales, contributes to the gradual enrichment of the neutral phase. Nitrogen behaves differently: $\Delta$(N/H) depends most strongly on outflow velocity, implying that stellar feedback regulates whether newly produced nitrogen remains trapped in the ionized phase.
    
    \item The nitrogen behavior in CLASSY mirrors the ``nitrogen-loud'' galaxies recently uncovered by JWST at $z>7$. If the high-$z$ N emission is predominantly nebular, our results indicate that WR-driven winds eject nitrogen straight from the ionized gas, bypassing the neutral phase and delaying mixing. CLASSY galaxies thus provide vital low-$z$ analogues for interpreting these extreme early chemical signatures.
\end{enumerate}

In summary, this work demonstrates that UV absorption-line spectroscopy provides a uniquely powerful probe of the chemical coupling between different ISM phases.
By linking multi-phase abundances to stellar population age, feedback strength, and gas dynamics, we show that the efficiency of metal mixing varies strongly over $\sim$10\,Myr timescales and depends on both the source and the transport mechanism of enriched material.
The tight connection between nitrogen enrichment and outflow strength implies that massive stars, not just supernovae, play a central role in powering and chemically shaping galactic winds, motivating future theoretical and observational efforts to quantify their role in driving early outflows.
Future UV-capable observatories such as the \textit{Habitable Worlds Observatory} will be essential for extending such multi-phase studies to statistically significant samples across cosmic time, thereby unveiling how feedback, outflows, and star formation jointly govern the chemical evolution of galaxies.

\begin{acknowledgments}
BLJ, MM, and SH are thankful for support from the European Space Agency (ESA).
DAB is grateful for the support for this program, HST-GO-15840 provided by 
NASA through a grant from the Space Telescope Science Institute, 
which is operated by the Associations of Universities for Research in Astronomy, 
Incorporated, under NASA contract NAS5-26555. 
We acknowledge the insightful discussions with Annalisa De Cia in conducting depletion corrections.
KZAC acknowledges support from a UKRI Frontier Research Guarantee Grant (PI Cullen; grant reference: EP/X021025/1). CK acknowledges funding from the UK Science and Technology Facility Council through grant ST/Y001443/1.
\end{acknowledgments}

\bibliography{CLASSY}

\appendix

\begin{sidewaystable}
% \hspace*{-2in}	
\begin{tiny}
\caption{Column density measurements and $b$-parameters. Uncertainties on $b$ parameters denoted as N/A correspond to cases where best fits were found by fixing the $b$-parameters to those derived by (\citetalias{Parker:2024}). Lines known to be saturated are shown as lower limits.}\label{tab:column_dens}
% \centering   
\begin{tabular}{lcccccccccccccc}
Target 	& 	z	&	N(HI)			&	N(\Ni)			&	$b$			&	N(\oi)			&	$b$			&	N(\pii)			&	$b$			&	N(\sii)			&	$b$			&	N(\feii)			&	$b$			&	N(\feiii)			&	$b$			\\
	\hline\hline			
	J0127-0619	&	0.0054	&	21.43	$\pm$	0.27	&	15.50	$\pm$	0.10	&	29.98	$\pm$	15.50	& $>$	15.24			&	23.67	$\pm$	2.41	&	13.80	$\pm$	0.10	&	29.69	$\pm$	11.34	&	15.77	$\pm$	0.05	&	40.59	$\pm$	5.68	&	15.19	$\pm$	0.11	&	27.53	$\pm$	3.56	&	\ldots	&	\ldots	\\
	J0144+0453	&	0.0052	&	20.18	$\pm$	0.02	&	14.35	$\pm$	0.09	&	55.90	$\pm$	N/A	& $>$	16.05			&	55.90	$\pm$	5.14	&	13.52	$\pm$	0.20	&	55.90	$\pm$	N/A	&	15.22	$\pm$	0.08	&	55.90	$\pm$	N/A	&	15.04	$\pm$	0.22	&	14.97	$\pm$	3.00	&	\ldots	&	\ldots	\\
	J0337-0502	&	0.0135	&	21.81	$\pm$	0.00	&	\ldots	&	33.14	$\pm$	2.84	& $>$	14.92			&	13.13	$\pm$	0.56	&	12.97	$\pm$	0.15	&	30.87	$\pm$	12.48	&	14.96	$\pm$	0.03	&	34.49	$\pm$	2.62	&	15.32	$\pm$	0.04	&	42.10	$\pm$	4.74	&	14.24	$\pm$	0.07	&	41.43	$\pm$	9.50	\\
	J0405-3648	&	0.0028	&	20.80	$\pm$	0.02	&	14.45	$\pm$	0.38	&	40.63	$\pm$	75.43	& \ldots			&	24.89	$\pm$	2.33	&	13.33	$\pm$	0.11	&	24.89	$\pm$	N/A	&	14.90	$\pm$	0.11	&	75.27	$\pm$	54.13	&	14.89	$\pm$	0.33	&	7.88	$\pm$	4.66	&	\ldots	&	\ldots	\\
	J0823+2806	&	0.0472	&	21.61	$\pm$	0.03	&	16.28	$\pm$	0.10	&	90.43	$\pm$	N/A	& $>$	16.41			&	90.44	$\pm$	8.44	&	\ldots	&	\ldots	&	16.03	$\pm$	0.03	&	101.89	$\pm$	7.65	&	15.80	$\pm$	0.34	&	86.73	$\pm$	32.39	&	14.69	$\pm$	0.16	&	90.44	$\pm$	N/A	\\
	J0934+5514	&	0.0025	&	21.24	$\pm$	0.00	&	14.54	$\pm$	0.06	&	33.49	$\pm$	26.32	& $>$	14.50			&	15.86	$\pm$	2.08	&	13.25	$\pm$	0.19	&	15.86	$\pm$	N/A	&	14.96	$\pm$	0.05	&	15.86	$\pm$	N/A	&	14.81	$\pm$	0.06	&	17.75	$\pm$	1.23	&	14.96	$\pm$	0.26	&	15.86	$\pm$	N/A	\\
	J0938+5428	&	0.1021	&	20.34	$\pm$	0.08	&	14.82	$\pm$	0.04	&	75.15	$\pm$	15.05	& $>$	15.26			&	126.74	$\pm$	8.05	&	13.85	$\pm$	0.08	&	126.74	$\pm$	N/A	&	15.26	$\pm$	0.07	&	126.74	$\pm$	N/A	&	14.73	$\pm$	0.11	&	126.74	$\pm$	N/A	&	\ldots	&	\ldots	\\
	J0940+2935	&	0.0017	&	21.26	$\pm$	0.01	&	14.72	$\pm$	0.05	&	58.84	$\pm$	N/A	& \ldots			&	58.84	$\pm$	12.55	&	13.60	$\pm$	0.12	&	58.87	$\pm$	N/A	&	15.38	$\pm$	0.38	&	12.51	$\pm$	10.31	&	15.42	$\pm$	0.41	&	13.89	$\pm$	3.27	&	\ldots	&	\ldots	\\
	J0944-0038	&	0.0048	&	21.67	$\pm$	0.05	&	15.29	$\pm$	0.06	&	110.22	$\pm$	32.70	& $>$	15.49			&	53.23	$\pm$	16.00	&	14.52	$\pm$	0.09	&	135.08	$\pm$	42.21	&	15.45	$\pm$	0.20	&	53.23	$\pm$	N/A	&	15.81	$\pm$	0.38	&	29.94	$\pm$	6.51	&	\ldots	&	\ldots	\\
	J0944+3442	&	0.0200	&	21.51	$\pm$	0.03	&	15.11	$\pm$	0.05	&	75.81	$\pm$	N/A	& $>$	15.83			&	75.81	$\pm$	11.52	&	13.92	$\pm$	0.13	&	75.81	$\pm$	N/A	&	\ldots	&	\ldots	&	14.73	$\pm$	0.21	&	71.70	$\pm$	11.97	&	15.38	$\pm$	0.08	&	82.39	$\pm$	13.12	\\
	J1024+0524	&	0.0332	&	20.58	$\pm$	0.07	&	14.46	$\pm$	0.10	&	238.91	$\pm$	108.45	& $>$	15.64			&	106.93	$\pm$	19.41	&	13.57	$\pm$	0.17	&	106.93	$\pm$	N/A	&	15.16	$\pm$	0.06	&	129.61	$\pm$	25.84	&	14.22	$\pm$	0.14	&	74.50	$\pm$	43.20	&	14.89	$\pm$	0.03	&	157.53	$\pm$	12.40	\\
	J1025+3622	&	0.1265	&	20.69	$\pm$	0.19	&	14.34	$\pm$	0.13	&	106.46	$\pm$	N/A	& $>$	15.84			&	106.46	$\pm$	9.23	&	\ldots	&	\ldots	&	15.26	$\pm$	0.09	&	106.46	$\pm$	N/A	&	15.55	$\pm$	0.13	&	106.46	$\pm$	N/A	&	14.57	$\pm$	0.17	&	88.50	$\pm$	30.91	\\
	J1044+0353	&	0.0129	&	21.85	$\pm$	0.44	&	15.47	$\pm$	0.32	&	15.32	$\pm$	4.56	& $>$	15.21			&	22.97	$\pm$	2.78	&	13.85	$\pm$	0.08	&	76.56	$\pm$	18.01	&	15.47	$\pm$	0.04	&	39.09	$\pm$	5.55	&	15.58	$\pm$	0.09	&	21.33	$\pm$	1.44	&	14.35	$\pm$	0.12	&	22.97	$\pm$	N/A	\\
	J1105+4444	&	0.0215	&	21.91	$\pm$	0.11	&	15.39	$\pm$	0.03	&	85.17	$\pm$	16.74	& $>$	15.74			&	110.33	$\pm$	3.62	&	14.10	$\pm$	0.05	&	77.98	$\pm$	22.25	&	15.93	$\pm$	0.04	&	65.23	$\pm$	7.77	&	15.73	$\pm$	0.11	&	46.42	$\pm$	5.40	&	15.00	$\pm$	0.06	&	143.41	$\pm$	62.79	\\
	J1119+5130	&	0.0045	&	20.77	$\pm$	0.03	&	14.22	$\pm$	0.10	&	56.88	$\pm$	N/A	& $>$	16.07			&	56.88	$\pm$	8.92	&	13.38	$\pm$	0.15	&	56.88	$\pm$	N/A	&	15.19	$\pm$	0.11	&	56.88	$\pm$	N/A	&	15.64	$\pm$	0.49	&	7.05	$\pm$	2.19	&	\ldots	&	\ldots	\\
	J1129+2034	&	0.0047	&	21.11	$\pm$	0.01	&	15.48	$\pm$	0.09	&	33.07	$\pm$	3.78	& $>$	15.78			&	59.10	$\pm$	14.27	&	13.82	$\pm$	0.09	&	56.32	$\pm$	19.70	&	15.74	$\pm$	0.09	&	36.30	$\pm$	12.34	&	15.27	$\pm$	0.10	&	26.99	$\pm$	3.20	&	\ldots	&	\ldots	\\
	J1132+5722	&	0.0050	&	21.24	$\pm$	0.02	&	14.78	$\pm$	0.05	&	64.90	$\pm$	N/A	& $>$	15.93			&	64.98	$\pm$	4.55	&	13.25	$\pm$	0.21	&	50.68	$\pm$	53.01	&	15.28	$\pm$	0.08	&	64.90	$\pm$	N/A	&	14.92	$\pm$	0.08	&	64.90	$\pm$	N/A	&	\ldots	&	\ldots	\\
	J1132+1411	&	0.0176	&	20.53	$\pm$	0.01	&	15.17	$\pm$	0.03	&	94.65	$\pm$	17.09	& $>$	14.74			&	44.94	$\pm$	14.24	&	13.74	$\pm$	0.11	&	44.94	$\pm$	N/A	&	15.98	$\pm$	0.09	&	44.94	$\pm$	N/A	&	15.28	$\pm$	0.07	&	49.86	$\pm$	4.48	&	15.12	$\pm$	0.07	&	205.51	$\pm$	42.01	\\
	J1144+4012	&	0.1269	&	20.52	$\pm$	0.06	&	15.08	$\pm$	0.09	&	117.71	$\pm$	49.22	& $>$	16.39			&	160.95	$\pm$	6.14	&	\ldots	&	\ldots	&	15.55	$\pm$	0.07	&	124.44	$\pm$	23.85	&	15.64	$\pm$	0.35	&	88.14	$\pm$	54.56	&	14.99	$\pm$	0.07	&	104.82	$\pm$	22.65	\\
	J1148+2546	&	0.0451	&	21.19	$\pm$	0.03	&	14.87	$\pm$	0.03	&	53.27	$\pm$	8.77	& $>$	16.51			&	63.45	$\pm$	5.90	&	13.91	$\pm$	0.09	&	126.75	$\pm$	33.00	&	15.45	$\pm$	0.03	&	63.45	$\pm$	N/A	&	15.62	$\pm$	0.07	&	63.45	$\pm$	N/A	&	14.88	$\pm$	0.08	&	140.63	$\pm$	26.07	\\
	J1150+1501	&	0.0024	&	21.04	$\pm$	0.01	&	15.45	$\pm$	0.26	&	22.27	$\pm$	5.42	& $>$	15.46			&	46.97	$\pm$	4.18	&	13.67	$\pm$	0.14	&	46.93	$\pm$	N/A	&	15.52	$\pm$	0.04	&	46.93	$\pm$	N/A	&	14.90	$\pm$	0.12	&	33.80	$\pm$	6.28	&	\ldots	&	\ldots	\\
	J1225+6109	&	0.0023	&	21.26	$\pm$	0.01	&	\ldots	&	19.88	$\pm$	5.87	& \ldots			&	23.25	$\pm$	0.49	&	13.64	$\pm$	0.14	&	23.25	$\pm$	N/A	&	15.43	$\pm$	0.05	&	40.40	$\pm$	11.28	&	15.18	$\pm$	0.12	&	20.72	$\pm$	2.33	&	\ldots	&	\ldots	\\
	J1253-0312	&	0.0227	&	21.41	$\pm$	0.04	&	15.78	$\pm$	1.18	&	9.85	$\pm$	5.24	& $>$	14.89			&	62.26	$\pm$	19.72	&	13.68	$\pm$	0.06	&	53.45	$\pm$	18.31	&	\ldots	&	\ldots	&	15.60	$\pm$	0.18	&	15.81	$\pm$	2.00	&	14.89	$\pm$	0.03	&	70.36	$\pm$	8.41	\\
	J1314+3452	&	0.0029	&	20.71	$\pm$	0.01	&	14.98	$\pm$	0.04	&	48.38	$\pm$	8.13	& $>$	15.70			&	45.47	$\pm$	4.70	&	13.72	$\pm$	0.05	&	45.46	$\pm$	N/A	&	15.41	$\pm$	0.04	&	39.63	$\pm$	7.13	&	14.78	$\pm$	0.06	&	45.23	$\pm$	6.07	&	\ldots	&	\ldots	\\
	J1359+5726	&	0.0338	&	21.46	$\pm$	0.19	&	\ldots	&	\ldots	& $>$	15.68			&	126.12	$\pm$	3.25	&	13.48	$\pm$	0.14	&	29.37	$\pm$	37.18	&	15.28	$\pm$	0.38	&	126.12	$\pm$	N/A	&	14.89	$\pm$	0.23	&	16.83	$\pm$	3.99	&	\ldots	&	\ldots	\\
	J1416+1223	&	0.1232	&	20.19	$\pm$	0.05	&	15.47	$\pm$	3.45	&	2.88	$\pm$	3.47	& $>$	15.95			&	262.30	$\pm$	7.71	&	13.20	$\pm$	0.79	&	N/A	$\pm$	N/A	&	15.29	$\pm$	0.08	&	111.89	$\pm$	29.37	&	15.30	$\pm$	0.20	&	83.01	$\pm$	59.15	&	\ldots	&	\ldots	\\
	J1418+2102	&	0.0086	&	21.30	$\pm$	0.03	&	14.85	$\pm$	0.04	&	38.22	$\pm$	6.07	& $>$	15.32			&	29.58	$\pm$	6.75	&	13.34	$\pm$	0.10	&	29.00	$\pm$	N/A	&	15.27	$\pm$	0.05	&	28.32	$\pm$	6.74	&	15.15	$\pm$	0.18	&	15.15	$\pm$	2.71	&	\ldots	&	\ldots	\\
	J1444+4237	&	0.0023	&	21.57	$\pm$	0.01	&	14.89	$\pm$	0.06	&	33.51	$\pm$	10.58	& $>$	14.40			&	11.40	$\pm$	3.06	&	13.38	$\pm$	0.20	&	17.11	$\pm$	51.11	&	15.63	$\pm$	0.18	&	11.40	$\pm$	N/A	&	15.51	$\pm$	0.72	&	9.54	$\pm$	3.27	&	\ldots	&	\ldots	\\
	J1448-0110	&	0.0274	&	21.56	$\pm$	0.01	&	15.30	$\pm$	0.04	&	50.50	$\pm$	5.91	& $>$	15.72			&	56.12	$\pm$	6.55	&	13.74	$\pm$	0.17	&	56.12	$\pm$	N/A	&	15.57	$\pm$	0.04	&	47.05	$\pm$	5.19	&	15.44	$\pm$	0.08	&	32.34	$\pm$	2.86	&	15.07	$\pm$	0.04	&	119.11	$\pm$	9.05	\\
	J1521+0759	&	0.0943	&	20.42	$\pm$	0.11	&	14.48	$\pm$	0.80	&	5.00	$\pm$	0.00	& $>$	16.03			&	112.87	$\pm$	-999.00	&	13.57	$\pm$	0.13	&	112.87	$\pm$	N/A	&	\ldots	&	\ldots	&	14.47	$\pm$	0.04	&	112.80	$\pm$	N/A	&	14.99	$\pm$	0.04	&	118.49	$\pm$	17.51	\\
	J1545+0858	&	0.0377	&	21.54	$\pm$	0.04	&	14.69	$\pm$	0.05	&	65.87	$\pm$	N/A	& $>$	15.57			&	65.87	$\pm$	8.44	&	13.54	$\pm$	0.40	&	20.12	$\pm$	41.17	&	\ldots	&	\ldots	&	14.97	$\pm$	0.10	&	73.59	$\pm$	25.10	&	14.75	$\pm$	0.05	&	94.45	$\pm$	13.75	\\
\tableline
\end{tabular}
\end{tiny}	
\end{sidewaystable}

	\begin{table}
	\begin{center}
	\begin{scriptsize}
	\caption{Neutral gas oxygen abundances derived from phosphorus (O(P)/H) and sulfur (O(S)/H), and the average from both P and S (O(P,S)/H), as described in Section~\ref{sec:abund}. Oxygen abundances derived from the saturated \oi~$\lambda$1302 line are provided as lower limits for comparison.}\label{tab:oxygen}
	\begin{tabular}{lcccc}
	\hline\hline
	Target	&	log(O(P)/H)			&	log(O(S)/H)			&	log(O(P,S)/H)			&	log(O/H)	\\
	\hline															
	J0127-0619	&	7.65	$\pm$	0.29	&	7.91	$\pm$	0.27	&	7.80	$\pm$	0.40	& $<$	5.81	\\
	J0144+0453	&	8.62	$\pm$	0.20	&	8.61	$\pm$	0.08	&	8.61	$\pm$	0.21	& $<$	7.87	\\
	J0337-0502	&	6.44	$\pm$	0.15	&	6.72	$\pm$	0.03	&	6.60	$\pm$	0.15	& $<$	5.11	\\
	J0405-3648	&	7.81	$\pm$	0.11	&	7.67	$\pm$	0.11	&	7.74	$\pm$	0.16	&	\ldots	\\
	J0823+2806	&	\ldots			&	7.99	$\pm$	0.04	&	7.99	$\pm$	0.04	& $<$	6.80	\\
	J0934+5514	&	7.29	$\pm$	0.19	&	7.29	$\pm$	0.05	&	7.29	$\pm$	0.20	& $<$	5.26	\\
	J0938+5428	&	8.79	$\pm$	0.12	&	8.49	$\pm$	0.11	&	8.67	$\pm$	0.16	& $<$	6.92	\\
	J0940+2935	&	7.62	$\pm$	0.12	&	7.69	$\pm$	0.38	&	7.66	$\pm$	0.40	&	\ldots	\\
	J0944+3442	&	7.69	$\pm$	0.13	&	\ldots			&	7.69	$\pm$	0.13	& $<$	6.32	\\
	J0944-0038	&	8.13	$\pm$	0.10	&	7.35	$\pm$	0.21	&	7.90	$\pm$	0.23	& $<$	5.82	\\
	J1024+0524	&	8.27	$\pm$	0.18	&	8.15	$\pm$	0.09	&	8.21	$\pm$	0.21	& $<$	7.06	\\
	J1025+3622	&	\ldots			&	8.14	$\pm$	0.21	&	8.14	$\pm$	0.21	& $<$	7.15	\\
	J1044+0353	&	7.28	$\pm$	0.45	&	7.19	$\pm$	0.44	&	7.24	$\pm$	0.63	& $<$	5.36	\\
	J1105+4444	&	7.47	$\pm$	0.12	&	7.59	$\pm$	0.12	&	7.53	$\pm$	0.17	& $<$	5.83	\\
	J1119+5130	&	7.89	$\pm$	0.15	&	7.98	$\pm$	0.12	&	7.94	$\pm$	0.19	& $<$	7.30	\\
	J1129+2034	&	7.99	$\pm$	0.09	&	8.20	$\pm$	0.09	&	8.11	$\pm$	0.12	& $<$	6.67	\\
	J1132+1411	&	8.49	$\pm$	0.11	&	9.02	$\pm$	0.09	&	8.83	$\pm$	0.15	& $<$	6.21	\\
	J1132+5722	&	7.29	$\pm$	0.21	&	7.61	$\pm$	0.09	&	7.48	$\pm$	0.23	& $<$	6.69	\\
	J1144+4012	&	\ldots			&	8.60	$\pm$	0.09	&	8.60	$\pm$	0.09	& $<$	7.87	\\
	J1148+2546	&	8.00	$\pm$	0.10	&	7.83	$\pm$	0.05	&	7.92	$\pm$	0.11	& $<$	7.32	\\
	J1150+1501	&	7.91	$\pm$	0.14	&	8.05	$\pm$	0.04	&	7.98	$\pm$	0.15	& $<$	6.42	\\
	J1225+6109	&	7.66	$\pm$	0.14	&	7.74	$\pm$	0.05	&	7.70	$\pm$	0.15	&	\ldots	\\
	J1253-0312	&	7.55	$\pm$	0.07	&	\ldots			&	7.55	$\pm$	0.07	& $<$	5.48	\\
	J1314+3452	&	8.29	$\pm$	0.05	&	8.27	$\pm$	0.04	&	8.28	$\pm$	0.07	& $<$	6.99	\\
	J1359+5726	&	7.30	$\pm$	0.23	&	7.39	$\pm$	0.42	&	7.35	$\pm$	0.48	& $<$	6.22	\\
	J1416+1223	&	8.29	$\pm$	0.79	&	8.67	$\pm$	0.10	&	8.52	$\pm$	0.80	& $<$	7.76	\\
	J1418+2102	&	7.32	$\pm$	0.10	&	7.54	$\pm$	0.06	&	7.45	$\pm$	0.12	& $<$	6.02	\\
	J1444+4237	&	7.09	$\pm$	0.20	&	7.63	$\pm$	0.18	&	7.44	$\pm$	0.27	& $<$	4.83	\\
	J1448-0110	&	7.46	$\pm$	0.17	&	7.58	$\pm$	0.04	&	7.53	$\pm$	0.17	& $<$	6.16	\\
	J1521+0759	&	8.43	$\pm$	0.17	&	\ldots			&	8.43	$\pm$	0.17	& $<$	7.61	\\
	J1545+0858	&	7.28	$\pm$	0.40	&	\ldots			&	7.28	$\pm$	0.40	& $<$	6.03	\\
	\tableline
	\end{tabular}
	\end{scriptsize}
	\end{center}
	\end{table}															
\section{Ionization Correction Factors}\label{sec:AICFs}
In this section we describe the derivation of neutral-gas ICFs, which are used to accurately determine the abundance of metals in the neutral ISM gas, the ionization state of the gas along the line of sight, as described in Section~\ref{sec:ICFs}. The  neutral-gas ICFs only apply to the column density measurements described in Section~\ref{sec:method} and should not be confused with ICFs used in ionized-gas chemical abundance measurements (described in Section~\ref{sec:ion_abund}).

Neutral‐phase ICFs (ICF$_{\rm neu.}$) account for unseen higher ionization states of metal ions that coincide with the H$^0$–dominated gas by adding the estimated column density of ions like $X^{i+1}$ to the observed dominant state $X^i$.
In practice, they correct the final abundance of the neutral gas by assuming $N(X)=N(X^i)+\sum_{j>0} N(X^{i+j})$, where $\sum_{j>0} N(X^{i+j})$ represents the species that are present in the neutral region but cannot be directly measured.

A ICF$_{\rm neu.}$ can be found by comparing the relative amount of $N(X^i)$ with $N(X^{i+1})$ (Fig.~\ref{fig:schematic}). Most of the elements within the neutral gas of galaxies with N$_{\rm HI} > $10$^{19}$ cm$^{-2}$ are either in the neutral (e.g., \ion{Si}{1}) or singly ionized (e.g., \sIii) phase. As such, the correction required to account for elements in higher ionization states (e.g., \sIiii) is relatively small. 
However, for \Ni, accounting for higher ionization species within the neutral gas (i.e., \nii, as it lacks a strong transition in the UV) is essential and can be up to 0.6~dex at high values of N$_{\rm HI}$ \citep{Hernandez:2020}. 

In addition to the neutral-phase ICF, we apply a second correction, ICF$_{\rm ion.}$ to remove the contribution of the dominant ionization state $N(X^i)$ that resides in ionized gas along the line of sight.
Although \ion{H}{2} regions are often approximated as stratified or shell-like, the thickness of each ionic layer varies with excitation energy (see Fig.~\ref{fig:schematic}).
As a result, while the neutral and ionized phases can be cleanly separated using hydrogen, a non-negligible fraction of the dominant metal ion in the neutral phase is still present in ionized hydrogen gas $[N(X^i)_{\rm HII}]$. This contribution must be subtracted from the total column density attributed to the neutral hydrogen dominated regions $[N(X^i)_{\rm HI}]$
For example, \ion{Si}{2} (IP = 16.34 eV) arises in both neutral and ionized gas regions. 
While it is predominantly a tracer of the neutral gas, we must correct for the fraction of the \sIii\ found in the \hii\ regions. 
Consequently, ICF$_{\rm ion.}$ has the net effect of reducing the inferred neutral-gas abundance of the affected species.

\begin{table}												
\begin{center}												
\begin{scriptsize}												
\caption{Input parameters for adhoc \textsc{cloudy} models used to derive the ICFs shown in Table~\ref{tab:ICFs} and discussed in Section~\ref{sec:ICFs}.}\label{tab:model_info}
\begin{tabular}{lccccc}												
	Target 	&	\feiii/\feii	&	$Z$	&	$F_{1500}$	&	$L_{UV}$	&	$N$(\hi)	\\
	\hline\hline											
	J0127-0619	&	-0.27	&	0.10	&	4.04	&	2.56E+38	&	21.43	\\
	J0144+0453	&	-0.27	&	0.12	&	1.87	&	1.18E+38	&	20.18	\\
	J0337-0502	&	-1.08	&	0.06	&	7.99	&	3.22E+39	&	21.81	\\
	J0405-3648	&	-0.27	&	0.02	&	0.96	&	1.39E+37	&	20.80	\\
	J0823+2806	&	-1.11	&	0.39	&	3.85	&	2.01E+40	&	21.61	\\
	J0934+5514	&	0.15	&	0.02	&	15.10	&	2.60E+38	&	21.24	\\
	J0938+5428	&	-0.27	&	0.36	&	3.56	&	9.41E+40	&	20.34	\\
	J0940+2935	&	-0.27	&	0.09	&	1.45	&	1.11E+37	&	21.26	\\
	J0944-0038	&	-0.27	&	0.14	&	1.40	&	7.39E+37	&	21.67	\\
	J0944+3442	&	0.65	&	0.09	&	0.69	&	6.25E+38	&	21.51	\\
	J1024+0524	&	0.67	&	0.14	&	4.50	&	1.13E+40	&	20.58	\\
	J1025+3622	&	-0.97	&	0.28	&	1.81	&	7.59E+40	&	20.69	\\
	J1044+0353	&	-1.23	&	0.06	&	1.70	&	6.15E+38	&	21.85	\\
	J1105+4444	&	-0.73	&	0.35	&	4.68	&	4.84E+39	&	21.91	\\
	J1119+5130	&	-0.27	&	0.08	&	2.63	&	1.26E+38	&	20.77	\\
	J1129+2034	&	-0.27	&	0.39	&	1.87	&	9.87E+37	&	21.11	\\
	J1132+5722	&	-0.27	&	0.08	&	2.57	&	1.63E+38	&	21.24	\\
	J1132+1411	&	-0.16	&	0.36	&	1.75	&	1.21E+39	&	20.53	\\
	J1144+4012	&	-0.65	&	0.55	&	1.20	&	5.07E+40	&	20.52	\\
	J1148+2546	&	-0.75	&	0.18	&	2.07	&	9.81E+39	&	21.19	\\
	J1150+1501	&	-0.27	&	0.28	&	12.60	&	1.82E+38	&	21.04	\\
	J1225+6109	&	-0.27	&	0.19	&	9.50	&	1.38E+38	&	21.26	\\
	J1253-0312	&	-0.71	&	0.23	&	9.11	&	1.07E+40	&	21.41	\\
	J1314+3452	&	-0.27	&	0.37	&	3.72	&	7.52E+37	&	20.71	\\
	J1359+5726	&	-0.27	&	0.19	&	6.34	&	1.66E+40	&	21.46	\\
	J1416+1223	&	-0.27	&	0.69	&	2.62	&	1.04E+41	&	20.19	\\
	J1418+2102	&	-0.27	&	0.11	&	1.17	&	1.92E+38	&	21.30	\\
	J1444+4237	&	-0.27	&	0.09	&	2.08	&	3.01E+37	&	21.57	\\
	J1448-0110	&	-0.37	&	0.28	&	4.08	&	6.91E+39	&	21.56	\\
	J1521+0759	&	0.52	&	0.42	&	3.52	&	7.86E+40	&	20.42	\\
	J1545+0858	&	-0.22	&	0.11	&	4.37	&	1.44E+40	&	21.54	\\									
\tableline												
\end{tabular}
\end{scriptsize}												
\end{center}												
\end{table}												

In order to obtain an accurate measure of ICF$_{neutral}$ and ICF$_{ionized}$, ad-hoc photoionization models were made for each of the CLASSY galaxies using \textsc{Cloudy} \citep{ferland17}. Simple spherical models were created using the metallicity of the source ($Z$), UV luminosity ($L_{\rm UV}$), and \hi\ column densities ($\log[N$(\hi)]). Since the effective temperature ($T_{\rm eff}$) of the stellar ionizing source was found to have a negligible effect on the ICF grids calculated by \citet{Hernandez:2020}, all models utilized a blackbody input spectrum fixed at $T_{\rm eff}=4\times10^4$~K. Each model was initially run for a grid of hydrogen volume densities ($n_{\rm H}$) with a range of $-3<\log(n_{\rm H}/cm^{-3})<3$. We then constrained the actual $n_{\rm H}$ value by comparing each model's level of ionization as a function of $n_{\rm H}$ against the galaxy's observed level of ionization level of the neutral gas derived from $N$(\feiii)/$N$(\feii). 

The ISM Fe$^{+2}$/Fe$^+$ absorption line ratio provides the cleanest way to measure the ionization state of the neutral gas due to the low oscillator strengths and multiple transitions involved. 

When the ionization ratio was unobtainable for a galaxy (15 galaxies in total, mostly due to lack of coverage of \feiii~$\lambda$1122), we utilized the average \feiii/\feii\ ratio from the sample of \feiii/\feii$=-0.27$. All model input parameters are given in Table~\ref{tab:model_info}.

The model of a given element was run twice, each with a different stopping criteria: 
(i) when the measured \hi\ column density was reached (essentially the edge of the neutral cloud) and 
(ii) when the amount of ionized hydrogen was less than the neutral hydrogen (essentially the edge of the photo-dissociation region). 

Since the models include neutral gas, cosmic rays are included in each run. The ICFs required for each element's column density within the neutral gas ($N(X)_{\rm HI}$) are then derived by sampling the column densities of the dominant neutral species $N(X^i)$ and the more highly ionized species $N(X^{i+1})$ from each of the two scenarios - firstly at the edge of the cloud ($N(X)_{\rm HI+HII}$) and secondly at the edge of the PDR ($N(X)_{\rm HII}$) such that:
\begin{equation}
  N(X)_{\rm HI} = N(X)_{\rm HI+HII} - N(X)_{\rm HII}.  
\end{equation}
We then obtained:
\begin{align}
      {\rm ICF}_{\rm ion.}&= \log[N(X^i)_{\rm HII}] \\
  &= \log[N(X^i)_{\rm HI+HII}] - \log[N(X^i)_{\rm HI}] \\
  \text{and:} & \\
 {\rm ICF}(X^i)_{\rm neu.} &= \log[\frac{N(X^i_{\rm HI}) + N(X^{i+1}_{\rm HI})}{N(X^i_{\rm HI})}].
\end{align}
The total ICF can then be defined as:
\begin{align}  
  {\rm ICF}_{\rm tot.} &= {\rm ICF}_{\rm neu.} - {\rm ICF}_{\rm ion.}  \\
  \log[N(X)_{\rm ICF} ] &= \log[N(X)] + ICF_{\rm tot.} \\
\end{align}
where $\log[N(X)]$ corresponds to the measured column density of element $X$. 

\begin{figure*}[ht!]
\centering

% ---------- Row 1 ----------
\begin{minipage}[b]{0.475\textwidth}
    \centering
    \includegraphics[width=\textwidth,page=2]{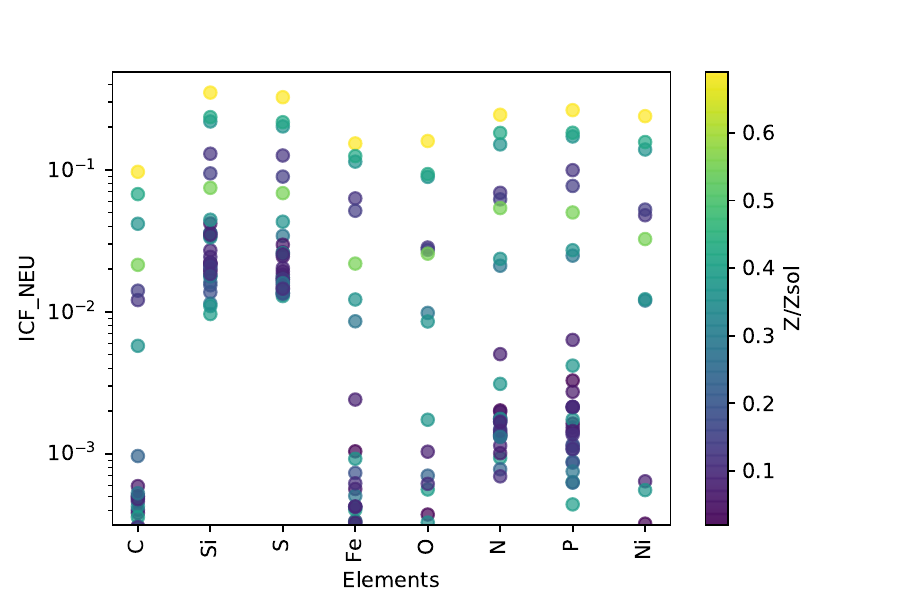}
    \textbf{(a) ICF$_{\rm neutral}$ vs. metallicity}
\end{minipage}%
\hfill
\begin{minipage}[b]{0.475\textwidth}
    \centering
    \includegraphics[width=\textwidth,page=2]{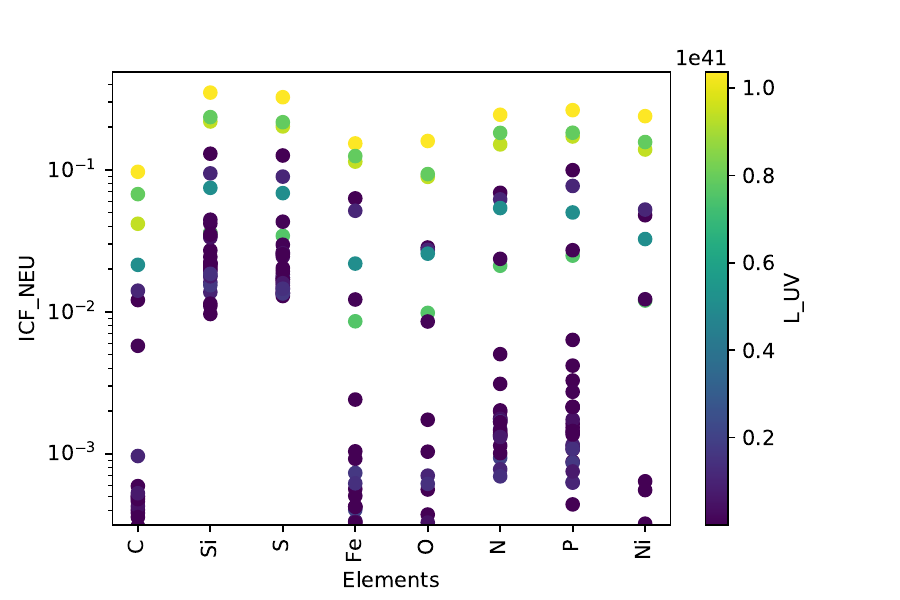}
    \textbf{(b) ICF$_{\rm neutral}$ vs. $L_{UV}$}
\end{minipage}

\vspace{1em}

% ---------- Row 2 ----------
\begin{minipage}[b]{0.475\textwidth}
    \centering
    \includegraphics[width=\textwidth,page=2]{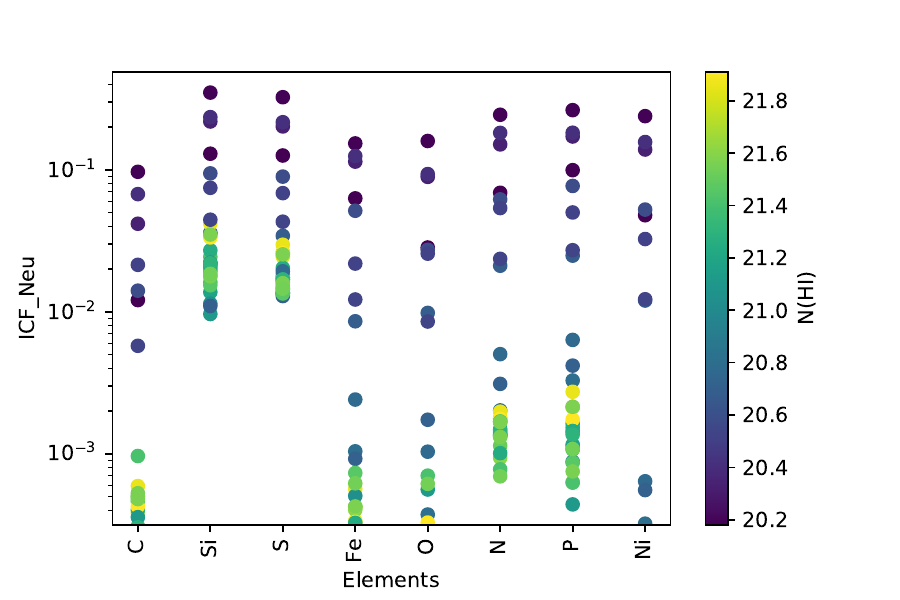}
    \textbf{(c) ICF$_{\rm neutral}$ vs. $N$(\hi)}
\end{minipage}%
\hfill
\begin{minipage}[b]{0.475\textwidth}
    \centering
    \includegraphics[width=\textwidth,page=2]{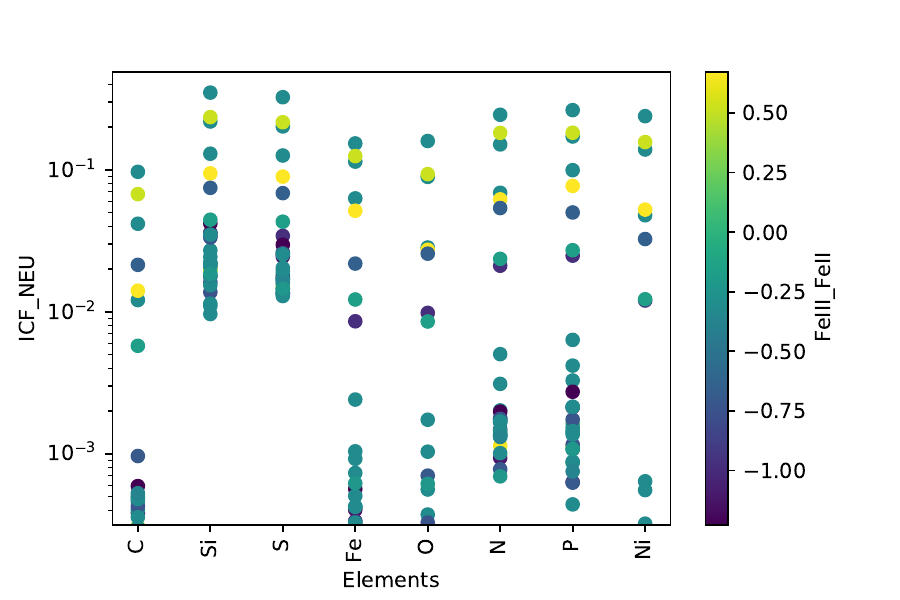}
    \textbf{(d) ICF$_{\rm neutral}$ vs. ionization factor}
\end{minipage}

\caption{\small ICF$_{\rm neutral}$ for each species, color-coded by the Cloudy model input parameters shown in Table~\ref{tab:model_info}, as described in Section~\ref{sec:AICFs}.}
\label{fig:ICFneu}
\end{figure*}

\begin{figure*}[ht!]
\centering

% ---------- Row 1 ----------
\begin{minipage}[b]{0.475\textwidth}
    \centering
    \includegraphics[width=\textwidth,page=3]{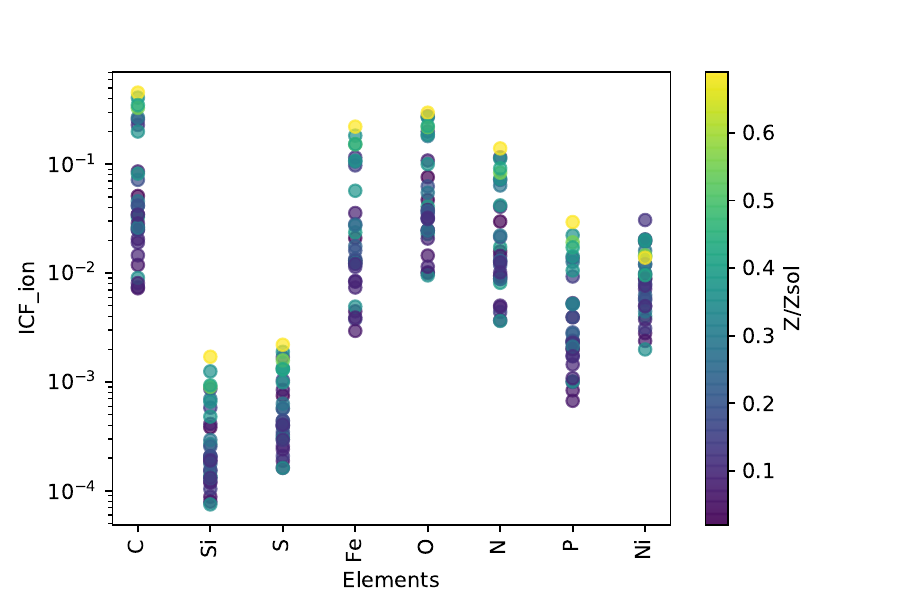}
    \textbf{(a) ICF$_{\rm ionized}$ vs. metallicity}
\end{minipage}%
\hfill
\begin{minipage}[b]{0.475\textwidth}
    \centering
    \includegraphics[width=\textwidth,page=3]{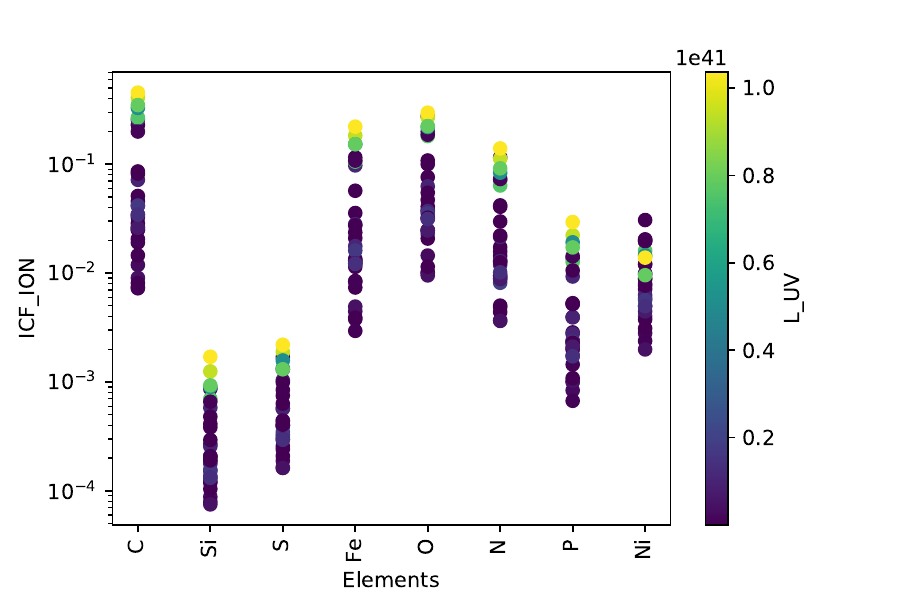}
    \textbf{(b) ICF$_{\rm ionized}$ vs. $L_{UV}$}
\end{minipage}

\vspace{1em}

% ---------- Row 2 ----------
\begin{minipage}[b]{0.475\textwidth}
    \centering
    \includegraphics[width=\textwidth,page=3]{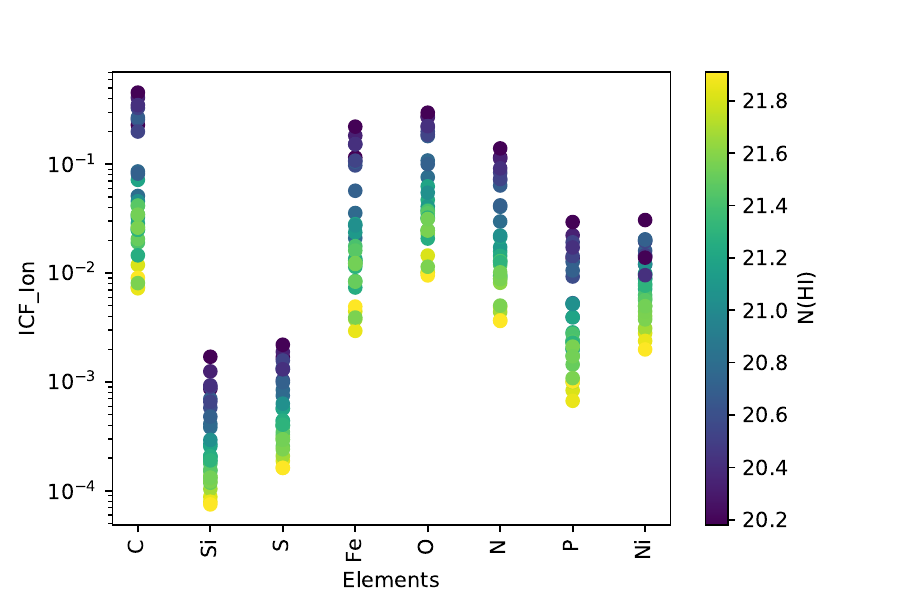}
    \textbf{(c) ICF$_{\rm ionized}$ vs. $N$(\hi)}
\end{minipage}%
\hfill
\begin{minipage}[b]{0.475\textwidth}
    \centering
    \includegraphics[width=\textwidth,page=3]{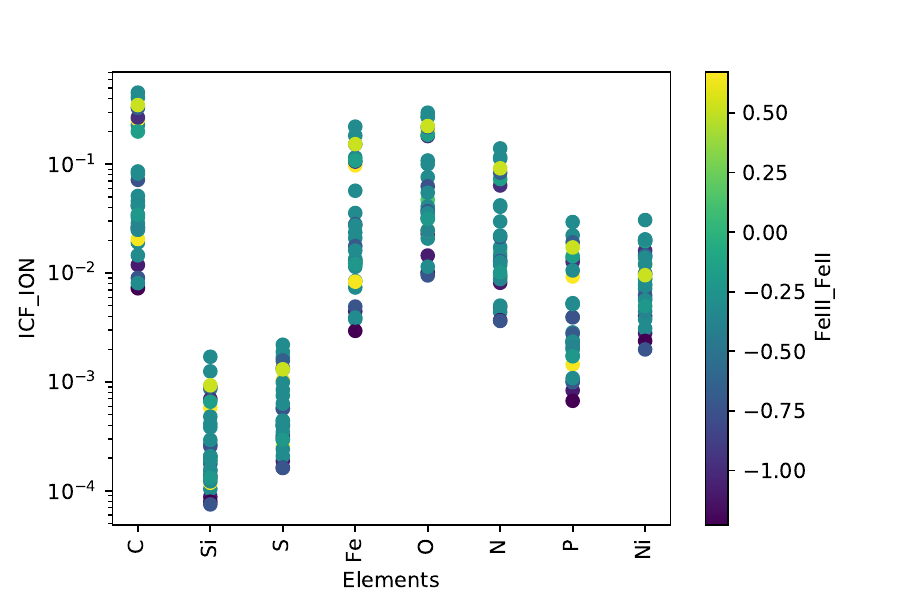}
    \textbf{(d) ICF$_{\rm ionized}$ vs. ionization factor}
\end{minipage}

\caption{\small ICF$_{\rm ionized}$ for each species, color-coded by the Cloudy model input parameters shown in Table~\ref{tab:model_info}, as described in Section~\ref{sec:AICFs}.}
\label{fig:ICFion}
\end{figure*}

We present the individual ICF components, ICF$_{\rm neu.}$ and ICF$_{\rm ion.}$ color-coded by $N$(\hi), UV luminosity ($L_{UV}$), metallicity, and ionization factor (\feiii/\feii) in Figures~\ref{fig:ICFneu} and ~\ref{fig:ICFion}, according to the input parameters shown in Table~\ref{tab:model_info}.  As seen in Figure~\ref{fig:ICFneu}, ICF$_{\rm neu.}$ are sensitive to all four parameters within the spherical model, with larger values (i.e., larger amounts of higher ionization gas to account for) accompanying models with higher metallicities, higher luminosities and lower \hi\ column densities. There is a weaker dependence of ICF$_{\rm neu.}$ on the ionization fraction utilized. For ICF$_{\rm ion.}$, there is a similar dependency in that the corrections for the amount of the neutral gas' dominant ion in the ionized gas is largest at high metallicities, high luminosities, and low \hi\ column densities (as seen in Figure~\ref{fig:ICFion}). 

	\begin{table*}											
	\begin{center}											
	\begin{scriptsize}											
	\caption{Total ICFs applied to the neutral gas column densities.}\label{tab:ICFs}
	\begin{tabular}{lccccc}											
	\hline\hline											
	Target	&	Nitrogen	&	Oxygen	&	Phosphorus	&	Sulfur	&	Iron	\\
	J0127-0619	&	2.42E-02	&	1.80E-02	&	-2.45E-02	&	-8.19E-03	&	-3.33E-04	\\
	J0144+0453	&	1.29E-01	&	1.24E-01	&	-2.47E-01	&	-4.68E-02	&	8.56E-02	\\
	J0337-0502	&	3.41E-02	&	2.43E-02	&	-1.43E-02	&	-3.14E-03	&	1.28E-03	\\
	J0405-3648	&	2.15E-02	&	1.67E-02	&	-7.57E-02	&	-2.77E-02	&	-6.56E-04	\\
	J0823+2806	&	1.77E-02	&	1.70E-02	&	-2.29E-02	&	-7.21E-03	&	-1.70E-03	\\
	J0934+5514	&	1.92E-02	&	1.43E-02	&	-4.70E-02	&	-1.44E-02	&	-3.98E-04	\\
	J0938+5428	&	2.18E-01	&	2.00E-01	&	-1.80E-01	&	3.86E-02	&	1.49E-01	\\
	J0940+2935	&	2.69E-02	&	1.99E-02	&	-2.05E-02	&	-7.99E-03	&	-4.65E-04	\\
	J0944-0038	&	3.58E-02	&	2.60E-02	&	-9.88E-03	&	-2.62E-03	&	1.13E-03	\\
	J0944+3442	&	1.98E-02	&	1.51E-02	&	-2.45E-02	&	-7.76E-03	&	-3.13E-04	\\
	J1024+0524	&	9.38E-02	&	8.85E-02	&	-1.72E-01	&	-9.92E-03	&	6.77E-02	\\
	J1025+3622	&	3.53E-02	&	3.29E-02	&	-1.71E-01	&	-4.26E-02	&	1.22E-02	\\
	J1044+0353	&	4.18E-02	&	2.95E-02	&	-9.76E-03	&	-1.71E-03	&	2.06E-03	\\
	J1105+4444	&	3.33E-02	&	2.57E-02	&	-9.19E-03	&	-1.87E-03	&	7.30E-04	\\
	J1119+5130	&	2.17E-02	&	1.84E-02	&	-1.07E-01	&	-3.56E-02	&	1.04E-03	\\
	J1129+2034	&	9.36E-03	&	1.31E-02	&	-4.03E-02	&	-1.60E-02	&	-4.77E-03	\\
	J1132+5722	&	2.13E-02	&	1.58E-02	&	-3.75E-02	&	-1.34E-02	&	-9.54E-04	\\
	J1132+1411	&	4.39E-02	&	4.18E-02	&	-1.78E-01	&	-4.95E-02	&	1.28E-02	\\
	J1144+4012	&	7.37E-02	&	6.69E-02	&	-1.94E-01	&	-2.98E-02	&	3.09E-02	\\
	J1148+2546	&	1.35E-02	&	1.28E-02	&	-6.26E-02	&	-2.00E-02	&	-2.77E-03	\\
	J1150+1501	&	1.11E-02	&	1.29E-02	&	-5.47E-02	&	-2.04E-02	&	-4.31E-03	\\
	J1225+6109	&	2.05E-02	&	1.64E-02	&	-3.15E-02	&	-1.17E-02	&	-1.78E-03	\\
	J1253-0312	&	1.60E-02	&	1.40E-02	&	-3.64E-02	&	-1.21E-02	&	-2.14E-03	\\
	J1314+3452	&	1.05E-02	&	1.19E-02	&	-9.82E-02	&	-3.87E-02	&	-6.40E-03	\\
	J1359+5726	&	1.52E-02	&	1.31E-02	&	-3.54E-02	&	-1.02E-02	&	-1.48E-03	\\
	J1416+1223	&	3.48E-01	&	3.23E-01	&	-1.38E-01	&	1.05E-01	&	2.34E-01	\\
	J1418+2102	&	2.18E-02	&	1.65E-02	&	-3.23E-02	&	-1.11E-02	&	-9.27E-04	\\
	J1444+4237	&	3.49E-02	&	2.51E-02	&	-1.12E-02	&	-3.36E-03	&	1.06E-03	\\
	J1448-0110	&	1.76E-02	&	1.56E-02	&	-2.46E-02	&	-7.50E-03	&	-1.36E-03	\\
	J1521+0759	&	2.34E-01	&	2.15E-01	&	-1.31E-01	&	9.05E-02	&	1.65E-01	\\
	J1545+0858	&	1.84E-02	&	1.42E-02	&	-3.07E-02	&	-9.47E-03	&	-6.41E-04	\\									
	\tableline											
	\end{tabular}
	\end{scriptsize}											
	\end{center}											
	\end{table*}											

\section{Stellar Population Ages} \label{sec:SPS_Age}
Two sets of stellar population ages for the CLASSY galaxies were provided in \citetalias{mingozzi:2022}, both obtained via UV continuum fitting using simple stellar population models. These ages correspond to the median age of the best-fitting model, weighted by the fraction of contributed light to that model (i.e, the light-weighted UV-based stellar population age, age$_{UV,L}$). As such, the resultant ages (ranging from $\sim$1-21~Myr, with a mean of 8$\pm$5\,Myr) represent the most recent burst of star-formation and thus provide only limited temporal information regarding chemical mixing on short timescales. In order to sample longer chemical-evolutionary timescales sampled by our data, we instead use the mass-weighted ages (age$_{UV,M}$)), which range from $\sim$1-36\,Myr with a mean of 15$\pm$12\,Myr, that also sample the older, more massive stellar populations. 

For comparison, we additionally compute the mass-weighted ages derived from fitting the optical continuum, age$_{opt,L}$, as detailed in \citetalias{mingozzi:2022} and \citetalias{arellano-cordova:2022}.  These instead sample \textit{only} the older stellar populations in each system and range from $\sim$24\,Myr - $\sim$12\,Gyr with a mean of $\sim$6$\pm$3\,Gyr. The methodology used to derive the optical-based ages is described in the following paragraphs.  We provide all of the stellar population ages in Table~\ref{tab:ages}. 

Optical stellar continuum fits were performed with \textsc{starlight20}, a stellar population synthesis (SPS) code developed by \citet{fernandes05} to analyze optical spectroscopic data and using the stellar models of \citet{bruzual03}, the extinction law of \citet{cardelli89} and  and the IMF of \citet{chabrier03}, with 25 stellar ages ranging between 1 Myr and 18 Gyr. 
We utilize the results of these optical fits to provide a comparison with the stellar population ages derived from SPS fitting to the UV stellar continuum (also described in \citetalias{mingozzi:2022}) because the optical continuum samples the older stellar populations and can thus sample much longer enrichment timescales.

In order to derive the optical stellar population ages, we use the \textsc{Starlight} outputs of the SPS models on the optical spectra in hand for our sample (SDSS, MUSE, MMT , KCWI and VIMOS spectra). \textsc{Starlight} uses a combination of spectral empirical evolution models of N simple stellar populations to fit the stellar continuum of an observed spectrum (O$_{\lambda}$). The emission lines, bad pixels and sky residuals are masked during the fitting. For the CLASSY optical sample we use a combination of N=150 simple stellar populations, each characterized by different combinations of ages and metallicities, to model each of the O$_{\lambda}$. In order to get the best model spectrum (M$_{\lambda}$), \textsc{Starlight}  employs a combination of simulated annealing and a Metropolis scheme. This minimization performs successive explorations of the parameter space, geometrically increasing the light weights ($x_{j}$) of the different stellar populations until it gets a minimum $\chi^{2}$ error/residual between O$_{\lambda}$ and M$_{\lambda}$ spectra along the wavelength range. As a refinement stage, the complete loop is re-executed discarding populations with $x_{j}$ =0. 

We use the individual \textsc{Starlight} outputs containing the statistical light ($x_{j}$) and mass ($m_{j}$) weights of the used stellar populations, that represent the light and mass contribution of each population to the final best spectral model. The $m_{j}$ values correspond to the current mass fraction of each stellar population at the end of the modeling run. The stellar ages and metallicities can be computed as the light-weighted or mass-weighted average contribution of all the stellar populations used in the best model, as follows: 

\begin{equation}
    S = \sum_{j}^{N} S_{j} \times W_{j},  
\end{equation}
where $S$ corresponds to the age or metallicity weighted average, $S_{j}$ corresponds to the age or metallicity of each stellar population, and N is the number of stellar populations used.
$W_{j}$ is the light ($W_{Lj}$) or mass ($W_{Mj}$) fraction of the stellar populations with the same age ($a$), defined as:

\begin{equation}
    W_{j} = \frac{\sum_{j}^{N} c_{j}[a]}{\sum_{j}^{N} c_{j}}, 
\end{equation}
where $c_{j}$ corresponds to $x_{j}$ or $m_{j}$, to compute the light and mass weighted ages respectively. 

	\begin{table*}
	\begin{center}
	\begin{scriptsize}
	\caption{Mass- and light-weighted ages of the stellar populations derived from UV and optical spectra.}\label{tab:ages}
	\begin{tabular}{lcccc}
	\hline\hline
		&	\multicolumn{2}{c}{UV}	&							\multicolumn{2}{c}{Optical}							\\
	Target	&	log(Age/Myr)$_M$			&	log(Age/Myr)$_L$			&	log(Age/Gyr)$_M$			&	log(Age/Gyr)$_L$			\\
	\hline																	\\
	J0127-0619	&	1.81	$\pm$	0.33	&	1.90	$\pm$	0.33	&	5.62	$\pm$	0.40	&	0.64	$\pm$	0.06	\\
	J0144+0453	&	8.95	$\pm$	1.19	&	7.05	$\pm$	0.88	&	0.34	$\pm$	0.05	&	0.12	$\pm$	0.02	\\
	J0337-0502	&	16.56	$\pm$	2.49	&	5.74	$\pm$	0.57	&	10.36	$\pm$	0.72	&	1.45	$\pm$	0.11	\\
	J0405-3648	&	36.11	$\pm$	5.96	&	21.37	$\pm$	3.36	&	0.89	$\pm$	0.10	&	0.29	$\pm$	0.03	\\
	J0823+2806	&	23.31	$\pm$	3.81	&	6.87	$\pm$	1.02	&	13.28	$\pm$	0.83	&	1.41	$\pm$	0.09	\\
	J0934+5514	&	28.18	$\pm$	4.59	&	13.48	$\pm$	2.01	&	0.08	$\pm$	0.01	&	0.04	$\pm$	0.01	\\
	J0938+5428	&	6.93	$\pm$	0.74	&	5.79	$\pm$	0.64	&	9.41	$\pm$	0.68	&	0.46	$\pm$	0.03	\\
	J0940+2935	&	31.95	$\pm$	5.25	&	15.57	$\pm$	2.41	&	4.63	$\pm$	0.50	&	0.58	$\pm$	0.07	\\
	J0944+3442	&	32.29	$\pm$	5.33	&	14.06	$\pm$	2.18	&	6.34	$\pm$	0.49	&	1.03	$\pm$	0.09	\\
	J0944-0038	&	34.23	$\pm$	5.70	&	13.51	$\pm$	2.18	&	12.44	$\pm$	0.70	&	1.38	$\pm$	0.08	\\
	J1024+0524	&	5.65	$\pm$	0.66	&	4.91	$\pm$	0.54	&	12.71	$\pm$	0.75	&	1.06	$\pm$	0.07	\\
	J1025+3622	&	7.71	$\pm$	0.93	&	6.82	$\pm$	0.83	&	14.03	$\pm$	0.88	&	1.21	$\pm$	0.08	\\
	J1044+0353	&	32.22	$\pm$	5.33	&	10.62	$\pm$	1.58	&	12.92	$\pm$	0.88	&	1.31	$\pm$	0.09	\\
	J1105+4444	&	26.39	$\pm$	4.41	&	7.41	$\pm$	1.27	&	6.22	$\pm$	0.48	&	0.75	$\pm$	0.06	\\
	J1119+5130	&	27.05	$\pm$	4.38	&	12.64	$\pm$	1.83	&	14.30	$\pm$	0.77	&	1.35	$\pm$	0.07	\\
	J1129+2034	&	3.14	$\pm$	0.42	&	2.56	$\pm$	0.36	&	12.45	$\pm$	0.76	&	1.74	$\pm$	0.12	\\
	J1132+1411	&	12.24	$\pm$	1.37	&	8.20	$\pm$	0.93	&	5.94	$\pm$	0.54	&	0.73	$\pm$	0.07	\\
	J1132+5722	&	18.50	$\pm$	1.56	&	13.85	$\pm$	1.22	&	14.88	$\pm$	0.83	&	0.89	$\pm$	0.05	\\
	J1144+4012	&	13.09	$\pm$	1.36	&	10.84	$\pm$	1.10	&	7.42	$\pm$	0.65	&	0.74	$\pm$	0.07	\\
	J1148+2546	&	6.85	$\pm$	0.84	&	5.75	$\pm$	0.70	&	12.69	$\pm$	0.77	&	1.39	$\pm$	0.09	\\
	J1150+1501	&	17.38	$\pm$	2.71	&	6.44	$\pm$	0.81	&	11.08	$\pm$	0.77	&	0.83	$\pm$	0.06	\\
	J1225+6109	&	23.51	$\pm$	3.82	&	6.96	$\pm$	0.95	&	0.56	$\pm$	0.07	&	0.09	$\pm$	0.01	\\
	J1253-0312	&	3.73	$\pm$	0.33	&	3.75	$\pm$	0.33	&	11.46	$\pm$	0.76	&	1.25	$\pm$	0.09	\\
	J1314+3452	&	2.25	$\pm$	0.33	&	2.31	$\pm$	0.33	&	9.65	$\pm$	0.70	&	1.35	$\pm$	0.10	\\
	J1359+5726	&	34.55	$\pm$	5.72	&	14.59	$\pm$	2.25	&	13.62	$\pm$	0.86	&	1.73	$\pm$	0.11	\\
	J1416+1223	&	7.80	$\pm$	0.56	&	6.06	$\pm$	0.47	&	2.13	$\pm$	0.27	&	0.16	$\pm$	0.02	\\
	J1418+2102	&	1.55	$\pm$	0.33	&	1.64	$\pm$	0.33	&	13.68	$\pm$	0.61	&	0.54	$\pm$	0.03	\\
	J1444+4237	&	34.69	$\pm$	5.74	&	15.44	$\pm$	2.36	&	8.96	$\pm$	0.52	&	0.52	$\pm$	0.04	\\
	J1448-0110	&	3.00	$\pm$	0.33	&	3.10	$\pm$	0.33	&	8.20	$\pm$	0.54	&	0.78	$\pm$	0.06	\\
	J1521+0759	&	9.30	$\pm$	0.68	&	8.46	$\pm$	0.64	&	11.47	$\pm$	0.78	&	0.91	$\pm$	0.07	\\
	J1545+0858	&	3.47	$\pm$	0.33	&	3.55	$\pm$	0.33	&	13.73	$\pm$	0.78	&	1.32	$\pm$	0.08	\\	
	\tableline
	\end{tabular}
	\end{scriptsize}
	\end{center}
	\end{table*}																	
\end{document}